\documentclass[11pt]{article}
\pdfoutput=1 
\usepackage{shorthand}

\usepackage{mathtools}
\usepackage{booktabs}
\usepackage[english]{babel}
\usepackage{amsmath,amssymb,amsbsy,amstext, amsthm, simplewick, amsfonts,braket}
\usepackage{graphicx}
\usepackage[small]{caption}
\usepackage{siunitx}
\usepackage{upgreek}
\usepackage{framed}
\usepackage{wrapfig}
\usepackage{multirow}
\usepackage{bbm}
\usepackage[numbers,sort&compress]{natbib}
\usepackage[svgnames,dvipsnames,x11names]{xcolor}
\usepackage[utf8x]{inputenc}
\usepackage{selinput}
\usepackage{bm}
\usepackage{float}
\usepackage{dsfont}
\usepackage{caption}
\usepackage{subcaption}
\usepackage{sidecap}
\usepackage{longtable}
\usepackage{anyfontsize}

\usepackage{epstopdf}
\usepackage{cancel}
\usepackage{tcolorbox}
\usepackage{latexsym,amsmath,amssymb,epsfig}
\usepackage{braket}
\usepackage{tensor}
\usepackage{tocloft}

\usepackage{ytableau}
\ytableausetup{centertableaux,boxsize=.8em}
 \usepackage{genyoungtabtikz}
\Yboxdim{11pt} 
\Ylinethick{.75pt}
\newcommand*{\medotimes}{\raisebox{-0.05ex}{\scalebox{1.25}{$\otimes$}}}
\newcommand*{\medoplus}{\raisebox{-0.05ex}{\scalebox{1.25}{$\oplus$}}}

\usepackage{tcolorbox}
\definecolor{greyish2}{rgb}{.96,.96,.96}

\def\xyma{\xymatrix@M.7em}
\def\xymas{\xymatrix@M.1em}

\newcommand{\Comment}[1]{{}}
\definecolor{darkblue}{rgb}{0.15,0.35,0.55}
\definecolor{reddish}{rgb}{0.65, 0.2, 0.2}
\definecolor{darkgreen}{RGB}{50,150,0}
\definecolor{greyish2}{rgb}{.96,.96,.96}
\usepackage[linktocpage=true]{hyperref}
\hypersetup{
colorlinks=true,
citecolor=darkblue,
linkcolor=reddish,
urlcolor=darkblue,
pdfauthor={},
pdftitle={},
pdfsubject={}
}

\DeclareFontFamily{OT1}{rsfs10}{}
\DeclareFontShape{OT1}{rsfs10}{m}{n}{ <-> rsfs10 }{}
\DeclareMathAlphabet{\mathscript}{OT1}{rsfs10}{m}{n}

\def\gsim{ \lower .75ex \hbox{$\sim$} \llap{\raise .27ex \hbox{$>$}} }
\def\lsim{ \lower .75ex \hbox{$\sim$} \llap{\raise .27ex \hbox{$<$}} }
\def\be{\begin{equation}}
\def\ee{\end{equation}}
\def\bea{\begin{eqnarray}}
\def\eea{\end{eqnarray}}
\def\KL{K\"all\'en--Lehmann }

\newcommand{\rd}{{\rm d}}

\newcommand{\tr}{\text{tr}}

\usepackage[a4paper,margin=1in]{geometry}

\usepackage{tikz}
\usetikzlibrary{decorations}
\pgfdeclaredecoration{complete sines}{initial}
{
    \state{initial}[
        width=+0pt,
        next state=upsine,
        persistent precomputation={\pgfmathsetmacro\matchinglength{
            \pgfdecoratedinputsegmentlength / int(\pgfdecoratedinputsegmentlength/\pgfdecorationsegmentlength)}
            \setlength{\pgfdecorationsegmentlength}{\matchinglength pt}
        }] {}
    \state{upsine}[width=\pgfdecorationsegmentlength,next state=downsine]{
        \pgfpathsine{\pgfpoint{0.25\pgfdecorationsegmentlength}{0.5\pgfdecorationsegmentamplitude}}
        \pgfpathcosine{\pgfpoint{0.25\pgfdecorationsegmentlength}{-0.5\pgfdecorationsegmentamplitude}}
    }
    \state{downsine}[width=\pgfdecorationsegmentlength,next state=upsine]{
        \pgfpathsine{\pgfpoint{0.25\pgfdecorationsegmentlength}{-0.5\pgfdecorationsegmentamplitude}}
        \pgfpathcosine{\pgfpoint{0.25\pgfdecorationsegmentlength}{0.5\pgfdecorationsegmentamplitude}}
}
    \state{final}{}
}

\definecolor{greyish}{rgb}{.90,.90,.90}
\definecolor{greyish2}{rgb}{.96,.96,.96}
\usepackage{xcolor,colortbl}
\usepackage{tcolorbox}

\usepackage[all]{xy}


\numberwithin{equation}{section}
\begin{document}
%
\renewcommand{\thefootnote}{\fnsymbol{footnote}}
\vspace{0truecm}
\thispagestyle{empty}

\begin{center}
{\fontsize{20}{24} \bf Gravity as a gapless phase}\\[5pt]
 {\fontsize{20}{24} \bf and biform symmetries}
\end{center}

\vspace{.15truecm}

\begin{center}
{\fontsize{12.7}{18}\selectfont
Kurt Hinterbichler,${}^{\rm a}$\footnote{\texttt{\href{mailto:kurt.hinterbichler@case.edu}{kurt.hinterbichler@case.edu}}}
Diego M. Hofman,${}^{\rm b}$\footnote{\texttt{\href{mailto:d.m.hofman@uva.nl}{d.m.hofman@uva.nl}}} 
Austin Joyce,${}^{\rm c}$\footnote{\texttt{\href{mailto:austinjoyce@uchicago.edu}{austinjoyce@uchicago.edu}}}
and Gr\'egoire Mathys,${}^{\rm d,b}$\footnote{\texttt{\href{mailto:gregoire.mathys@cornell.edu}{gregoire.mathys@cornell.edu}}} 
}
\end{center}

\vspace{.2truecm}

 \centerline{{\it ${}^{\rm a}$CERCA, Department of Physics,}}
 \centerline{{\it Case Western Reserve University, 10900 Euclid Ave, Cleveland, OH 44106, USA}} 
 
\vspace{.3cm}
 
 \centerline{{\it ${}^{\rm b}$Institute for Theoretical  Physics,}}
 \centerline{{\it University of Amsterdam, Amsterdam, 1098 XH, NL}}
 
\vspace{.3cm}
  
   \centerline{{\it ${}^{\rm c}$Kavli Institute for Cosmological Physics, Department of Astronomy and Astrophysics}}
 \centerline{{\it University of Chicago, Chicago, IL 60637, USA} } 
 
\vspace{.3cm}

\centerline{{\it ${}^{\rm d}$Department of Physics,}}
 \centerline{{\it Cornell University, Ithaca, NY 14850, USA} } 
 \vspace{.25cm}

\vspace{.3cm}
\begin{abstract}
\noindent
We study effective field theories (EFTs) enjoying (maximal) biform symmetries. These are defined by the presence of a conserved (electric) current that has the symmetries of a Young tableau with two columns of equal length.  When these theories also have a topological (magnetic) biform current, its conservation law is anomalous. We go on to show that this mixed anomaly uniquely fixes the two-point function between the electric and magnetic currents. We then perform a \KL spectral decomposition of the current-current correlator, proving that there is a massless mode in the spectrum, whose masslessness is protected by the anomaly. Furthermore, the anomaly gives rise to a universal form of the EFT whose most relevant term---which resembles the linear Einstein action---dominates the infrared physics. As applications of this general formalism, we study the theories of a Galileon superfluid and linearized gravity. Thus, one can view the masslessness of the graviton as being protected by the anomalous biform symmetries. The associated EFT provides an organizing principle for gravity at low energies in terms of physical symmetries, and allows interactions consistent with linearized diffeomorphism invariance. These theories are not ultraviolet-complete---the relevant symmetries can be viewed as emergent---nor do they include the nonlinearities necessary to make them fully diffeomorphism invariant, so there is no contradiction with the expectation that quantum gravity cannot have any global symmetries.

\end{abstract}

\newpage

\setcounter{page}{2}
\setcounter{tocdepth}{2}
\tableofcontents
\newpage
\renewcommand*{\thefootnote}{\arabic{footnote}}
\setcounter{footnote}{0}




\section{Introduction}
\label{sec:intro}

Symmetry is a powerful concept in physics, providing an organizing principle that dictates how we construct mathematical theories describing nature. Symmetries control the properties of phase transitions and they explain why certain phases are robust, with Goldstone's theorem serving as perhaps the oldest example of this line of reasoning \cite{Nambu:1960tm,Goldstone:1961eq,Goldstone:1962es}. In the last decade, our conceptions of what symmetries are, or can be, have been extensively broadened. Vast generalizations of symmetry have been proposed, starting with higher-form symmetries \cite{Gaiotto:2014kfa}---which act on extended objects---and including more exotic structures, like 2-groups and non-invertible symmetries.
Anomalies for these new structures have been understood and incorporated into the general program of studying the possible phases of quantum field theory. These new viewpoints have both brought new results to light and have placed old results into sharper focus. For example, one can understand Goldstone's theorem as a consequence of anomalies, which protect the gaplessness of certain phases \cite{Delacretaz:2019brr}.

Symmetries play an essential role in the phenomenological study of both condensed matter and particle physics. Experimentally, we only ever have access to the physics of a system below some energy scale (the resolution of our apparatus, or the amount of energy we can generate). In this case, a complete microscopic description of the system is not necessarily the most useful parameterization of the physics. Instead, it is often effective field theory (EFT) that provides the most practical organizing principle. In this way we can understand a wide range of problems---all the way from the Standard Model to the Ising model---in the same language. This perspective has been extremely successful in understanding much of the physics of the possible phases of matter at low energies, and the transitions between them.

As an important example of the utility of this paradigm, one can understand the emergence of gauge theories as a consequence of global symmetries and their effective descriptions. For example, the magnetic 1-form symmetry associated to the conservation of magnetic flux leads, at low energies, to EFTs described by the usual Maxwell theory coupled to electric matter. When this matter is gapped, the deep infrared is described by free photons that emerge as the Goldstones for the original magnetic 1-form symmetry \cite{Gaiotto:2014kfa,Hofman:2018lfz,Lake:2018dqm}, which is spontaneously broken. The explanatory power of this perspective suggests that one should take this physical symmetry (as opposed to the unphysical redundancy that is gauge invariance) as the organizing principle underlying these theories as a guide to finding a more microscopic description from which electromagnetism is emergent.\footnote{Early approaches to this problem involved the formulation of theories in loop space~\cite{Polyakov:1980ca} which has been courageously revisited recently in~\cite{Iqbal:2021rkn}.}

What about gravity? How does gravity fit into this paradigm?\footnote{See~\cite{Hartnoll:2013sia} for an interesting attempt at understanding this problem from a related point of view.} Unfortunately, there is good evidence (see for example~\cite{Harlow:2018tng}) that gravity cannot have any exact global symmetries that survive all the way to the ultraviolet (UV). 
If one had a unique UV-complete description of the universe which was under control, we could take this to be the end of the story and just view gravity to be a low energy accident of this complete description. 
However, we do not have full non-perturbative control of string theory, or even know if it is the unique UV-complete theory that can describe nature. And further, there is plausibly a full landscape of solutions in string theory which can lead to very different low energy physics. Some progress has been made in addressing these challenges in the form of the Swampland program (see e.g.,~\cite{vanBeest:2021lhn} for a review). Nevertheless, it is fair to say that it is still unclear whether this approach can fully explain the characteristics of our low energy phase of nature, which we usually associate with Einstein gravity (and its irrelevant corrections). Therefore, it seems useful to search for an organizing principle with which to construct gravity as an EFT.  In other words: what is the gauge-invariant language that characterizes gravity?

In this work we make a modest attempt to answer this question in the language of EFT. We show that linearized gravity has a global continuous symmetry that fits into a category of symmetries that can be called maximal higher-biform symmetries. We show that this provides a guiding principle that is sufficient to construct the linearized Einstein theory and its irrelevant corrections. Furthermore, we show how an anomaly between the magnetic and an electric version of this biform symmetry is responsible for protecting the masslessness of the graviton and keeping the low-energy phase gapless. This line of reasoning extends previous results for usual 0-form symmetries~\cite{Delacretaz:2019brr} and higher form symmetries~\cite{Hofman:2018lfz} to the present context. This result amounts to a generalized Goldstone theorem for this type of system.

We begin by presenting a general framework to study theories with biform symmetries, focusing mostly on the description of the gapless phase of these systems. There are two alternative, but equivalent, perspectives on how to think about these phases. From the more abstract perspective, it is only the symmetries and their anomalies that determine the low energy physics, and everything else follows from this symmetry structure. Given this input, one can constrain the general form of correlation functions of conserved currents. From these universal correlation functions, one can deduce the spectrum of the system, and see that it must necessarily have a gapless degree of freedom. We use this perspective to obtain the low energy physics of the simplest systems exhibiting biform symmetries: a Galileon superfluid and linearized gravity. 
A complementary viewpoint is to realize these universal results via a concrete instantiation by building an EFT Lagrangian that describes the relevant physics and reproduces the anomaly structure.
Interestingly, the anomaly that is responsible for the gaplessness of the phase ends up being coded in a general kind of interaction similar to the linearized Einstein action. This low energy action shares many features with Chern--Simons \cite{Chern:1974ft,Deser:1981wh} and BF theories---its equations of motion are a partial flatness condition on the gauge-invariant curvature---but, crucially, gives rise to propagating massless degrees of freedom. We study this phenomenon in detail and construct the associated EFTs realizing higher-biform symmetries.

After having constructed the EFTs of interest, one can notice that they always involve some form of generalized gauge theory. Indeed, massless degrees of freedom with spin $s \geq 1$ always display emergent gauge symmetries, which are crucial for removing redundant degrees of freedom while keeping a manifestly Lorentz invariant description. Historically, these gauge symmetries have been used to ``explain" the gaplessness of the low energy phase, for example the photon is typically said to be massless because of gauge symmetry. However, from this perspective they are merely an inevitable consequence of our desire to write a manifestly Lorentz-invariant EFT  that realizes the physical biform symmetries. Still, one can ask: what are these massless modes gauging? This is equivalent to asking what the charged degrees of freedom are that we gap and integrate out to obtain a theory purely of massless gauge fields. In the case of electromagnetism, these massive degrees of freedom just correspond to charged particles such as electrons. Interestingly, it turns out that for linearized gravity, the gapped degrees of freedom that carry charge under the emergent gauge symmetry of the graviton can be viewed as fractonic particles. This provides a precise construction of the ideas advocated in~\cite{Pretko:2017fbf,Pretko:2018jbi,Benedetti:2021lxj}.\footnote{There has been a substantial amount of recent work related to the physics of fractons and their relation to these general ideas. More details can be found for example in~\cite{Banerjee:2022unj, Gorantla:2022eem, Distler:2021bop,Banerjee:2021qcj, Grosvenor:2021hkn, Jain:2021ibh, Bidussi:2021nmp, Moinuddin:2021eox, Seiberg:2020bhn,Moinuddin:2022frp} and references therein.}

It is important to emphasize that the symmetries described here do not only apply to free linearized gravity. As in any EFT, it is possible to classify and add an infinite number of irrelevant deformations which---while trivial in the infrared---introduce interactions at higher energies. However, these interactions are still {\it linear} gauge invariant, of the type described in~\cite{Wald:1986bj,Li:2015vwa,Chatzistavrakidis:2016dnj,Bai:2017dwf,Bonifacio:2018van}, and so do not include the types of nonlinear interactions that allow the theory to become nonlinearly diffeomorphism invariant. Much like Einstein gravity, these theories are also not UV complete. Therefore, there is still the important question of why or how these biform symmetries arise at low energies from a microscopic UV-complete theory of gravity. For example, are there associated biform Ward identities in string theory, and what are their consequences? It is also interesting to understand what if anything the Swampland program has to say about the possible UV completions of these EFTs.
While we return to discuss these questions in section~\ref{sec:conclusion}, we hope that this is a first step toward understanding the emergence of gravity in our universe.

The rest of this paper is organized as follows: in section~\ref{sec:higherbiformsyms} the general theory of biform symmetries is presented, gapless field theories displaying these biform symmetries are constructed, and the connection to the physics of fractons is discussed. In section~\ref{sec:superfluid} we review the EFT of relativistic superfluids, its symmetries and anomalies, and generalize these results to the Galileon superfluid. This system constitutes the simplest theory enjoying a (scalar) biform symmetry. In section~\ref{sec:LinGrav} we study linearized gravity and its irrelevant corrections in detail and frame this physics in the language of biform symmetries. We show how the symmetry and anomaly structure are sufficient to completely fix the two-point function of conserved currents and, as a consequence, protect the masslessness of the graviton. In section~\ref{sec:conclusion} we synthesize these results and discuss future directions. Several technical details somewhat outside the main line of development are presented in appendices: In appendix~\ref{ap:KLrep} we describe the systematics of the spectral decompositions required to prove that various phases have gapless excitations. In appendix~\ref{ap:GalSup} we provide some additional details about the anomaly structure of the Galileon superfluid. In appendix~\ref{sec:EM} we present Maxwell electromagnetism in the same language as the rest of the paper, realizing the photon as a Goldstone. Finally, in appendix~\ref{app:lingravanom} we present more technical details about the anomaly structure and spectral decomposition in the case of linearized gravity.

\paragraph{Disclaimer:} While this work was in preparation~\cite{Benedetti:2021lxj,Benedetti:2022zbb} appeared, which have some overlap with this work.

\paragraph{Notation and conventions:} We work in flat $d$-dimensional Euclidean space with metric $\eta_{\mu \nu}$ which we use to raise, lower, and contract indices. We (anti-)symmetrize indices with unit weight, e.g., $ \partial_{(\mu}A_{\nu)} = \frac{1}{2}(\partial_\mu A_\nu +\partial_\nu A_\mu)$.  We use the manifestly antisymmetric convention for mixed-symmetry tensors and Young tableaux. We label two-column Young diagrams as $(i|j)$ where $i,j$ are the lengths of the columns from left to right.

\newpage
\section{Field theories with higher-biform symmetries}
\label{sec:higherbiformsyms}

We begin by discussing the structure of theories with higher biform symmetries, of which linearized gravity will be the prime example. We will start by keeping the discussion as general as possible, and then in the following sections we will elaborate on the most important examples in excruciating detail. One of the benefits of this unified exposition is that it will allow us to treat scalars, $p$-forms, and linearized gravity with similar techniques and elucidate their commonalities.

We will describe the global symmetries of these theories and show how they naturally give rise to presentations in terms of gauge fields and their associated gauge invariances.
The resulting framework will allow for the construction of EFT Lagrangians for these theories, which describe their different phases. 
We will concentrate on gapless phases, but we expect that the geometric approach we describe can be extended to other interesting cases. We also discuss the coupling of these theories to sources and how they relate to background gauge potentials.

\subsection{Conserved currents and EFTs}
\label{sec:conservcurrentEFTs}

When constructing universal EFTs, the guiding principle should always be the physical symmetries of the system.
Once we know the algebra of conserved charges and the way that charged operators represent these symmetries we can then build the EFT. 
This is what makes electromagnetism, for example, universal: it is the local EFT of a theory with a 1-form symmetry in its symmetry broken (i.e., gapless) phase. 
Notice that this approach is completely agnostic about what the UV microphysics that gave rise to this EFT is, or even whether the EFT  can be UV completed. 
The theory might be formulated microscopically in terms of the dynamics of extended objects (as is often the case for higher-form symmetries), or a quantum field theory, or even a lattice model. This is, at this stage, of little consequence. All we demand is that the theory is a local field theory at low energies.

In local theories, conserved quantities corresponding to continuous symmetries are generally given by integrals of local current operators. In all of the following we will concentrate on abelian continuous symmetries, and will assume the existence of associated conserved currents. We want to trace the logic that leads from these conserved quantities to local degrees of freedom in a field-theoretic description. The physics that we review in this section is elementary, but we want to introduce it in a language that
will be appropriate for generalization to cases enjoying higher biform symmetries, like linearized gravity.

One way to proceed is to start with the desired current---for example for a 0-form symmetry the current is $J^\mu$. Next, we express this operator in terms of elementary fields as $J^\mu[\phi]$. Lastly, we construct an EFT Lagrangian such that equations of motion for the fields $\phi$ lead to the conservation law for the current: $\partial_\mu J^\mu =0$. This construction makes manifest the symmetry for which $J^\mu$ is a Noether current, in the sense that this symmetry acts locally on the fields $\phi$.
The associated conserved charge can be written as:
\be
Q = \int_{\Sigma_{d-1}} * J \, ,
\ee
where the integral is over a codimension 1 surface $\Sigma_{d-1}$. {As is well-known, the conservation of $J$ implies that $Q$ depends only topologically on the surface $\Sigma$.}  These symmetries are often called dynamical, or electric.

An alternative option is to realize our desired current as a magnetic symmetry. In this case, we would start with a $(d-1)$-form electric current, $J^{\mu_1 \cdots \mu_{d-1}}$, where $d$ is the dimension of spacetime. If we want a magnetic 0-form symmetry to emerge from this operator, it must be associated to the dual current 1-form current $K_{(1)}\equiv \ast J_{(d-1)}$ which must be conserved:
\be
\partial_\mu K^\mu = \partial_\mu \left(\epsilon^{\mu \nu_1 \cdots \nu_{d-1}} J_{\nu_1 \cdots \nu_{d-1}} \right)=0\, .
\label{eq:mkcons}
\ee
The conservation condition~\eqref{eq:mkcons} implies that $J_{(d-1)}$ is a closed $(d-1)$-form: $\rd J_{(d-1)} =0 $.
Assuming  that there are no topological obstructions, we can always parameterize this as an exact form:
\be
J_{(d-1)} = \rd \mathcal{A}_{(d-2)} \, ,
\label{eq:AJparam}
\ee
\noindent where $\mathcal{A}_{(d-2)}$ is a $(d-2)$-form. This parameterization necessarily introduces a gauge symmetry $\mathcal{A}_{(d-2)} \mapsto\mathcal{A}_{(d-2)} + \rd\xi_{(d-3)}$, where $\xi_{(d-3)}$ is a $(d-3)$-form. This gauge symmetry is of course unphysical, and only a consequence of our choices. We can now construct any EFT Lagrangian in terms of ${\cal A}_{(d-2)}$, provided it is gauge invariant. The conservation law we were interested in is then guaranteed by~\eqref{eq:AJparam}, independent of the equations of motion. Sometimes we call these symmetries topological or magnetic, but there is no real physics in this distinction. Topological symmetries can become dynamical symmetries in a different parameterization of the EFT in terms of different local fields.

The breaking of a magnetic symmetry can only happen through the inclusion of topological defects which invalidate the equation~\eqref{eq:AJparam}. On the other hand, the electric symmetry is not guaranteed and it depends on the choice of EFT Lagrangian. If there is light electric matter, this symmetry will be explicitly broken.\footnote{Notice that in this case where we use $\mathcal{A}_{(d-2)}$ as the field in the Lagrangian, electric matter is given by extended objects with space-time dimension $d-2$. While this complicates the construction of a Lagrangian, it does not appreciably affect the physics discussed. } This is certainly allowed, as we were only building an EFT for the symmetry associated to $K_{(1)}$. However, if electric matter is gapped, we expect the electric current to be conserved and the original magnetic symmetry to be spontaneously broken. That is, conservation of the electric current, which is equivalent to {\it dual conservation} of $K_{(1)}$
\be
\rd K_{(1)} = 0\, ,
\ee 
will imply a wave equation for ${\cal A}_{(d-2)}$ of the form
\be
\square\mathcal{A}_{(d-2)} =0\,,
\ee
where $\square$ is the gauge-invariant Laplacian operator (e.g., $\square \mathcal{A}_\nu = \partial_\mu \partial^\mu \mathcal{A}_\nu - \partial_\nu \partial^\mu \mathcal{A}_\mu$ if $\mathcal{A}_\nu$ is a 1-form). This is the gapless phase of the theory. Coming back to our starting point, we see that we could have easily started from the electric realization of the symmetry and reached an equivalent destination.

In order to make the physics of this gapless phase more explicit, note that
if we want to build an EFT with this symmetry, without any additional matter (i.e., all other matter is gapped), we see that the most relevant term we can write whose Lagrangian is  gauge invariant is
\be
S_{\rm EFT}[\mathcal{A}] =\int \rd^dx\, J_{\mu_1 \cdots \mu_{d-1}} J^{\mu_1 \cdots \mu_{d-1} }\, .\label{eq:QFTaction1}
\ee
This theory is nothing other than a free abelian $(d-2)$-form gauge theory.  
From some perspective, this reasoning is the actual origin of gauge symmetries in physics. 

As desired, the action~\eqref{eq:QFTaction1} yields the equation of motion $\rd K_{(1)} =0$, corresponding to conservation of the electric symmetry. Had we chosen to represent $K_{(1)} = \rd \phi$, for a scalar $\phi$, we would have obtained the same result, making it clear that the above theory is just a free scalar with a shift symmetry $\phi \mapsto \phi + c$ for any constant $c$. Notice that this choice of field variables would trivialize $\rd  K_{(1)} =0$ (corresponding to conservation of the electric current), while conservation of the (1-form) magnetic current would be a consequence of the equations of motion, further illustrating that what we call a topological and what we call a dynamical symmetry depends on our choice of field variables.

The magnetic charge associated with the original 0-form symmetry that we were interested in  can be written as:
\be
Q_m(\Sigma_{d-1}) =  \int_{\Sigma_{d-1}} *K_{(1)} \, ,
\ee
while the emergent electric charge associated with a $(d-2)$-form symmetry in the gapless phase is given by:
\be
Q_e(\Sigma_{1}) =  \int_{\Sigma_{1}} * J_{(d-1)} \, .
\ee
We see from this discussion in the magnetic language how a symmetry broken phase gives rise to a gapless scalar Goldstone mode.

All of this can be straightforwardly generalized to any higher form symmetry associated to a conserved magnetic $(d-p-1)$-form current $K_{(d-p-1)}$. (The previous example corresponds to  $p = d-2$.) In this case, the starting point is an electric 
$(p+1)$-form current $J_{(p+1)}$ that must satisfy $\rd J_{(p+1)}=0$---which is equivalent to conservation of the magnetic current. We choose to parametrize this electric current in terms of a $p$-form gauge field $\mathcal{A}_{(p)}$ as
\be
J_{(p+1)} = \rd\mathcal{A}_{(p)}\, ,
\ee
which trivializes the constraint that $J_{(p+1)}$ be closed.
Following the same path as before we find the effective theory in the phase where electric matter is gapped
\be
S_{\rm EFT}[\mathcal{A}] = \int \rd^dx\, J_{\mu_1 \cdots \mu_{p+1}} J^{\mu_1 \cdots \mu_{p+1} }\, ,
\ee
which gives rise to the magnetic $(d-p-2)$-form charge
\be
Q_m(\Sigma_{p+1}) =  \int_{\Sigma_{p+1}} * K_{(d-p-1)} \, ,
\ee
along with an emergent electric $p$-form charge
\be
Q_e(\Sigma_{d-p-1}) =  \int_{\Sigma_{d-p-1}} * J_{(p)} \, .
\ee
Operators charged under these physical symmetries are electric $p$-dimensional and magnetic $(d-p-2)$-dimensional extended objects which couple directly to $\mathcal{A}_{(p)}$ and $\phi_{(d-p-2)}$, respectively, where $\phi$ is a $(d-p-2)$-form defined as $K_{(d-p-1)} = \rd\phi_{(d-p-2)}$.

More irrelevant terms in the effective theory can be constructed from $J_{(p+1)}$ and its derivatives and added to the action. For example, one could imagine adding to the action terms like $J^4, J^6, \cdots$. These terms would keep the magnetic symmetry unmodified, since $\rd J_{(p+1)}=0$ still, but would change the electric symmetry current operator.   {This is an important point---in general in the gapless phases of interest the currents $J_{(p+1)}$ and $K_{(d-p-1)}$ are just Hodge duals of each other in the deep IR (i.e., at the free level). However, these two currents are not equivalent once we begin to introduce interactions.} Still, the electric symmetry would not be disturbed provided we do not add light electrically charged matter to our Lagrangian.
One way see this is to invoke Noether's theorem. The actions above, including their irrelevant deformations,  have a $p$-form symmetry $\delta \mathcal{A}_{(p)} = \Lambda_{(p)}$, where $\Lambda_{(p)}$ is a constant $p$-form satisfying $\rd\Lambda_{(p)} = 0$. Noether's theorem then guarantees that, if we consider a general perturbation $\mathcal{A}_{(p)} \mapsto \mathcal{A}_{(p)} + \delta\mathcal{A}_{(p)}$, we must have:
\be
\delta S \sim \int \rd^dx \, \delta\mathcal{A}_{\mu_1 \cdots \mu_p} \partial_\mu \hat{J}^{\mu \mu_1 \cdots \mu_p} \, ,
\ee
\noindent where $\hat{J}_{(p+1)}$ must be conserved on-shell, and is therefore the desired electric current. This reasoning is valid for all our effective actions and thus $\hat{J}_{(p+1)}$ can be computed in each case in terms of $J_{(p+1)}$. Its precise form will depend on the details of the theory, but the existence of some conserved current is robust.

We claimed, rather carelessly, that in the absence of matter the EFT associated to the higher form symmetries above describes gapless phases of matter. This is not strictly true. When integrating out matter, the effect of their anomalies cannot be neglected. For example, for $d=2p+1$, there exist Chern--Simons terms that can appear in the low energy effective action if the integrated-out matter breaks parity invariance. These Chern--Simons terms are more relevant than the Maxwell-type terms described above. They would give a theory in the deep IR of the form
\be
S_{\rm CS}[\mathcal{A}] = \int \rd^{2p+1}x\,  \epsilon^{\mu_1 \cdots \mu_{2p+1}} \mathcal{A}_{\mu_1 \cdots \mu_{p}} J_{\mu_{p+1} \cdots \mu_{2p+1}}\, .
\ee
The equations of motion are fundamentally different in this case---being a flatness condition, $J_{(p+1)}=0$, rather than a wave equation---and the phase of matter becomes gapped. If the original continuous symmetry is compact, the gapped phase just has a discrete symmetry. If it is non-compact, the lack of local degrees of freedom ensures one can still build topological charge operators (i.e., they are not built from a gauge-invariant local current) of the form:
\be
Q_{\text{top}}(\Sigma_p) = \int_{\Sigma_p} \mathcal{A}_{(p)}\, .
\ee
In both cases, we could say the original symmetry we were aiming to describe by $J_{(p+1)}$ (and its dual $K_{(d-p-1)}$) got gauged and new symmetries appear.

A more general version of this phenomenon is given by BF-type Lagrangians. In that case, if we have two gauge fields, say $\mathcal{A}_{(p)}$ and $\mathcal{B}_{(d-p-1)}$ we can write the Lagrangian
\be
S_{\rm BF} =\int \rd^{d}x\,  \epsilon^{\mu_1 \cdots \mu_{d}} \mathcal{A}_{\mu_1 \cdots \mu_{p}} \left(\rd {\cal B}\right)_{\mu_{p+1} \cdots \mu_{d}}\, .
\ee
 Like the Chern--Simons example, this theory is gapped and our would-be currents $\rd\mathcal{A}$ and $\rd\mathcal{B}$ vanish as a consequence of the equations of motion.

The systematics of the above structures are built on the theory of differential forms and their exterior calculus. The Poincar\'e lemma plays an important role by guaranteeing that, in the absence of topological defects, closed forms can be written in terms of exact ones, i.e., that $\rd J_{(p+1)} = 0$ implies $J_{(p+1)} = \rd\mathcal{A}_{(p)}$. This is of course always locally true.  
In terms of representation theory, all the currents and gauge fields transform in antisymmetric tensor representations of GL$(d, \mathbb{R})$ labeled by single-column Young diagrams.
In the next subsection we will generalize this by considering Young tableaux made up of two vertical columns.
This may seem to be an extravagance, but is actually necessary to cast (linearized) gravity in the same framework.

\subsection{Higher-biform symmetries\label{sec:hbs}}

We now want to generalize the type of conserved currents we consider when building EFTs. We will be interested in currents transforming in irreducible representations of GL$(d,{\mathbb R})$ labeled by two-column Young diagrams (see \cite{Curtright:1980yk,Labastida:1986gy,Labastida:1987kw,Hull:2000zn,Hull:2000rr,Hull:2001iu,Alkalaev:2003hc,Bekaert:2004dz,Boulanger:2004rx,Joung:2016naf,Zinoviev:2016mxh} for other work involving these representations). We will call these mixed-symmetry representations biforms, since they are, in a sense, the natural generalization of differential forms to having two sets of antisymmetric indices. Under the action of a properly defined exterior derivative, these form a complex. The mathematics associated to its differential calculus, Poincar\'e lemma, and cohomology has been developed \cite{Dubois-Violette:1998tea,Dubois-Violette:1999iqe,Dubois-Violette:2001wjr,Dubois-Violette:2005bep,Dubois-Violette:2009lnn,Chatzistavrakidis:2019len}, see \cite{Dubois-Violette:2000fok,Bekaert:2002cz} for reviews and a complete list of references and  \cite{Bekaert:2002dt} for the results of interest to our discussion. Here, in order to make the discussion self contained, we present the very little technology that we need, with some small modifications with respect to \cite{Bekaert:2002dt}, in order to move forward. Equivalent biform field theories were also constructed in~\cite{deMedeiros:2002qpr,Francia:2004lbf}

Let us start with the simplest (and most important) example. Consider a theory with an electric current $J_{(p+1 | p+1)}$ that transforms in an irreducible representation of GL$(d,\mathbb{R})$ labeled by a Young diagram with two identical vertical columns of length $p+1$:
\be 
J_{(p+1|p+1)} ~\in ~ \left.\raisebox{2.5ex}{\Yboxdimx{13pt}
\Yboxdimy{13pt}\gyoung(~;~,\vdts;\vdts,~;~)}\,\right\}\,p+1\,.
\label{eq:electriccurrent}
\ee
We call such an object a $(p+1 | p+1)$-biform.\footnote{We adopt the convention for mixed-symmetry tensors that they are manifestly antisymmetric in the indices associated to a column. The only other constraint is that antisymmetrizing all the indices from the first column plus any index from the second column causes the tensor to vanish. When necessary, we refer to two-column Young diagrams using the notation $(p\,|\,q)$, where $p$ and $q$ label the column lengths from left to right.}   We wish to build a conserved magnetic current, $K$, from $J_{(p+1|p+1)}$. The Hodge star operator can now act on either of the two columns of our current. In this case, since $J_{(p+1|p+1)}$ has two identical columns, it is of no consequence which column we dualize. 
It is actually most convenient to dualize {\it both} columns and define
\be
K_{(d-p-1|d-p-1)}= \ast J_{(p+1|p+1)}\ast \, : \quad K^{\mu_1 \cdots \mu_{d-p-1} |  \nu_1 \cdots \nu_{d-p-1}} = \epsilon^{\mu_1 \cdots \mu_{d}} J_{\mu_{d-p} \cdots \mu_d | \nu_{d-p} \cdots \nu_{d}} \epsilon^{\nu_1 \cdots \nu_{d}} \, .
\ee
Here both $J$ and $K$ are antisymmetric under the interchange of any of the indices inside a block separated by $|$. Since the two sets of indices are of equal length, they are additionally symmetric under the interchange of these two blocks of indices.

We would like $K_{(d-p-1|d-p-1)}$ to satisfy the conservation equation
\be
\partial_{\mu_1} K^{\mu_1 \cdots \mu_{d-p-1} |\nu_1 \cdots \nu_{d-p-1}}  =0 \, ,
\ee
which is equivalent to imposing the condition
\be
\left(\rd J\right)_{(p+2 | p+1)}=0\, .
\ee
The exterior derivative above is defined with the correct (anti-)symmetrization  properties so that $(\rd J)_{(p+2 | p+1)}$ is a $(p+2 | p+1)$-biform. More generally, let us define the set of {\it maximal biforms}, denoted by $\hat{\Omega}$. These are given by the direct sum of the space of $(p | p)$-biforms, $\Omega_{(p|p)}$,  and $(q+1|q)$-biforms, $\Omega_{(q+1|q)}$ for all $  p, \,q$.
\be
\hat{\Omega} = \left(\medoplus_p\, \Omega_{(p|p)} \right)\, \bigoplus \,\left(\medoplus_q \,\Omega_{(q+1|q)} \right)\, .
\ee
We call the first set above $\left(\medoplus_p\, \Omega_{(p|p)} \right)$ \textit{even} biforms and the second set $ \left(\medoplus_q\, \Omega_{(q+1|q)} \right)$ \textit{odd}  biforms. In this section we will be interested in an even current $J_{(p+1|p+1)}$. More explicitly, the symmetry types of elements of these vector spaces are
\be 
\Omega_{(p|p)}~\in~ \raisebox{3ex}{\Yboxdimx{13pt}
\Yboxdimy{13pt}\gyoung(|3p|3p)}~ ,\qquad \qquad \Omega_{(q+1|q)}~ \in~ \raisebox{3ex}{\Yboxdimx{13pt}
\Yboxdimy{13pt}\gyoung(|3q|3q,~)}~ .
\ee

 On this set we define the action of the exterior derivative by (anti-)symmetrization such that it takes elements of $\Omega_{(p|p)}$ to $\Omega_{(p+1|p)}$ and elements of $\Omega_{(p+1|p)}$ to $\Omega_{(p+1|p+1)}$:
 \be
\rd : \Omega_{(p|p)}\, \rightarrow \,  \Omega_{(p+1|p)} \, , \qquad \qquad \qquad  \rd: \Omega_{(p+1|p)} \, \rightarrow \,  \Omega_{(p+1|p+1)}\, .
 \ee
Note that the operator $\rd$ is nilpotent of degree 3, i.e., $\rd^3=0$.  This follows because acting with $\rd$ three times guarantees that we will be antisymmetrizing over at least two partial derivatives.
It is clear, from the present discussion, why we chose to define $K$ by the double action of the Hodge operator $*$. This choice implies that the dual of a maximal biform is itself a maximal biform.
 \be
 *  \cdot \, *: \Omega_{(p | p)} \rightarrow \Omega_{(d-p | d-p)} \, ,\qquad \qquad \qquad  * \cdot \, *: \Omega_{(p+1 | p)} \rightarrow \Omega_{(d-p | d-p -1)}\, .
 \ee

Returning to  our current,  we would like to introduce a gauge field that makes $\rd J_{(p+1 | p+1)}=0$ automatic in the absence of defects. From the fact that $\rd^3 = 0$, it is clear that we can take\footnote{More formally, this follows from the Poincar\'e lemma for this complex~\cite{Bekaert:2002dt}.}
\be
J_{(p+1 | p+1)}= \rd^2 h_{(p | p)}\, ,
\ee
where we have introduced a $(p | p)$-biform $h_{(p | p)}$ as a new gauge field. 
With this definition $h_{(p | p)}$ enjoys a gauge symmetry $h_{(p | p)} \mapsto h_{(p | p)} + \rd\xi_{(p | p-1)}$, where $\xi_{(p | p-1)}$ is a $(p | p-1)$-biform.

Any effective theory constructed in a gauge-invariant manner from $h_{(p | p)}$ will necessarily have a new type of symmetry arising from the conservation of the magnetic current $K_{(d-p-1 | d-p-1)}$. We will call this symmetry a {\it maximal $(d-p-2)$-biform symmetry}.

We can construct magnetic charges in this theory by contracting $K_{(d-p-1 | d-p-1)}$ with a $(d-p-1)$-form, $\varrho_{(d-p-1)}$ using the flat metric to define:
\be
K^{(\varrho)}_{\mu_1 \cdots \mu_{d-p-1}} = K_{\mu_1 \cdots \mu_{d-p-1}| \nu_1 \cdots \nu_{d-p-1}} \varrho^{\nu_1 \cdots \nu_{d-p-1}}\, .
\ee
With this definition, the conserved charges are given by
\be\label{magcha1}
Q^{(\varrho)}_m(\Sigma_{p+1}) = \int_{\Sigma_{p+1}} * K^{(\varrho)}_{(d-p-1)}\, .
\ee
Conservation of this charge is guaranteed if $\varrho_{(d-p-1)}$ satisfies appropriate conditions, which are described further in section~\ref{sec:charges}.
Notice that $K^{(\varrho)}$ is a $(d-p-1)$-form and as such its Hodge dual is given by the single action of the $*$ operator.

We are now ready to write down the simplest (i.e. most relevant) action for an EFT enjoying this symmetry. The natural guess is that the action we are looking for is:
\be
S_{\rm EFT}[h] = \int \rd^dx\,  J_{\mu_1 \cdots \mu_{p+1}|\nu_1 \cdots \nu_{p+1} } J^{\mu_1 \cdots \mu_{p+1} |\nu_1 \cdots \nu_{p+1}}\, .\label{eq:EFTaction1}
\ee
The equations of motion following from~\eqref{eq:EFTaction1} indeed produce a gapless phase, where the equation of motion is
\be
\rd^2 K_{(d-p-1 | d-p-1)} = \rd^2 * J_{(p+1 | p+1)}\,* = \rd^2 * \rd^2 h_{(p | p)} * =0 \implies \square ^2 h_{(p | p)} = 0\, ,
\ee
with $\square$  again a gauge-invariant version of the Laplacian. However, this equation is {\it fourth} order in derivatives! Furthermore, $K_{(d-p-1 | d-p-1)}$ satisfies a different equation from $J_{(p+1 | p+1)}$ ($\rd^2 K =0$ vs. $\rd J = 0$) which means that the emergent electric symmetry is of a slightly different character than the original symmetry.

What equation of motion would give the desired two derivative equations and the corresponding electric conservation law, $\rd K=0$? Whatever it is, must be gauge invariant so we should produce it from $J_{(p+1 | p+1)}$. But $J_{(p+1 | p+1)}$ already involves two derivatives of $h_{(p | p)}$, so we can't add any extra ones. Additionally, since ultimately we will vary an action with respect to $h_{(p | p)}$---which is a $(p | p)$-biform---we expect that the equation of motion will also be in that representation. Sadly,  $J_{(p+1 | p+1)}$ is a $(p+1 | p+1)$-form, so we need to add one more piece of structure that allows us to reduce the size of the representations. In a spacetime with a metric (which we have taken to be flat), we can always trace over a pair of indices, one from each column in a biform. We therefore define a trace operation:
\be
\tr\left( \cdot\right)\, : \Omega_{(q | p)} \rightarrow \Omega_{(q-1 | p-1)}  \, ,\qquad \textrm{as} \qquad \tr \left(X\right) =  \eta^{\mu_q \nu_p} X_{\mu_1 \cdots \mu_q | \nu_1 \cdots \nu_p} \, .\label{eq:tracedef}
\ee
Conveniently, this trace operation~\eqref{eq:tracedef} maps maximal biforms to other maximal biforms.

The candidate equation of motion for an EFT with a higher biform symmetry is therefore:
\be
\tr\left(J_{(p+1 | p+1)}\right) = \tr \left( \rd^2 h_{(p | p)}\right) = 0\, .
\label{eq:biformeinsteineq}
\ee
For $p=1$, this is nothing other than the linearized Einstein equation.
Furthermore, it turns out this condition exactly guarantees the conservation of the electric current. We have
\be
\tr \left(\rd J\right) =a  * \left(\rd * J *\right) * + b \, \big(\rd\,  \tr \,\left( J\right)\big)\, ,
\ee
\noindent where $a$ and $b$ are constants that depend on conventions for the normalization of biforms which we have not specified. Using this equation, we infer that the conditions
\be
 \rd J_{(p+1 | p+1)} = 0 \quad {\rm and} \quad \tr\left( J_{(p+1 | p+1)}\right)=0\,,
 \ee
 imply the following equations for the magnetic current
 \be
\quad \rd K_{(d-p-1 | d-p-1)} =0 \quad {\rm and} \quad \rd\,\tr\left( K_{(d-p-1 | d-p-1)}\right)=0 \, .
 \ee
In other words, the electric current is conserved (which is equivalent to $\rd K = 0$).

The simple equation of motion~\eqref{eq:biformeinsteineq} describes a gapless phase, which we associate with linearized gravity for $p=1$, and  gives rise to an emergent electric maximal $(p | p)$-biform symmetry with current $J_{(p+1 | p+1)}$ and charges constructed as:
\be
Q_e^{(\zeta)}(\Sigma_{d-p-1}) = \int_{\Sigma_{d-p-1}} * J^{(\zeta)}_{(p+1)} \, ,\quad\quad \textrm{with} \quad\quad J^{(\zeta)}_{\mu_1 \cdots \mu_{p+1}} = J_{\mu_1 \cdots \mu_{p+1}| \nu_1 \cdots \nu_{p+1}} \zeta^{\nu_1 \cdots \nu_{p+1}}\, ,
\label{eq:ecgh}
\ee
for an appropriately defined Conformal Killing tensor-like $(p+1)$-form $\zeta_{(p+1)}$ which we further discuss in section~\ref{sec:charges}.

What is the action that produces this interesting equation of motion? We, of course, know the answer for $p=1$: it is the Fierz--Pauli action of linearized gravity. If we did not already know about it we might be surprised by the existence of such a theory, since we have argued that the most relevant term we can write in terms of $J_{(p+1  | p+1)}$ leads to fourth-order equations of motion. In order to produce second-order equations of motion, we require an action that contains a more relevant operator.
The resolution of this apparent tension is that our desired action is actually more like the Chern--Simons example discussed in section~\ref{sec:conservcurrentEFTs} than the Maxwell-like construction~\eqref{eq:EFTaction1}. The surprising fact is that for EFTs of higher biform symmetries, this type of action still produces gapless phases (in high enough dimension). This is because $\tr\left(J\right)=0$ does {\it not} imply that the full current $J=0$, for dimensions $d \geq 2 p +2$. This leaves some degrees of freedom left in the current to propagate at low energies.  These Chern--Simons-like actions can be built in any dimension.

In order to construct an action with equation of motion~\eqref{eq:biformeinsteineq}, we first have to point out the existence of an important object: a generalized Einstein $(p|p)$-biform, which is defined as:
\be
G_{(p | p)} \equiv \sum_{k=0}^p \frac{(-1)^k}{k+1!\,  k!}\, \eta^k_{(1|1)} \wedge \tr^{k+1} J_{(p+1 | p+1)}\, .
\ee
Here $ \eta_{(1|1)}$ is the flat metric thought of as a fixed background $(1|1)$ biform, the notation $\eta_{(1|1)}^k$ denotes a repeated wedge product of $k$ copies of~$\eta$, and we have used the obvious wedge product of biforms given by $\wedge$ which produces new, higher rank biforms by appropriate (anti)symmetrization.\footnote{We have chosen this product to be unnormalized, so for example  $X_{\mu | \nu} \wedge Y_{\rho | \sigma} = X_{\mu | \nu} Y_{\rho | \sigma} -X_{\rho | \nu} Y_{\mu | \sigma}-X_{\mu | \sigma} Y_{\rho | \nu}+X_{\rho | \sigma}Y_{\mu | \nu}$.} 
The defining property of the Einstein biform is that it is automatically conserved as consequence of the original magnetic symmetry. That is, the dual conservation $\rd J_{(p+1 | p+1)}=0$ implies
\be
 \rd * G_{(p | p)} * =0\,,
 \ee
which is equivalent to the component expression $\partial_{\mu_1}  G^{\mu_1 \cdots \mu_p | \nu_1 \cdots \nu_p}=0$. The Einstein biform is also
obviously gauge invariant, since it is constructed from $J_{(p+1 | p+1)}$.

We now have enough ingredients to write what we call the {\it Einstein EFT action} (also discussed in \cite{deMedeiros:2002qpr}):
\begin{tcolorbox}[colframe=white,arc=0pt,colback=greyish2]
\be\label{einsact}
S_{\rm E}[h] = \int \rd^d x \, h_{\mu_1 \cdots \mu_p | \nu_1 \cdots \nu_p} G^{\mu_1 \cdots \mu_p | \nu_1 \cdots \nu_p}\, .
\ee
\end{tcolorbox}
\vspace{-6pt}
\noindent
This action is gauge invariant, up to a boundary term---just like Chern--Simons theory. The variation of the action $h_{(p | p)} \mapsto h_{(p | p)} + \rd\xi_{(p | p-1)}$ indeed vanishes due to the conservation of $G_{(p | p)}$. The equation of motion following from~\eqref{einsact} is easy to obtain:
\be
G_{(p | p)} = 0\, , \qquad \implies \quad \tr\left(J_{(p+1 | p+1)}\right) = 0\, .
\ee
This is a partial flatness condition on the (gauge-invariant) curvature, $J_{(p+1 | p+1)}$, further emphasizing the similarity between the action~\eqref{einsact} and Chern--Simons theory.

We saw from~\eqref{eq:biformeinsteineq} that setting the trace of  $J_{(p+1 | p+1)}$ to zero implies a wave equation for the biform gauge potential $h_{(p|p)}$, and that this equation also implies the conservation of an electric biform current.
Therefore the action~\eqref{einsact} describes the gapless phase of a theory with both a magnetic maximal $(d-p-1)$-biform  and an electric maximal $p$-biform symmetry. For $p=1$, this is the action of linearized Einstein gravity, which we will study in detail further below. 
Here we see what the general structure of the theory is, and how it is a member of a larger set of theories.
Another interesting example is provided by $p=0$, which is a theory of a free scalar
that falls in this same class of theories, and therefore shares many features with gravity. 
We will also explore the details of this theory in section~\ref{sec:superfluid}, and will point out the subtle differences that make the EFT of this scalar somewhat different from the usual Goldstone theory for a 0-form symmetry: in a nutshell, the IR Lagrangians agree while their irrelevant corrections need not.

It is also interesting to consider defects charged under these symmetries. One can couple the field $h_{(p | p)}$ to a $p$-dimensional defect by contracting half of  its indices with velocity vectors defined by the surface and thinking of the remaining indices as producing a $p$-form, which can then be integrated. It is a simple exercise to see that the resulting operator can only be gauge invariant if the defect is an extremal surface (e.g., a geodesic for $p=1$) of the background geometry. This is well-known in linearized gravity \cite{Bunster:2006rt}---and generalizes to higher $p$---and is a strong hint that biform symmetries are related to properties of spacetime. The restriction on the possible types of charged objects can also be related to the physics of fractons \cite{Pretko:2018jbi}, which we will comment on further below.

More irrelevant corrections to~\eqref{einsact} would be of the form $J^2, J^3, J^4, \cdots$. These terms\footnote{The precise allowed interactions depend on the value $p+1$, corresponding to the number of indices the biform has. For example, for $p$ odd, cubic terms are allowed for a single species of biform field $h_{(p|p)}$, while for $p$ even such interactions vanish as a consequence of permutation symmetry.} would preserve both the gaplessness of the theory, and conservation of both the electric and magnetic currents, up to some deformation of the electric current $J \rightarrow \hat J$ similar to the ordinary higher-form case. 
Notice that, if one fine-tunes the most relevant Einstein term in this action to zero, we go back to the Maxwell-type actions discussed in~\eqref{eq:EFTaction1}. Linearized conformal gravity belongs to this class of theories.

It is worth underlining both the commonalities and differences between these theories and Chern--Simons theories. While they share many features, these theories remain gapless (for $d \geq 2p+2$). One way to understand this is to note that if we think of $J_{(p+1|p+1)}$ as a curvature, not all of it is required to vanish on-shell, only its trace. This point of view can be related to the physics of fractons, which we explore next.

\subsection{Biform gauge symmetries and fractons}\label{fracgau}

The previous discussion focused primarily on the physical symmetries of the problem, which were used to construct an effective theory that describes the gapless phase where the magnetic symmetry is nonlinearly realized. In doing this, we have somewhat glossed over the discussion of the gauge symmetries.
What are we really gauging here?  In order to make contact with familiar physics, it is useful to repackage the previous construction in terms of curvatures built by acting with a single derivative on the gauge fields. This will amount to a first-order formulation of the theories of interest. Let's build a first-order curvature for our gauge field $h$ as:
\be
Q_{(p+1|p)} = \rd h_{(p|p)}\, .
\ee
In this setup, $h_{(p|p)}$ is invariant under a gauge transform of the form: $h_{(p|p)} \mapsto  h_{(p|p)} + \rd^2\omega_{(p-1|p-1)}$, where $\omega_{(p-1|p-1)}$ is a $(p-1|p-1)$-biform. What is this gauging? For $p>1$, this symmetry necessarily acts on extended objects,  but for $p=1$ we can give an intuitive description: $h_{(1|1)}$ is a gauge field for the spatial part of a fracton field \cite{Pretko:2018jbi}.

Consider a scalar field $\Phi$ transforming under an abelian symmetry as 
\be
\Phi \mapsto e^{i \alpha(x)} \Phi\,,
\label{eq:fracsym}
\ee
where $\alpha(x)$ is a linear function of the coordinates: $\alpha(x) = \alpha + \alpha_\mu x^\mu$,
with $\alpha$ a constant and $\alpha_\mu$ a constant vector.
This symmetry is necessary to have both charge and dipole moment conservation.\footnote{Theories that have higher conserved moments have also been considered. They can be constructed along similar lines to what we describe here, and would fit into a family of theories with a generalization of the biform symmetries discussed in section~\ref{sec:hbs} to {\it multiforms}, involving tensors labeled by Young diagrams with more columns~\cite{deMedeiros:2002qpr,Francia:2004lbf}. The relevant representatives of that class of theories with conserved multipole $\ell$ would be the ones with symmetric tensor gauge fields, corresponding to $\ell+1$ columns of length 1.} 
It turns out it is possible to build a quadratic combination of $\Phi$ and its derivatives that transforms covariantly under this scalar fractonic symmetry.\footnote{Here we are only concerned with the space-like part of fractonic theories. In condensed matter applications, a non-relativistic time direction is also included.} In detail, the combination
\be
\left(\partial^2 \Phi^2\right)_{\mu \nu} \equiv \Phi \partial_\mu \partial_\nu \Phi -\partial_\mu \Phi \partial_\nu \Phi \, ,\label{eq:FracComb}
\ee
transforms as
\be 
\left(\partial^2 \Phi^2\right)_{\mu \nu} ~~\longmapsto~~ e^{i 2 \alpha(x)} \left(\partial^2 \Phi^2\right)_{\mu \nu}\, ,
\ee 
under~\eqref{eq:fracsym} with $\alpha(x) = \alpha + \alpha_\mu x^\mu$.
If one wishes to gauge this symmetry, so that the combination~\eqref{eq:FracComb} transforms covariantly for all functions $\alpha(x)$, this can be done by introducing a symmetric gauge field $h_{\mu | \nu}$ and defining
\be
\left(D^2 \Phi^2\right)_{\mu \nu} \equiv \Phi \partial_\mu \partial_\nu \Phi -\partial_\mu \Phi \partial_\nu \Phi - i h_{\mu | \nu} \Phi^2\, .
\label{eq:fracgaugephi2}
\ee
If the gauge field transforms as
\be
h_{\mu | \nu } \mapsto h_{\mu| \nu} + \partial_\mu \partial_\nu \alpha\, ,
\ee
then~\eqref{eq:fracgaugephi2} will transform covariantly:
\be
\left(D^2 \Phi^2\right)_{\mu \nu}  ~~\longmapsto~~ e^{i 2 \alpha(x)} \left(D^2 \Phi^2\right)_{\mu \nu}\, ,
\ee
where $\alpha(x)$ is now an arbitrary function.
In the language of section~\ref{sec:hbs}, $h_{\mu|\nu}$ is a $(1|1)$-biform, which transforms as $h_{(1|1)} \mapsto h_{(1|1)} + \rd^2 \alpha$. Notice this is {\it not} the gauge transformation discussed in the previous section, nor is it the gauge symmetry enjoyed by linearized gravity.

The gauge invariant curvature associated to $h_{(1|1)}$ we call $Q_{(2|1)}$, which is defined as:
\be
Q_{(2|1)} = \rd h_{(1|1)}\, .
\label{eq:Qcurv1}
\ee
A natural question to ask is: how do we enlarge the gauge symmetry so that transformations $h_{(1|1)} \mapsto h_{(1|1)} +\rd \xi_{(1|0)}$ are also unphysical? First,  notice that Lagrangians constructed from the curvature $Q_{(2|1)}$ would enjoy an electric symmetry related to shifts of $h_{(1|1)}$ that satisfy $\rd^2 \xi_{(1|0)}=0$.  We can gauge this putative symmetry by introducing a new gauge field $\Gamma_{(2|1)}$ and deforming the curvature $Q_{(2|1)}$ as:
\be
\mathcal{Q}_{(2|1)} = \rd h_{(1|1)} - \Gamma_{(2|1)}\, .
\ee
We do not want to gauge totally generic shifts of $h_{(1|1)}$, but just a small part. This is achieved by considering only transformations of the form 
\be
h_{(1|1)} \mapsto  h_{(1|1)} + \rd \xi_{(1|0)}\, ,  \qquad\qquad\qquad  \Gamma_{(2|1)} \mapsto \Gamma_{(2|1)} + \rd^2\xi_{(1|0)}\, . \label{eq:shift2}
\ee
The shift \eqref{eq:shift2} guarantees that a small part of the electric global symmetry will survive in the EFT. Notice that there is another curvature that we can construct purely from $\Gamma_{(2|1)}$:
\be
J_{(2|2)} = \rd \Gamma_{(2|1)}\, ,
\ee
which is gauge invariant as a consequence of the fact that $\rd^3 = 0$.

It seems that we are left with a new fractonic-type gauge symmetry for $\Gamma_{(2|1)}$, since this gauge field also transforms by a term with two derivatives in~\eqref{eq:shift2}. In order to recover the Einstein-like theories discussed in section~\ref{sec:hbs}, all we need to do is demand a flatness condition on $\mathcal{Q}_{(2|1)}$ to relate $\Gamma_{(2|1)}$ and $h_{(1|1)}$, which we could interpret as coming from an equation of motion of a BF-like action:
\be
\mathcal{Q}_{(2|1)} = 0 \qquad \implies \qquad J_{(2|2)}= \rd^2h_{(1|1)}\, .
\ee
As advertised, this construction recovers the familiar gauge invariance $h_{(1|1)} \mapsto h_{(1|1)} + \rd\xi_{(1|0)}$ of linearized gravity.  This type of gauge field was also introduced in~\cite{Pretko:2018jbi}, appearing in theories that violate rotational invariance, with a clear origin in lattice models. Here we have obtained this gauge theory from a further gauging of the original fracton-like symmetry. This is completely analogous to the Green--Schwarz mechanism \cite{Green:1984sg} in string theory.

While we have set $p=1$ in our example above, the whole construction goes through for any $p$, provided that the fractonic matter field $\Phi$ is gapped. If this is not the case, we would be forced to deal with dynamical extended objects when $p>1$. It would be interesting to construct UV complete theories for $p=1$ using these ingredients.

Even for $p=1$, extended objets make an appearance. Notice that the Green--Schwarz mechanism implies that $\rd h_{(1|1)}$ is no longer gauge invariant. As such, we can't couple it to worldlines in the usual way. Let's examine this. Before introducing $\Gamma_{(2|1)}$, the natural coupling between $h_{\mu | \nu}$ and an external worldline  is
\be\label{mono}
S^{({\rm monopole})}_{\rm line} = \int h_{\mu | \nu} \,  \frac{\dot{x}^\nu}{\sqrt{\dot{x}^2}}   \, \rd x^\mu\, ,
\ee
\noindent where $\dot{x}^\mu = \frac{\rd  x^\mu}{\rd \tau}$ for some parameterization of the worldline given by $\tau$. The expression above is manifestly diffeomorphism invariant on the worldline, as it must be.  However, this coupling is
only gauge invariant under $h_{\mu |\nu} \mapsto h_{\mu |\nu} + \partial_\mu \partial_\nu \xi$ if $\frac{\rd}{\rd \tau}\left( \frac{\dot{x}^\nu}{\sqrt{\dot{x}^2}}\right)=0$. This is the statement that monopoles are not mobile in fractonic theories. A dipole coupling is however possible:
\be
S^{({\rm dipole})}_{\rm line} = \int h_{\mu | \nu} \,  x^\nu  \, \rd x^\mu\, .
\ee
This coupling is gauge invariant and possible for any trajectory $x^\mu(\tau)$. Once we include the enhanced gauge symmetry by introducing $\Gamma_{(2|1)}$, this coupling needs to be reconsidered. Just as in string theory, gauge invariance can only be preserved by realizing the worldline as a boundary of  some two-dimensional surface, $\Sigma$. If that is the case, we can write:
\be\label{eq:dipoleact}
S^{({\rm dipole})}_{\rm surface} = \int_{\partial\Sigma} h_{\mu | \nu} \,  x^\nu  \, \rd x^\mu - \int_\Sigma \Gamma_{\mu \nu | \sigma} x^\sigma \rd x^\mu \wedge \rd x^\nu\, .
\ee

We therefore see that 
the theory that includes the new gauge field $\Gamma_{(2|1)}$ naturally includes higher dimensional defects. 
In the gapless gravity phase, $\Gamma_{(2|1)}=\rd h_{(1|1)}$ as a consequence of the flatness condition $\mathcal{Q}_{(2|1)}=0$. In this situation, the coupling above disappears as it becomes identically zero. The upshot is that the extra gauging has the physical effect of removing the dipole particles from our EFT, they no longer couple to the gauge fields. All that remains are the monopoles~\eqref{mono}, which are forced to move on geodesics. This builds the connection between biform symmetries and spacetime. From the perspective advocated here, this is a natural consequence of the properties of the original fractonic matter.

What is the EFT which agrees with the Einstein action~\eqref{einsact} in these variables? Schematically, it takes the form:
\be
S = \int \rd^dx\,  h \, G[\Gamma] + \Gamma^2\, .
\label{eq:biformEHaction}
\ee
We will write the precise form of this action for theories of interest in the following sections. The Einstein biform $G$ above is written in terms of $\Gamma$, rather than $h$. Because of this, it is not automatically conserved and gauge invariance requires the introduction of $\Gamma^2$ terms, with a precise structure that we will write explicitly in the later sections. The only possible gauge-invariant equations of motion are of course the flatness conditions
\be
G_{(1|1)} = 0\, , \qquad\qquad \mathcal{Q}_{(2|1)}=0\, .
\ee
This amounts to a first order formalism for the theories described in the previous section.
When describing linearized gravity, there is a simple and familiar interpretation for these biforms: $\Gamma_{(2|1)}$ is related to the usual Christoffel connection and $\mathcal{Q}_{(2|1)}$ is the nonmetricity tensor, which is set to zero here.

\subsection{Beyond maximal symmetries \label{sec;bms}}

Within the space of maximal biforms $\hat{\Omega}$, there is one more family of interesting theories that can be studied.
Consider as our starting point a $(p+1|p)$-biform current $H_{(p+1|p)}$ satisfying
\be
\rd H_{(p+1|p)}=0\, .
\label{eq:oddbiformcurv1}
\ee
This example differs from the previous case in that $H_{(p+1|p)}$ is what we called an odd biform. The important change is that the naive form of the Poincar\'e lemma that we used above does not hold~\cite{Bekaert:2002dt}. In order to see this, we need to leave the space of maximal forms.

While the interested reader can find the full details in~\cite{Bekaert:2002dt}, we will sketch here only the necessary machinery. We have to refine the concept of exterior derivative and allow it to act on either the left or right column of the Young tableaux, when this is possible. As such we define both left and right differentials:
\be
\rd_L : \Omega_{(p|q)} \rightarrow  \Omega_{(p+1|q)} \, ,\quad \forall p\geq q \, , \qquad\qquad \rd_R : \Omega_{(p|q)} \rightarrow  \Omega_{(p|q+1)} \, ,\quad \forall p\geq q+1\, .
\ee
The fact that $\rd_R$ does not act on even biforms, and at the same time even forms are not in the image of $\rd_L$ is why we did not need this technology in the previous section. 

With this refinement, rather than~\eqref{eq:oddbiformcurv1} what we need to impose is the two equations
\be\label{Hd1}
\rd_L H_{(p+1|p)}=0 \, ,\qquad \qquad \rd_R H_{(p+1|p)}=0\, ,
\ee
in order to guarantee that we can write $H_{(p+1|p)}$ in terms of a $(p | p-1)$-biform gauge potential as
\be\label{Hdef}
H_{(p+1|p)} = \rd_L \rd_R \,a_{(p|p-1)}\, .
\ee
The conditions~\eqref{Hd1} just guarantee that $* H_{(p+1|p)} *$ is conserved when taking the divergence with respect to an index in either of its two columns. 
This family of theories enjoys a more general gauge symmetry of the form $a_{(p|p-1)} \mapsto a_{(p|p-1)} + \rd_L \xi_{(p-1|p-1)} + \rd_R \chi_{(p|p-2)}$. Having already left the space of maximal biforms $\hat\Omega$, we see that these theories are not very different conceptually from having started with $H$ being a totally generic $(p|q)$-biform. We will not study this totally general case here, nevertheless
there is a very simple and interesting example of this family of theories that we will treat in more detail in appendix~\ref{sec:EM}, that we briefly describe. 

Consider the case where $H_{(2|1)}$ is a $(2|1)$-biform, which is the simplest nontrivial example. In this case, notice that $a_{(1|0)}$ has the simple gauge symmetry $a_{(1|0)} \mapsto a_{(1|0)}+ \rd_L \xi$, with $\xi$ a scalar, since no $(1|0)$-biform sits in the image of $\rd_R$. Given this simple gauge transformation, it is natural to suspect we are dealing with a familiar theory. 
Notice that the version of the Einstein biform which is relevant in this case is
\be
S_{(1|0)} = \tr\left( H_{(2|1)} \right)\, .
\ee
In components this reads
\be
S_\mu = \partial_\rho\partial^\rho a_\mu - \partial_\mu \partial_\rho a^\rho\, .
\ee
It can be easily checked that $S_\mu$ is automatically conserved, $\partial_\mu S^\mu =0$. From this, we can immediately write the Einstein action for these objects:
\be
S_E[a] = \int \rd^dx \, a_\mu S^\mu \, ,
\ee
\noindent which yields the equation of motion
\be
S_\mu =0\, .
\label{eq:sec2smeq}
\ee
What is this equation? Notice that, because of the reduced gauge symmetry of this particular case, there is another gauge invariant curvature:
\be
F_{(2|0)} =\rd_L a_{(1|0)}\,,
\ee
such that we have $H_{(2|1)} = \rd_R F_{(2|0)}$. Then, 
the equations of motion~\eqref{eq:sec2smeq} are nothing other than the Maxwell equations
\be
\partial_\mu F^{\mu \nu} =0\, .
\ee

We have thus obtained electromagnetism as a simple example of a non-maximal biform theory with an Einstein-type Lagrangian! Notice that this similarity is lost once the theory is written directly in terms of $F$. Of course, the irrelevant terms that we can add to this EFT are not the same if we restrict ourselves to observables that can only be expressed in terms of $H$. This is analogous to the different treatments of the scalar theory as a biform (i.e., Galileon superfluid) or single form (i.e., ordinary superfluid) theory discussed in section~\ref{sec:superfluid}. We comment further on this in appendix~\ref{sec:EM} below.

\subsection{Conserved charges and anomalies}\label{sec:charges}

Now that we have set up the notation for the exterior calculus of general biforms, let us return to the issue of conserved quantities in theories with maximal higher-biform symmetries. 
Consider a theory with a  $(p+1|p+1)$-biform conserved electric current $J_{(p+1|p+1)}$ satisfying:
\be
\rd * J_{(p+1|p+1)} * =  0\, .
\ee
We would like to study under which circumstances the $(p+1)$-form current $J^{(\zeta)}_{(p+1)}$ is conserved, where
\be
J^{(\zeta)}_{\mu_1 \cdots \mu_{p+1}} = J_{\mu_1 \cdots \mu_{p+1}| \nu_1 \cdots \nu_{p+1}} \zeta^{\nu_1 \cdots \nu_{p+1}} \, ,
\ee 
with $\zeta_{(p+1)}$ a $(p+1)$-form.
Taking the divergence of the equation above and using the conservation of $J_{(p+1|p+1)}$ we find:
\be\label{conform}
\partial^{\mu_1} J^{(\zeta)}_{\mu_1 \cdots \mu_{p+1}} = 0 \quad \implies \quad J_{\mu_1 \cdots \mu_{p+1}| \nu_1 \cdots \nu_{p+1}} \partial^{\mu_1} \zeta^{\nu_1 \cdots \nu_{p+1}}=0\, .
\ee
From the point of view of representation theory,  $J_{(p+1|p+1)}$ is a an irreducible $(p+1| p+1)$-biform. Therefore, equation~\eqref{conform} carries no information about the totally antisymmetric $(p+2)$-form $\rd_L \zeta_{(p+1)}$, its contraction with $J_{(p+1|p+1)}$ vanishes automatically by symmetry.  It does contain, however, information about the other irreducible possibility $\rd_R \zeta_{(p+1)}$, which is a $(p+1 |1)$-biform.

We will further require our current to be traceless, as is the case on-shell in the EFTs we studied above. In biform language we can then write the following constraint on $\zeta_{(p+1)}$, that will ensure conservation of $J^{(\zeta)}$:
\be
\left(\rd_R \zeta\right)_{(p+1 |1)} = \left(\eta \wedge \gamma_{(p)} \right)_{(p+1|1)}\, , \label{eq:zetaconstraintw}
\ee
where $\gamma_{(p)}$ is some $p$-form---which we write below---that, when combined via the wedge product with the metric, produces a $(p+1|1)$-biform. In component notation we can write:
\be
\partial_\mu \zeta_{\nu_1 \cdots \nu_{p+1}} + \frac{1}{p+1} \sum_{i=1}^{p+1} \partial_{\nu_i} \zeta_{\nu_1 \cdots \mu \cdots \nu_{p+1}} = \frac{p+2}{(d-p) (p+1)}\sum_{i=1}^{p+1} g_{\mu \nu_i} \partial^\rho \zeta_{\nu_1 \cdots \rho \cdots \nu_{p+1}}\, .
\label{eq:zetaconstraint}
\ee
Here we have determined  $\gamma_{(p)}$ by taking the trace of~\eqref{eq:zetaconstraintw} and demanding consistency. The equation~\eqref{eq:zetaconstraint} is the generalization of the conformal Killing equation to a $(p+1)$-form.  The $(p+1)$-forms that solve this equation are sometimes referred as conformal Killing--Yano tensors.  In flat space, this equation can be solved by acting with derivatives and contracting appropriately to obtain the equation
\be
\partial_\rho \partial_\sigma \partial_\mu \zeta_{\nu_1 \cdots \nu_{p+1}} =0\label{eq:3derivativesvector}\, ,
\ee
that a conformal Killing $(p+1)$-form must satisfy.
The most general solution to this auxiliary equation that is compatible with the original conformal Killing equation \eqref{eq:zetaconstraint} is given by
\be
\begin{aligned}
 \zeta_{\mu_1 \cdots \mu_{p+1}} = A^{(p+1)}_{\mu_1 \cdots \mu_{p+1}}  &+ B^{(p+2)}_{\mu_1 \cdots \mu_{p+2}} x^{\mu_{p+2}}  + \left(\sum_{i=1}^p C^{(p)}_{\mu_1 \cdots \mu_{i-1} \mu_{p+1} \mu_{i+1}\cdots \mu_{p}}  x_{\mu_i} - C^{(p)}_{\mu_1 \cdots \mu_{p}}  x_{\mu_{p+1}} \right) \\
 &+ \left(D^{(p+1)}_{\mu_1 \cdots \mu_{p+1}} x^2 -2 \sum_{i=1}^{p+1} x_{\mu_i}  D^{(p+1)}_{\mu_1 \cdots \mu_{i-1}\sigma\mu_{i+1} \cdots \mu_{p+1}} x^\sigma\right)\, ,
\end{aligned}
\label{eq:chargedsolntensors}
\ee
where  $A^{(q)}, B^{(q)}, C^{(q)}$, and $D^{(q)}$ are constant $q$-forms. Notice that $A$ is analogous to translations, $B$ to rotations, $C$ to dilations, and $D$ to special conformal transformations. This general solution can be cast in a nicer form in terms of the standard CFT embedding of ${\mathbb R}^d$ into  ${\mathbb R}^{d+2}$ with metric $\eta_{AB}={\rm diag}(-1,1,\eta_{\mu\nu})$,  $X^A(x)=\left({1+x^2\over 2},{1-x^2\over 2},x^\mu\right)$.
In these variables, the conformal Killing tensors can be written as
\be  \zeta_{\mu_1 \cdots \mu_{p+1}}=K_{A_1\cdots A_{p+1}A_{p+2}}{\partial X^{A_1}\over \partial x^{\mu_1}}\cdots {\partial X^{A_{p+1}}\over \partial x^{\mu_{p+1}}}X^{A_{p+2}},\ee
where $K_{A_1\cdots A_{p+2}}$ is a constant $(p+2)-$form in the $(d+2)-$dimensional embedding space.   Upon dimensional reduction to the physical $d$-dimensional space, the $(p+2)-$form $K$ breaks up into $(p+2)-$form, 2 different $(p+1)-$forms and a $p-$form, which are precisely the coefficients $B^{(p+2)}, A^{(p+1)}, D^{(p+1)}, C^{(p)}$.

We are interested in these equations when $d \geq 2 p+2$, since in lower dimensions our theories of interest are gapped. Something special happens in the equations above when $d=2p +2$. For $p=0$, we have a free scalar CFT$_2$ and there exist an infinite number of solutions to the conformal Killing equation giving rise to a Kac--Moody current algebra. For higher values of $p$, the solutions \eqref{eq:chargedsolntensors} are still the full set of solutions. Nevertheless, the structure of the equations is different in these special dimensions and the solutions further satisfy the Laplace equation:
\be
\partial_\rho\partial^\rho \zeta_{\nu_1 \cdots \nu_{p+1}} =0 \quad \quad \textrm{when} \qquad d=2 p+ 2\, .
\ee
For electromagnetism in $d=4$, it was found that the special structure of the equations leads to the existence of 0-form symmetries which also imply the existence of Kac--Moody algebra~\cite{Hofman:2018lfz}. This suggests that $d=4$ linearized gravity might also enjoy a symmetry enhancement for lower codimension charges. An explicit construction of these charges could possibly arise from the double copy techniques in gravity~\cite{Bern:2019prr}.  Soft graviton theorems can already be viewed as an example of this enhanced symmetry structure~\cite{He:2014laa}.

In conclusion, we have seen that charges for theories with maximal higher biform symmetries are labeled by conformal Killing--Yano forms. Given a current satisfying $\rd * J_{(p+1|p+1)} * =  0$ and $\tr (J_{(p+1|p+1)})=0$, conserved charges are given by:
\begin{tcolorbox}[colframe=white,arc=0pt,colback=greyish2]
\be\label{ech}
Q^{(\zeta)}(\Sigma_{d-p-1}) = \int_{\Sigma_{d-p-1}} * J^{(\zeta)}_{(p+1)}\, .
\ee
\end{tcolorbox}
\vspace{-6pt}
\noindent
Notice that some of these charges might be trivial. Indeed, provided there are not topological features and that the fundamental fields are globally well defined, it is not hard to see that only $\zeta$s satisfying $\partial^\rho \zeta_{\nu_1 \cdots \nu_p \rho} \neq 0$ will yield nontrivial charges. The rest just give total derivatives. In the language of the conformal group, the nontrivial generators are the ones that are not part of the Poincar\'e group, i.e., the dilations and special conformal transformations.   For us, it is the $C^{(p)}$ and $D^{(p+1)}$ that parameterize the non-trivial charges.\footnote{We stress again that this is the case if there are no topological defects. In many cases these defects would not be allowed. But in the critical dimension $d= 2p+2$, they can appear as part of the magnetic symmetries. See section \ref{sec:chargedsol} for such an example.}

It is relatively easy to see the origin of these conserved quantities.
The Einstein type actions we consider turn out to be invariant under the global symmetry generated by
\be\label{globtran}
h_{\mu_1 \cdots \mu_p | \nu_1 \cdots \nu_p} \mapsto h_{\mu_1 \cdots \mu_p | \nu_1 \cdots \nu_p} + {\cal Y}_{(p|p)}\, \partial^\rho \left(\zeta_{\mu_1 \cdots \mu_p \rho} \epsilon_{\nu_1 \cdots \nu_p} \right)\, ,
\ee
where ${\cal Y}_{(p|p)}$ is the Young projector onto a $(p|p)$-biform, and $\zeta_{(p+1)}$ is a conformal Killing $(p+1)$-form. It is clear that this transformation is only nontrivial when $\partial^\rho \zeta_{\mu_1 \cdots \mu_p \rho}\neq 0$. The divergence of a conformal Killing $(p+1)$-form is at most linear in the coordinates, so the current $J_{(p+1|p+1)}=\rd^2 h_{(p|p)}$ is manifestly invariant under this global transformation. It is then easy to see that this implies all Einstein and Maxwell-type terms are invariant under this symmetry.

Magnetic charges can be constructed in similar fashion. Notice, however, that our magnetic current is not generically traceless. In that case, looking at (\ref{magcha1}), we must demand a more constraining Killing type equation for $\varrho$:
\be
\partial_\mu \varrho_{\nu_1 \cdots \nu_{p+1}} + \frac{1}{p+1} \sum_{i=1}^{p+1} \partial_{\nu_i} \varrho_{\nu_1 \cdots \mu \cdots \nu_{p+1}} =0\, .
\label{eq:rhoconstraint}
\ee
which has the smaller set of solutions:
\be
\begin{aligned}
 \varrho_{\mu_1 \cdots \mu_{p+1}} = A^{(p+1)}_{\mu_1 \cdots \mu_{p+1}}  &+ B^{(p+2)}_{\mu_1 \cdots \mu_{p+2}} x^{\mu_{p+2}}  \, .
\end{aligned}
\label{eq:chargedsolntensorsrho}
\ee

\paragraph{Coupling to gauge fields and anomalies:}
In what follows, we would like to couple these theories to background gauge fields that gauge the symmetries~\eqref{globtran}. In order to that, we let our fundamental field $h_{(p|p)}$ shift by the divergence of an arbitrary $(p+1|p)$-biform as:
\be
h_{\mu_1 \cdots \mu_p | \nu_1 \cdots \nu_p} \mapsto h_{\mu_1 \cdots \mu_p | \nu_1 \cdots \nu_p} +{\cal Y}_{(p|p)}\, \partial^\rho \Lambda_{\rho\mu_1 \cdots \mu_p  |\nu_1 \cdots \nu_p}  \, ,
\ee
and demand our actions are invariant by the addition of appropriate compensating gauge fields. We do this in detail in the following sections for the cases of interest, including linearized gravity. Schematically, the expectation is that the curvatures get nonlinearly gauged by the addition of background $(p+1|p+1)$-biform fields $C_{(p+1 | p+1)}$:\footnote{Throughout this work, we will denote currents before the introduction of background gauge fields with capital roman letters and their gauge-invariant counterparts with calligraphic capital letters.}
\be\label{naiveg}
J_{(p+1 | p+1)} \rightarrow {\mathcal{J}}_{(p+1 | p+1)} \equiv J_{(p+1 | p+1)} - C_{(p+1 | p+1)}\, .
\ee
This background gauge field transforms as $C_{(p+1 | p+1)} \mapsto C_{(p+1 | p+1)} + \rd^2 \,  \partial \cdot \Lambda_{(p+1|p)}$ under gauge transformations. Already at this level, it is clear that this will lead to a mixed anomaly with the magnetic conservation equation since:
\begin{tcolorbox}[colframe=white,arc=0pt,colback=greyish2]
\be
\rd {\mathcal{J}}_{(p+1 | p+1)} = - \rd {\cal C}_{(p+1 | p+1)} \neq 0\, .
\ee
\end{tcolorbox}
\vspace{-6pt}
\noindent
This anomaly is directly responsible for protecting the gaplessness of this phase, as was studied for superfluids in~\cite{Delacretaz:2019brr}. In the following, we extend the arguments to linearized gravity, showing how it can be interpreted as a gapless phase defined by a particular structure of mixed anomalies.

It turns out that~\eqref{naiveg} is not quite the end of the story. There are two interlocked reasons for this. The first is that we would expect gauge fields that act as sources for the electric current~\eqref{eq:electriccurrent} to transform by something that has a single derivative of the gauge parameter $\Lambda_{(p+1|p)}$, in order to couple to a conserved current satisfying $\rd * J_{(p+1|p+1)} * =0$. The second is that the Einstein term is no longer invariant under the original symmetries, parameterized by conformal Killing $(p+1)$-forms, once we gauge as in~\eqref{naiveg}. We expect this not to be allowed for abelian symmetries. It turns out that the way out of both puzzles is that $C_{(p+1|p+1)}$ must be constructed as:
\be
C_{(p+1|p+1)} \sim \rd \left(\partial \cdot A_{(p+1|p+1)}\right)\, ,
\label{eq:cintermsofA}
\ee
\noindent where $A_{(p+1|p+1)}$ is also a $(p+1|p+1)$-biform. The equation~\eqref{eq:cintermsofA} is just schematic; explicit expression are written for the examples of interest below. For transformations of the form~\eqref{globtran}, $A_{(p+1|p+1)}$ shifts by a term proportional to the conformal Killing equation, guaranteeing both invariance under global symmetries and that the gauge field transforms by something linear in derivatives.


\section{Superfluids\label{sec:superfluid}}

We begin by describing the simplest theory that exhibits a biform symmetry. This is a scalar theory which can be thought of as a somewhat peculiar kind of superfluid that has a 0-biform symmetry. 
We first review the ordinary superfluid case from the perspective that it is a gapless phase protected by anomalies, as was considered in~\cite{Delacretaz:2019brr}. We then add a new twist to the superfluid discussion by considering a 0-biform superfluid, which is related to the physics of Galileons~\cite{Nicolis:2008in} which share many of the features of gravity~\cite{Bonifacio:2019rpv}. This galileon superfluid theory belongs to the same class of theories as gravity, and shares a number of features with linearized gravity, making it a useful illustrative example.

Both the ordinary superfluid and the galileon superfluid are members of a broader class of theories that could be called fractonic superfluids. All of these theories agree in the deep infrared, and correspondingly have the same scalar gapless degree of freedom. The theories differ in their symmetries, which are of the form $\delta \phi = c_{\mu_1\cdots\mu_N}x^{\mu_1}\cdots x^{\mu_N}$, where $c_{\mu_1\cdots\mu_N}$ is a traceless tensor. (The ordinary superfluid has the $N=0$ symmetry and the galileon superfluid additionally has the $N=1$ symmetry.)  The theories have different allowed irrelevant deformations depending on the maximal value of $N$ which is a symmetry. From the operator perspective, the fundamental conserved currents in these theories are $J_{\mu_1\cdots \mu_{N+1}}$, and in the deep infrared are given by $J_{\mu_1\cdots \mu_{N+1}} = \partial_{\mu_1}\cdots \partial_{\mu_{N+1}}\phi$. The fact that all of these theories agree in the deep IR---or equivalently that all of these currents propagate a gapless scalar---is a manifestation of the so-called inverse Higgs effect, where a single Goldstone degree of freedom nonlinearly realizes several spacetime symmetries \cite{Ivanov:1975zq}.

\subsection{Ordinary superfluid}

As a first example in order to orient ourselves, we describe the effective field theory of an ordinary superfluid \cite{Son:2002zn}.\footnote{We are slightly abusing terminology, strictly speaking the hydrodynamics of a superfluid is described by expanding the effective field theory that we will construct around a finite density configuration~\cite{Son:2002zn}. See~\cite{Delacretaz:2019brr} for the details of how this works in this context.} This is a rephrasing of the discussion that appeared in~\cite{Delacretaz:2019brr}, emphasizing the aspects that will be important for the extension to the gravity case. The perspective that we take begins with the global symmetries that define a superfluid phase: $0$-form $U(1)$ symmetry and a $(d-2)$-form $U(1)^{(d-2)}$ symmetry in $d$ spacetime dimensions. The gapless superfluid phonon can be thought of as an inevitable consequence of the mixed anomaly between these two global symmetries---sourcing one of the conserved currents via a background gauge field necessarily causes a non-conservation of the other current.

\subsubsection{Conservation equations\label{sec:superfluidconsequation1}}

An interesting and recurring theme in physics is that phases of matter can be classified by anomalies. In~\cite{Delacretaz:2019brr} it was shown how this classification scheme can also be usefully applied to systems with spontaneously broken global symmetries, providing an alternative viewpoint on Goldstone's theorem.
In this spirit, following~\cite{Delacretaz:2019brr} we {\it define}
a superfluid as a phase involving two conserved currents with a mixed 't Hooft anomaly. 

The effective field theory has two conserved currents: one is
 an ordinary 1-form Noether current, $J^\mu$, associated to a $0$-form symmetry. 
 The other current is a  $(d-1)$-form $
K^{\mu_1\cdots\mu_{d-1}}$, and is associated to a $(d-2)$-form symmetry. In the standard presentation of the theory, this latter current is usually called topological, but---as was discussed in section~\ref{sec:higherbiformsyms}---this depends on the variables that we choose to parameterize the theory.

We now want to gauge one of these symmetries, and we can choose which one. The standard way to do this is to introduce a background gauge field that couples to the symmetry currents. Upon turning on a background gauge field for one symmetry current, the other current will cease to be conserved. The fact that its conservation equation becomes anomalous captures the impossibility of simultaneously gauging both symmetries.

If we choose to gauge the ordinary $U(1)$ shift symmetry by introducing a $1$-form background gauge field $A_\mu$, we need to improve both the ordinary $U(1)$ current and the higher-form current in order to make them gauge invariant. We will denote these gauge-invariant currents by ${\cal J}^\mu$ and ${\cal K}^{\mu_1\cdots\mu_{d-1}}$, respectively. Because we have decided to source ${\cal J}^\mu$ by $A_\mu$, conservation of this current will be maintained: $\partial_\mu {\cal J}^\mu =0$. On the other hand, this gauge improvement will spoil the conservation of ${\cal K}^{\mu_1\cdots\mu_{d-1}}$; we instead have
\be
\partial_\mu \mathcal{J}^\mu = 0\, ,\qquad \qquad \partial_{[\mu}*\mathcal{K}_{\nu]} = -\frac{1}{2}F_{\mu\nu}\, ,\label{eq:superfluidcons1}
\ee
where
$F_{\mu\nu} = 2\partial_{[\mu}A_{\nu]}$ is the usual field strength, so that the anomaly is gauge invariant, as it must be.\footnote{Note that we are normalizing the $J^\mu$ and $K^{\mu_1\cdots\mu_{d-1}}$ currents in the way that is appropriate for compact symmetries. This means that even in the deep infrared they are not exactly Hodge duals of each other, because their normalizations differ.
}
Here we have dualized the ${\cal K}$ current for convenience, so that its conservation corresponds to the vanishing of the exterior derivative of $*{\cal K}$.
Note that---unlike axial-type anomalies---this mixed anomaly between $U(1)$ and $U(1)^{(d-2)}$ symmetries is not dimension-dependent, and can occur in any spacetime dimension.

The equation where the anomaly appears is always a choice, and so we can choose to put the anomaly in the $\mathcal{J}^\mu$ conservation equation if we like. This is achieved by gauging the $(d-2)$-form symmetry by introducing a $(d-1)$-form background gauge field, $B^{\nu_1\cdots \nu_{d-1}}$. In this case, the conservation equations instead read
\be 
\partial_\mu \mathcal{J}^\mu =  \epsilon_{\mu_1\cdots\mu_d}\partial^{\mu_1}B^{\mu_2\cdots\mu_d}\, , \qquad \qquad \partial_{[\mu}* \mathcal{K}_{\nu]} = 0\, ,\label{eq:superfluidcons2}
\ee
where $\epsilon_{\mu_1\cdots\mu_d}\partial^{\mu_1}B^{\mu_2\cdots\mu_d}$ is the (Hodge dual of the) field strength associated to the higher-form gauge field.
Either presentation of the anomaly---eq.~\eqref{eq:superfluidcons1} or eq.~\eqref{eq:superfluidcons2}---is sufficient to completely fix the current-current correlation function $\braket{J_\mu*K_\nu}$, as we now review.

\subsubsection{Current-current correlator\label{sec:SFcor}}

The current-current two-point function  $\braket{{ J}_\mu*{ K}_\nu}$ is completely fixed by the anomalous conservation conditions in the presence of a background field that sources one of the currents. The way that this manifests at the level of correlation functions with the background fields turned off is that it is impossible to impose conservation of both ${J}^\mu$ and ${K}^{\mu_1\cdots\mu_{d-1}}$ at coincident points. There will always be contact terms that violate one of the two conservation conditions. Since we will be interested in the properties of the current-current correlator at coincident points, it is natural to Fourier transform, so that position space contact terms appear as analytic terms in the momentum $p$ in Fourier space.

We can see this structure by parameterizing the most general possible correlator. 
The Fourier transform of the mixed correlator is constrained by Lorentz invariance to take the form
\be 
\braket{{ J}_\mu(p)*{ K}_\nu(-p)}\equiv \int \rd^dx\, 	 e^{ip\cdot x}\, \braket{{J}_\mu(x)* {K}_\nu(0)} = c_1(p^2)p_\mu p_\nu + c_2(p^2) p^2 g_{\mu\nu}\, ,\label{eq:Pimunudebut}
\ee
where $c_1$, $c_2$ are at this point arbitrary functions.  Note that even though the two currents in the correlator are different, this is symmetric in $\mu$ and $\nu$. 

We now want to impose conservation of the currents.   We will see that we can pick one of the two currents to be conserved everywhere, and that the mixed anomaly will then constrain the other current to not be conserved at coincident points. This failure of conservation by contact terms is the hallmark of an anomaly.
The final form of  $\braket{ J_\mu(p)* K_\nu(-p)}$ depends on which current we choose to be conserved.
The different choices that we can make shift the final two-point function by local contact terms. However, the nonlocal part of the correlator that describes physics at separated points is completely fixed, and it is this part that we will use to deduce facts about the spectrum of the theory.

\begin{itemize}

\item {\bf The $\boldsymbol 1$-form current is conserved:} We first consider the case where we demand that the current $J^\mu$ is conserved everywhere---including at coincident points. This  requires that the two free functions in~\eqref{eq:Pimunudebut} are related as $c_2(p^2) = -c_1(p^2)$. The residual freedom to select $c_1(p^2)$ is then fixed by the anomaly equation~\eqref{eq:superfluidcons1},  which when differentiated with respect to the background fields to get the two-point function and written in momentum space, requires the failure of conservation to be purely analytic in $p$ (corresponding to a position space contact term),
\be p_{[\rho}\braket{J_{|\mu|}(p)*K_{\nu]}(-p)}=-p_{[\rho}\eta_{\nu]\mu}\,. \ee
Imposing this completely fixes the two-point correlator to be~\cite{Delacretaz:2019brr}
\be 
\braket{J_\mu(p)*K_\nu(-p)} = \frac{p_\mu p_\nu -p^2\eta_{\mu\nu}}{p^2}\, .\label{eq:superfluidcor1}
\ee
The conservation of $J^\mu$ holds exactly (i.e., even at coincident points), while the dual conservation of $*K_\nu$ holds only at separated points. The failure of $K^{\mu_1\cdots\mu_{d-1}}$ to be conserved at coincident points due to a contact term is precisely the statement that there is an 't~Hooft anomaly between the symmetries that these currents correspond to.

\item
{\bf The $\boldsymbol{(d-1)}$-form current is conserved:} We can instead require that $K^{\mu_1\cdots \mu_{d-1}}$ is conserved even at coincident points. This forces us to set $ c_2(p^2) = 0$ in~\eqref{eq:Pimunudebut}. It is then not possible to make $J^\mu$ identically conserved. The anomaly equation~\eqref{eq:superfluidcons2}, when differentiated with respect to the background fields to get the two-point function and written in momentum space, requires the failure of conservation to be the particular contact term
\be p^\mu\braket{J_\mu(p)*K_\nu(-p)}=p_\nu\,. \ee
Imposing this completely fixes the two-point correlator to be
\be 
\braket{J_\mu(p)*K_\nu(-p)} = \frac{p_\mu p_\nu}{p^2}\, .\label{eq:superfluidfinalnonlocal}
\ee

\end{itemize}

We see that the difference between the two correlators \eqref{eq:superfluidcor1} and \eqref{eq:superfluidfinalnonlocal} is a local contact term $\propto\eta_{\mu\nu}$, which is what we anticipated.
However, the nonlocal part of the correlator, which controls the behavior at separated points, is identical. This part is universal in any theory with these conserved currents, and will be the part that tells us the theory has a massless mode.

\subsubsection{\KL for a superfluid}

We now perform a \KL \cite{Kallen:1952zz,Lehmann:1954xi} spectral decomposition of the correlators in order to prove that there is a massless scalar mode in the spectrum~\cite{Delacretaz:2019brr}. As we saw in the previous section, our two possible choices for which current is conserved merely serve to shift the contact term appearing in the correlator, but in both cases the nonlocal part is the same. In order to perform a \KL decomposition, it is convenient to dualize one of the currents so that the two operators in the correlator have the same number of Lorentz indices. It is more convenient to dualize $K^{\mu_1\cdots\mu_{d-1}}$ to a vector, so we will do that in the following.
As we elaborate on in appendix~\ref{ap:KLrep}, only the nonlocal part is necessary to deduce the spectrum of the theory, the spectral decomposition is insensitive to what contact terms are present in the theory. We can therefore take the correlator in the form~\eqref{eq:superfluidfinalnonlocal}.

Our starting point is the spectral representation of the current two-point function  (see appendix~\ref{ap:KLrep} for details of its derivation)
\be
\braket{J_\mu(p)*K_\nu(-p)}= 
\int_0^\infty \rd s\, \frac{s}{p^2 + s}	 \left[\rho_1(s)\tl{\Pi}^{(1)}_{\mu\nu}-\rho_0(s)\tl{\Pi}^{(0)}_{\mu\nu}\right]\label{eq:KL11}\, .
\ee
Here $\rho_0(s)$ and $\rho_1(s)$ are the spin-$0$  and spin-$1$ spectral densities, which tell us about the presence of spin 0 and spin 1 states that couple to the currents  (note that only a massive spin-1 field can couple to a conserved current~\cite{Weinberg:1980kq,Weinberg:2020nsn,Distler:2020fzr}, so the spectral density $\rho_1$ must go to zero as $s\to 0$), and the projectors are defined as 
\be 
\tl{\Pi}^{(0)}_{\mu\nu} = -\frac{p_\mu p_\nu}{s}\, , \qquad\qquad \tl{\Pi}^{(1)}_{\mu\nu}  =\eta_{\mu\nu}+\frac{p_\mu p_\nu}{s}\, . \label{eq:superfluidproj2}
\ee
which are related to the usual projectors
\be 
\Pi^{(0)}_{\mu\nu} = \frac{p_\mu p_\nu}{p^2}\, , \qquad\qquad \Pi^{(1)}_{\mu\nu}  = \eta_{\mu\nu}-\frac{p_\mu p_\nu}{p^2}\, , \label{eq:superfluidproj}
\ee
by the replacement $p^2\rightarrow -s$, and agree with them on-shell where $p^2=-s$. 

We can then rewrite \eqref{eq:KL11} as 
\be 
\braket{J_\mu(p)*K_\nu(-p)}= \int_0^\infty \frac{\rd s}{p^2+s}\bigg[s\rho_1(s)\eta_{\mu\nu}+  \,                                                                                                                                                                                                                                                                                                                                                                                                                                                                                                        \big(\rho_1(s) + \rho_0(s)\big)\, p_\mu p_\nu\bigg]\,.\label{eq:KLsup1}
\ee

In order to reproduce~\eqref{eq:superfluidfinalnonlocal}, the spectral densities must be given by\footnote{Note that in performing the spectral decomposition, the distinction between the actual projectors given in \eqref{eq:superfluidproj} and the off-shell projectors~\eqref{eq:superfluidproj2}, along with the constraint that $\rho_1 \to 0$ as $s\to 0$, are crucial. 
Notice that the
two-point function with different contact terms can be written directly in terms of projectors. 
For example, we can write~\eqref{eq:superfluidfinalnonlocal} as $\braket{J_\mu* K_\nu} = \Pi^{(0)}_{\mu\nu}$ and~\eqref{eq:superfluidcor1} as $\braket{J_\mu*K_\nu} = -\Pi^{(1)}_{\mu\nu}$, respectively.
This would seem to suggest that when we perform the spectral decomposition, we could actually decompose this latter correlator in terms of spin-1 states, implying a massless spin-1 mode instead of a spin-0 one. However, because of the constraint that $\rho_1$ has no support at $s=0$, it is impossible to reproduce the two-point function with $\rho_1(s) \sim \delta(s)$.
}
\be 
\rho_1(s) = 0\, , \qquad \qquad \rho_0(s) = \delta(s)\, .
\ee
That is, we see that there is a massless spin-0 mode in the spectrum as a consequence of the structure of anomalies, and furthermore, no other states contribute to the two-point function.\footnote{The same conclusion can be reached with less work by noting that in the spectral decomposition the only massless particle that can couple to a conserved current is a scalar~\cite{Weinberg:1980kq,Weinberg:2020nsn}, and since the two-point function has a pole at $p^2\to 0$, there must be a massless particle in the spectrum.} The local contact term does not affect the spectral functions (see appendix~\ref{ap:KLrep}), so regardless of the choices made in where to put the anomaly, there is a gapless mode, establishing a version of a Goldstone theorem~\cite{Delacretaz:2019brr}.

\subsubsection{The superfluid EFT}

The previous discussion did not rely on any action to derive the current conservation equations. However, the universality of the result suggests 
that one can build an effective field theory  that captures this physics in the form of an infrared Lagrangian with the appropriate degree of freedom, which further includes irrelevant corrections via higher dimension operators.

Indeed, the superfluid EFT is given by the action
\be 
S = a\int\,  \rd^dx \, \left[\frac{1}{2}\partial_\mu\phi(x)\partial^\mu\phi(x)\right]\, + \cdots ,\label{eq:scalaraction}
\ee
where $\cdots$ represents the possible inclusion of irrelevant corrections which we discuss below, and $a$ is the normalization of the action. The action \eqref{eq:scalaraction} is invariant under the usual constant shift 
\be 
\phi(x)\mapsto \phi(x) + c\, ,
\ee
for which $\phi$ serves as the Goldstone boson. The corresponding conserved current is $J_\mu = \partial_\mu\phi$.
This global symmetry can be gauged in the free theory by introducing a $1$-form background gauge field $A_\mu$ as 
\be 
S =a \int\,  \rd^dx \, \left[\frac{1}{2}(\partial_\mu\phi-A_\mu)^2\right]\, ,\label{eq:Actionsuperfluid1}
\ee
with gauge transformations $\delta\phi(x) = c(x)$ and $\delta A_\mu = \partial_\mu c(x)$. The equation of motion derived from \eqref{eq:Actionsuperfluid1} is 
\be 
\square\phi = \partial_\alpha A^\alpha\, .\label{eq:eomsuperfluid1}
\ee
The gauge invariance of the action guarantees that the current
\be 
\mathcal{J}_\mu \equiv \frac{\delta S}{\delta A^{\mu}} = a\left(\partial_\mu \phi-A_\mu\right)\, ,
\ee
is conserved on-shell.

There is additionally a topological current
\be
K_{\mu_1\cdots\mu_{d-1}} = \epsilon_{\mu_1\cdots\mu_{d-1}\alpha}\partial^{\alpha}\phi\, ,
\ee
which in the absence of $A_\mu$ is identically conserved. However, in the presence of the background gauge field for the shift symmetry, we have
\be 
\partial_{[\mu}*{\cal K}_{\nu]} = -\frac{1}{2}F_{\mu\nu}\, ,
\ee
where $*{\cal K}_\mu = \partial_\mu \phi-A_\mu$ is the (dual of) the gauge-invariant version of $K$. (Notice that ${\cal J}_\mu$ and $*{\cal K}_\mu$ differ only by normalization at the free level.)

The action~\eqref{eq:scalaraction} produces the two-point function derived in equation~\eqref{eq:superfluidfinalnonlocal}. 
Typical analytic corrections for an ordinary superfluid are of the form $(\partial_\mu\phi \,\partial^\mu\phi )^{n} ,\, n\in \mathbb{N}$. In this sense, the superfluid EFT is a Maxwell action in the language of sections~\ref{sec:hbs} and~\ref{sec;bms}, which has a $0$-form symmetry. 

Less obvious is the fact that  the most relevant term of this EFT enjoys another symmetry of the form $\delta\phi = c_\mu x^\mu$, which is the defining feature of a Galileon theory \cite{Nicolis:2008in}, as well as an infinite tower of higher symmetries with higher powers of $x^\mu$ \cite{Hinterbichler:2014cwa,Griffin:2014bta,Hinterbichler:2015pqa}. If we want to describe this EFT and its irrelevant corrections that preserve this enhanced symmetry, we must reinterpret the leading quadratic term as an Einstein action for a maximal $0$-biform symmetry. This theory then shares many features with gravity. We describe this perspective in the following section.

\subsection{Galileon superfluid\label{sec:GalSup}}

We now want to consider a slightly different superfluid theory, which differs from the usual superfluid in that it has a larger set of symmetries. In order to make these symmetries more intuitive, we show how one could ``discover" them in the IR action~\eqref{eq:scalaraction} below. Once we have a handle on this theory and its symmetry structure, we invert the logic and  study its anomalies and universal current-current two-point function.

Let us start by rewriting the action \eqref{eq:scalaraction} in a slightly different form by simply integrating by parts 
\be 
S = a \int \rd^d x\left[-\frac{1}{2}\phi\, \partial_\mu \partial^\mu \phi\right]\, .\label{eq:scalaraction2}
\ee
While the action \eqref{eq:scalaraction} was manifestly invariant under the usual $U(1)$ shift symmetry, by writing it in the form~\eqref{eq:scalaraction2} we see it admits a larger set of symmetries which include the more general shifts:\footnote{In fact, the action~\eqref{eq:scalaraction2} has an infinite number of nonlinearly realized symmetries where $\phi$ shifts by any harmonic function of the form $c_{\mu_1\cdots\mu_N}x^{\mu_1}\cdots x^{\mu_N}$, where $c_{\mu_1\cdots\mu_N}$ is traceless (see, e.g.,~\cite{Hinterbichler:2014cwa}). The symmetry~\eqref{eq:fancyvar} is the one whose EFT fits into the class of biform theories that we are interested in presently, but EFTs that preserve more of the shift symmetries would describe other fractonic superfluids.
}
\be 
\phi(x)\mapsto \phi(x) + \partial_\mu \xi^\mu\, ,\label{eq:fancyvar}
\ee
where $\xi^\mu$ is a conformal Killing vector satisfying \eqref{eq:3derivativesvector}. The most general solution to \eqref{eq:3derivativesvector} is given in \eqref{eq:chargedsolntensors}.  In particular, any vector $\xi^\mu$ that satisfies \eqref{eq:3derivativesvector} has a divergence $\partial_\mu \xi^\mu$ which is either a constant, $c$, or linear in coordinates, i.e., $\partial_\mu \xi^\mu = c + c_\mu x^\mu$, with $c, c_\mu$ constant. It is straightforward to check that these two cases are symmetries of the action \eqref{eq:scalaraction2}. These symmetries are the ones relevant for theories of Galileons~\cite{Nicolis:2008in}, so we will refer to the type of superfluid that has this symmetry as a Galileon superfluid. It is also simple to see that Killing vectors that satisfy $\partial_\mu \xi^\mu=0$ cannot generate non-trivial symmetries if the fundamental field $\phi(x)$ is globally well defined and topologically trivial. These Killing vectors are associated with translations and rotations in (\ref{eq:chargedsolntensors}).

In the deep infrared, where we only need the leading order term in the EFT, this theory coincides with the usual superfluid \eqref{eq:scalaraction}. 
However, the allowed irrelevant corrections are different in each case. We have already described the subleading corrections to~\eqref{eq:scalaraction}, but the subleading corrections to~\eqref{eq:scalaraction2} involve more derivatives and are of the form $(J_{\mu|\nu})^n$ (with appropriate index contractions), where $J_{\mu|\nu} = a\,  \partial_\mu \partial_\nu \phi$ is the conserved current associated to \eqref{eq:fancyvar}.\footnote{\label{ft:galWZ}In addition to these interactions---which are exactly invariant under the relevant global symmetries---there are interactions with fewer derivatives that are also invariant up to a total derivative. These are known as the Galileon interactions~\cite{Nicolis:2008in}, and can be thought of as Wess--Zumino (WZ) terms for the spontaneous breaking of the symmetries~\cite{Goon:2012dy}. Since they have the fewest derivatives per field, they will be the least irrelevant terms in the infrared, though they are not renormalized or induced by integrating out matter that is coupled in a manner that preserves the symmetries we are discussing \cite{Goon:2016ihr}.} These terms are more restricted than the generic interactions of a superfluid and, hence, the two theories agree solely at leading order, i.e., in the deep infrared.

\subsubsection{First-order EFT Lagrangian}

In order to couple this theory to background gauge fields for the symmetry \eqref{eq:fancyvar}, it is convenient to rewrite the action~\eqref{eq:scalaraction2} in first-order form, along the lines of the discussion in section~\ref{fracgau}. For the free scalar theory, the first-order formulation is 
\be 
S = -a\int \rd^d x\left(\phi\,\partial^\mu s_\mu + \frac{1}{2}s^\mu s_\mu\right)\, .\label{eq:superfluid1order1}
\ee
The $\phi$ equation of motion sets 
\be 
\partial^\mu s_\mu = 0\, , \label{eq:s1}
\ee
which we can think of as a flatness condition. The equation of motion for $s_\mu$ allows for us to solve for it in terms of $\phi$:
\be 
s_\mu = \partial_\mu \phi\, .\label{eq:s2}
\ee
Substituting~\eqref{eq:s2} into~\eqref{eq:s1}, we obtain the Klein--Gordon equation $\square \phi =0$. We could also integrate out $s_\mu$ using its equation of motion~\eqref{eq:s2} in the action~\eqref{eq:superfluid1order1}, which yields the usual free scalar action~\eqref{eq:scalaraction2}.

Returning to the discussion in section~\ref{fracgau}, we would like to identify the relevant curvatures. In this case, the gauge symmetry can't show all its muscle, as $\phi$ is only a 0-form.  Nevertheless, we can ask which objects are invariant under the global symmetries, which act as $\delta\phi = c + c_\mu x^\mu$ and $\delta s_\mu = c_\mu$ . We consider the two objects,\footnote{Notice that $\partial_{[\mu}s_{\nu]}$ is also invariant under the symmetries, which is an accident of the scalar theory. In any case, this tensor is set to zero by equations of motion.}
\begin{align}
J_{\mu|\nu} &= a\, \partial_{(\mu}s_{\nu)}\, ,\label{eq:FirstR}\\
\mathcal{Q}_\mu &= a \left(\partial_\mu\phi-s_\mu\right)\, .
\label{eq:SecondR}
\end{align}
These are the same curvatures discussed in section~\ref{sec:hbs}, provided we identify $\phi$ with $h_{(0|0)}$ and $s_\mu$ with $\Gamma_{(1|0)}$. We have introduced a normalization factor for the currents, $a$, which follows from the normalization of the action. 
With these identifications we recognize~\eqref{eq:superfluid1order1} as an Einstein-like action.
We can think of the field $s_\mu$ as gauging the 0-form symmetry present in the ordinary superfluid current.
The EFT expansion is then organized in terms of $J_{\mu|\nu}$ via contractions of the form $\left( \mathcal{J}_{\mu | \nu}[s] \right)^n$ (excepting WZ terms---see Footnote~\ref{ft:galWZ}), distinguishing this theory from the ordinary superfluid EFT.

The equations of motion actually set most of the curvatures~\eqref{eq:FirstR} and~\eqref{eq:SecondR} to zero. The equation~\eqref{eq:s1} written in terms of $J_{\mu|\nu}$ reads
\be
 \tr\left(J\right) =0\,,
 \label{eq:eom10}
\ee
while~\eqref{eq:s2} corresponds to 
\be
 \mathcal{Q} = 0\, .\label{eq:eom11}
\ee
Importantly, the fact that $\tr\left(J\right) =0$ does {\it not} imply that all of $J_{\mu|\nu}$ vanishes---in particular the trace-free part can be nonzero. This is the crucial difference between a BF type Lagrangian and the Einstein form in~\eqref{eq:superfluid1order1}, the relevant curvature only obeys a partial flatness condition so there is enough room for degrees of freedom to propagate.
The equations of motion further imply the conservation of the electric current:
\be
\partial^\mu J_{\mu | \nu} = \partial_\nu\square \phi =0\, ,
\ee
as expected. In addition to the conserved electric current~\eqref{eq:FirstR} there is a conserved 
(topological) magnetic symmetry current 
\be\label{dualc1}
K^{\mu_1 \cdots \mu_{d-1}| \nu_1 \cdots \nu_{d-1}} = \epsilon^{\mu_1 \cdots \mu_d} \epsilon^{\nu_1 \cdots \nu_d} \partial_{\mu_d} \partial_{\nu_d} \phi\, ,
\ee
which is (identically) conserved as a consequence of the fact that partial derivatives commute.

\subsubsection{Coupling to background gauge fields}

We now turn to the study of the fate of these symmetries in the presence of background sources for the conserved currents in the theory.
The most obvious way to gauge the global symmetries is to improve the curvatures by introducing a background gauge field as
\be
J_{\mu | \nu} \quad \longrightarrow \quad \mathcal{J}_{\mu | \nu} = a \left(\partial_{(\mu}s_{\nu)}-g \, \eta_{\mu\nu} \partial^\alpha s_\alpha-C_{\mu | \nu}\right)\, ,
\label{eq:Jgaugeimp}
\ee
where we have additionally shifted $J$ by its trace with a free coefficient $g$ for later convenience, and where $C_{\mu | \nu}$ is a symmetric two-index background gauge field (i.e. a $(1|1)$-biform), which corrects the shift of $s_\mu$ from the gauged version of the global symmetry of interest. 
The curvature $\mathcal{J}_{\mu | \nu}$ is invariant under the gauge transformations 
\begin{align}
\delta\phi &= \partial_\alpha \Lambda^\alpha\, ,\label{gaugsg}\\
\delta s_\mu &= \partial_\mu \partial_\alpha  \Lambda^\alpha\, ,\label{gaugsg2}\\
\delta C_{\mu | \nu} &= \partial_\mu \partial_\nu\partial_\alpha  \Lambda^\alpha-g \,\eta_{\mu\nu}\square \partial_\alpha\Lambda^\alpha\, ,\label{gaugsg3}
\end{align}
for any $1$-form $ \Lambda$. Notice that $\mathcal{Q}$ is already invariant and so does not need an additional background gauge field improvement. 

There is, however, a surprise. We cannot build an Einstein-type action using $C_{\mu | \nu}$ as our background gauge field. We anticipated this problem in section~\ref{sec:charges}. The resolution is that we have to build $C_{\mu | \nu}$ itself from two derivatives of a symmetric gauge field $A_{\mu | \nu}$ as 
\be 
\begin{aligned}
C_{\mu|\nu} = 2\kappa\Big(\partial^\alpha \partial_{(\mu}A_{\nu)|\alpha}-\frac{1}{2}\square A_{\mu|\nu}\Big)&-(\kappa+g-1)\Big(\partial_{(\mu}\partial_{\nu)}A+\eta_{\mu\nu}\partial^\alpha \partial^\beta A_{\alpha|\beta}\Big)\\
&+(\kappa+g^2-1)\eta_{\mu\nu}\square A\, ,
\end{aligned}
\label{eq:Ctensdefgalfluid}
\ee
where we have defined $A\equiv \eta^{\mu\nu} A_{\mu|\nu}$ as the trace $\text{tr}\left(A_{\mu|\nu}\right)$, and $\kappa$ is another free coefficient.
Provided that under a gauge transformation, $A_{\mu|\nu}$ transforms as
\be
\delta A_{\mu|\nu} = \frac{1}{(1-g)}\partial_{(\mu} \Lambda_{\nu)}\,,
\label{gaugsg4}
\ee
then~\eqref{eq:Ctensdefgalfluid} will shift as in~\eqref{gaugsg3}. Note that there is a two-parameter family of tensors (parameterized by $g$ and $\kappa$) that all have the same transformation properties, so that the current~\eqref{eq:Jgaugeimp} is gauge invariant. Noting that $J_{(1|1)} = a \ast\hspace{-2pt} K\ast_{(1|1)}$ in the free theory, it is also straightforward to promote the magnetic current to its gauge-invariant version:
\be
\ast {\cal K}\ast_{\mu|\nu} \equiv \ast K\hspace{-2.5pt}\ast_{\mu|\nu}-g\,\eta_{\mu\nu}\square\phi - C_{\mu|\nu}\,.
\ee
We now want to investigate the properties of these gauge-invariant currents.

\vspace{6pt}
\noindent
{\bf Anomalies:} Recall that the currents $J_{\mu|\nu}$ and $K_{\mu_1\cdots\mu_{d-1}|\nu_1\cdots\nu_{d-1}}$ satisfy the following equations on-shell (and at separated points)
\be
\begin{aligned}
\tr\,J &= 0 \,,  & \qquad\qquad\quad   &\\ 
\partial^{\mu} J_{\mu|\nu} & = 0\,, \qquad\quad\quad  &  \partial_{[\rho}\ast K\ast_{\mu]|\nu}  &= 0\,,
\end{aligned}
\label{eq:galfluidcurrenteoms}
\ee
where the condition on $ \ast K\ast$  is the dualization of the conservation of the magnetic current $K_{(d-1|d-1)}$.
In the presence of the background gauge field $A_{\mu|\nu}$, it is impossible to satisfy all of these conditions on shell. This is the expression of a mixed anomaly between the electric and magnetic symmetries.

The two anomalous conservation conditions are the trace condition on the electric current
\begin{tcolorbox}[colframe=white,arc=0pt,colback=greyish2]
\be
\tr\,{\cal J} =  a \big[1-d+\kappa(d-2)\big]\Big(\partial_\alpha \partial_\beta A^{\alpha|\beta}-\square A\Big)\, ,
\label{eq:galfluidtraceanom}
\ee
\end{tcolorbox}
\vspace{-8pt}
\noindent
and the conservation of the magnetic current, which can be expressed as
\begin{tcolorbox}[colframe=white,arc=0pt,colback=greyish2]
\be
\partial_{[\rho} \ast {\cal K}\ast_{\mu]|\nu} =-\partial_{[\rho} {\cal C}_{\mu]|\nu} \,,
\label{eq:galfluiddualconsanom}
\ee
\end{tcolorbox}
\vspace{-8pt}
\noindent
where we have defined the tensor
\be
\begin{aligned}
{\cal C}_{\mu|\nu} \equiv  2\kappa\Big(\partial^\alpha \partial_{(\mu}A_{\nu)|\alpha}-\frac{1}{2}\square A_{\mu|\nu}\Big)-(\kappa-1)\eta_{\mu\nu}\left(\partial^\alpha\partial^\beta A_{\alpha|\beta}-\square A\right)\, .
\end{aligned}
\label{eq:kdivrhsgal}
\ee
The condition 
$\partial^{\mu} {\cal J}_{\mu|\nu}  = 0$ continues to hold even in the presence of the background gauge field.
It is straightforward to check that the right hand sides of both~\eqref{eq:galfluidtraceanom}
and~\eqref{eq:galfluiddualconsanom} are gauge invariant, as they must be. From these expressions, we see first that there is no choice of $g,\kappa$ that makes these conservation conditions exact, which will end up being an anomaly in the theory.  We further see a conceptual difference between $d=2$ and $d>2$. In generic dimension, it is possible to choose $\kappa$ to make ${\cal J}_{\mu|\nu}$ traceless, while this is not possible in $d=2$. We therefore treat these two cases separately in the following.
\begin{itemize}

\item $\boldsymbol{d >2:}$ In generic dimension, it is convenient to set
\be
\kappa = \frac{d-1}{d-2}\,,
\label{eq:kingendim2}
\ee
along with $g = 1/d$. In this case, we find both that the right-hand side of~\eqref{eq:galfluidtraceanom} vanishes and that the anomalous conservation equation~\eqref{eq:galfluiddualconsanom} can be written in terms of the {\it traceless} part of $A_{\mu|\nu}^{(T)}$:
\be
\begin{aligned}
{\cal C}_{\mu|\nu}^{(d)} =   \frac{d-1}{d-2}\bigg[2\partial^\alpha \partial_{(\mu}A_{\nu)|\alpha}^{(T)}-\square A_{\mu|\nu}^{(T)}-\frac{1}{d-1}\eta_{\mu\nu}\partial^\alpha\partial^\beta A_{\alpha|\beta}^{(T)}\bigg]\, ,\label{eq:genctens2}
\end{aligned}
\ee
where $ A^{(T)}_{\mu|\nu} \equiv  A_{\mu|\nu} - \frac{1}{d}\eta_{\mu\nu}A$. This is the minimal presentation of the anomaly, where the only anomalous equation is the conservation of ${\cal K}_{(d-1|d-1)}$. This makes clear that the mixed anomaly is between the electric and magnetic symmetries responsible for the conservation of these currents.

\item $\boldsymbol{d =2:}$ In two spacetime dimensions, there is a further trace anomaly. Since $\kappa$ multiplies $d-2$ in~\eqref{eq:galfluidtraceanom}, we can no longer choose it to make the trace of ${\cal J}_{\mu|\nu}$ vanish.\footnote{One way to understand this is to note that the tensor~\eqref{eq:genctens2} (after multiplying through by $d-2$) is accidentally gauge {\it invariant} in $d=2$, and so cannot be used to gauge the current $J_{(1|1)}$.}  In this case, the currents satisfy the equations
\be
\begin{aligned}
\tr\,{\cal J} &= a\,{\cal C}^{(2)} \,,  & \qquad\qquad\quad   &\\ 
\partial^{\mu} {\cal J}_{\mu|\nu} & = 0\,, & \qquad\qquad\quad \partial_{[\rho}\ast {\cal K}\ast_{\mu]|\nu}  &= -\partial_{[\rho} {\cal C}^{(2)}_{\mu]|\nu}\,,
\end{aligned}
\ee
where the field strengths appearing  on the right hand sides are
\be
{\cal C}^{(2)} =  -\partial^\alpha \partial^\beta A_{\alpha|\beta}+\square A\,,
\label{eq:ctracetens3}
\ee
along with~\eqref{eq:kdivrhsgal}, and there is no choice of the free parameters that makes either of them vanish. Notice that the gauge-invariant tensor~\eqref{eq:ctracetens3}
is proportional to the linearized Ricci tensor in gravity if we interpret $A_{\mu |\nu}$ as a metric perturbation, so that we have effectively coupled our Galileon theory to a background linearized geometry.

\end{itemize}
In both cases, the combined (anomalous) conservation equations are sufficient to completely fix the two-point function involving the conserved currents $J_{(1|1)}$ and $K_{(d-1|d-1)}$, as we demonstrate explicitly in section~\ref{sec:cur2ptfunctiongalileon}. However, we first show how one can reproduce the  structure of anomalies described here starting directly from an action principle for the free scalar.

\vspace{6pt}
\noindent
{\bf Gauging the action:} We can now write a gauge-invariant action that reproduces this universal physics as
\be
\begin{aligned}
S &= -a\int \rd^dx\bigg[\phi\left(\partial^\mu s_\mu - \partial^\mu\partial^\nu A_{\mu|\nu} +g\,\square A\right) + \frac{1}{2}s_\mu s^\mu -\kappa \partial^\mu A_{\mu|\alpha}\partial_\nu A^{\nu|\alpha}\\
&\hspace{2cm}
+\frac{\kappa}{2} \partial_\mu A_{\nu|\alpha}\partial^\mu A^{\nu|\alpha}
+(\kappa+g-1)\partial^\mu A\, \partial^\nu A_{\mu|\nu}- \frac{1}{2}\left(\kappa+g^2-1\right)\partial_\mu A\partial^\mu A\bigg] ,
\end{aligned}
\label{eq:ActionSuperfluidGauged2}
\ee
which is the most general action coupling $A_{\mu|\nu}$ to $\phi$ that is invariant under the gauge transformations~\eqref{gaugsg},~\eqref{gaugsg2} and~\eqref{gaugsg4}. It is worth noting that the equation of motion for $s_\mu$ is unchanged by the presence of the background gauge field (reflecting the fact that the curvature ${\cal Q}$ is already gauge invariant), and still sets $s_\mu = \partial_\mu\phi$.

From the action~\eqref{eq:ActionSuperfluidGauged2} we can extract gauge-invariant versions of the analogue of the Einstein tensor (which is a scalar in this case) and the conserved current $J_{(1|1)}$ as 
\begin{align}
\mathcal{G}& = \frac{\delta S}{\delta \phi}\, ,\\
\mathcal{J}_{\mu|\nu} &= \frac{\delta S}{\delta A^{\mu|\nu}}\, .
\end{align}
Explicitly, we find that the Einstein scalar is given by
\begin{tcolorbox}[colframe=white,arc=0pt,colback=greyish2]
\be
\begin{aligned}
{\cal G} = -a\Big(\partial^\mu s_\mu - \partial_\mu \partial_\nu A^{\mu|\nu}+g\, \square A\Big)\, ,
\label{eq:gaugeinveins2}
\end{aligned}
\ee
\end{tcolorbox}
\vspace{-6pt}
\noindent
while the gauge-invariant current is
\begin{tcolorbox}[colframe=white,arc=0pt,colback=greyish2]
\be
{\cal J}_{\mu|\nu} = a\Big( \partial_{(\mu}\partial_{\nu)}\phi -g\, \eta_{\mu\nu}\square\phi-C_{\mu|\nu}\Big)\,,
\label{eq:Jcurrentgauge2}
\ee
\end{tcolorbox}
\vspace{-6pt}
\noindent
where the tensor $C_{\mu|\nu}$ is the same as in~\eqref{eq:Ctensdefgalfluid}. This current is the same as the one in~\eqref{eq:Jgaugeimp}.
The gauge invariance of the action guarantees that ${\cal J}_{\mu|\nu}$ will be conserved on-shell. But its trace does not vanish for generic parameter choices, and is instead given by~\eqref{eq:galfluidtraceanom}. In this simple example, we can see that  ${\cal J}_{(1|1)} = a \ast {\cal K}\ast_{(1|1)}$, so the topological magnetic current is essentially the same as the electric current. We can then read off the anomalous conservation condition from the definition of ${\cal J}$, which reproduces exactly~\eqref{eq:galfluiddualconsanom}.

For $d\not=2$, there is a further simplification of the action~\eqref{eq:ActionSuperfluidGauged2} that can be effected by choosing $\kappa$ as in~\eqref{eq:kingendim2} and setting $g = 1/d$. With this choice, the trace $A$ completely decouples and the 
action~\eqref{eq:ActionSuperfluidGauged2} becomes
\begin{align}
S &= -a\int \rd^dx\bigg[\phi\left(\partial^\mu s_\mu -\partial^\mu\partial^\nu A_{\mu|\nu}^{(T)}\right) + \frac{1}{2}s_\mu s^\mu +\frac{d-1}{d-2}\left(\frac{1}{2} (\partial_\mu A_{\nu|\alpha}^{(T)})^2 - (\partial^\mu A_{\mu|\alpha}^{(T)})^2\right) \bigg]\,,\ (d\not=2), \nn\\
\label{eq:ActionSuperfluidGauged4}
\end{align}
where $ A^{(T)}_{\mu|\nu}$ is the traceless part of $A_{\mu|\nu}$. As a consequence of the fact that only the traceless part of $A_{(1|1)}$ couples to the dynamical fields, the current  ${\cal J}_{\mu|\nu}$ is now traceless {\it off-shell}. This simplification does not take place in $d=2$, and is an action-level manifestation of the trace anomaly discussed above.

There is an elegant simplification of the action~\eqref{eq:ActionSuperfluidGauged2} in second-order form. If we eliminate $s_\mu$ using its equation of motion, we can write 
\be 
S = \frac{1}{2}\int \rd^d x\left(\phi \,\mathcal{G}+ A^{\mu|\nu}\mathcal{J}_{\mu|\nu}\right)\, .
\ee
Varying this action with respect to $\phi$ produces \eqref{eq:gaugeinveins2}, where we write $s_\mu = \partial_\mu \phi$, while varying with respect to $A^{\mu|\nu}$ produces \eqref{eq:Jcurrentgauge2}.

As emphasized before, these results, although obtained from an EFT, are completely universal. In the next section we compute the current-current correlation functions using the symmetry structure alone and reproduce the pattern of anomalies discovered above. We then go on to prove that this symmetry structure implies the gaplessness of the phase, which is therefore protected by the anomalies found.

\subsubsection{Current two-point function\label{sec:cur2ptfunctiongalileon}}

We now want to consider the two-point function between the electric and magnetic currents in the theory of a Galileon superfluid. We wish to show that---much like the case of the ordinary superfluid---the structure of anomalies completely fixes this two-point function, and mandates that the system is in a gapless phase.
The relevant two-point function of interest is between the electric current ${\cal J}_{\mu|\nu}$ and the (dualized) magnetic current $* \mathcal{K} *_{\mu|\nu}$.

We first treat the generic dimension case ($d\neq 2$), where the currents satisfy the equations 
\be
\begin{aligned}
\tr\,{\cal J} &= 0 \,,  & \qquad\quad  &\\
\partial^{\mu} {\cal J}_{\mu|\nu} & = 0\,, \qquad\quad &  \partial_{[\rho}\ast{\cal K}\ast_{\mu]|\nu}  &= -\partial_{[\rho}{\cal C}_{\mu]|\nu}^{(d)} \,,
\end{aligned}
\label{eq:galfluidcurrentanoms}
\ee
with the tensor ${\cal C}_{\mu|\nu}^{(d)}$ given by~\eqref{eq:genctens2}.
In momentum space, the most general possible form of the current-current two-point function is 
\be
\begin{aligned}
\braket{{J}_{{\mu_1}|{\mu_2}}* {K}*_{{\nu_1}|{\nu_2}}} &=\, 2c_1(p^2)p^2\eta_{{\mu_1}({\nu_2}}\eta_{{\nu_1}){\mu_2}} + c_2(p^2)p^2 \eta_{{\mu_1}{\mu_2}}\eta_{{\nu_1}{\nu_2}} + c_3(p^2)\eta_{{\mu_1}{\mu_2}}p_{\nu_1} p_{\nu_2}
\\
&\phantom{=}~ + c_4(p^2)\eta_{{\nu_1}{\nu_2}}p_{\mu_1} p_{\mu_2}+ 2c_5(p^2)\Big(\eta_{{\mu_2}({\nu_2}}p_{{\nu_1})}p_{\mu_1} +\eta_{{\mu_1}({\nu_2}}p_{{\nu_1})}p_{\mu_2}\Big)\\
&\phantom{=}~ + c_6(p^2) \frac{p_{\mu_1} p_{\mu_2} p_{\nu_1} p_{\nu_2}}{p^2}\, ,
\end{aligned} 
\label{eq:SymCurSymCur}
\ee
where $c_{1},\cdots,c_6$ are arbitrary functions. We have assumed that the two-point function is separately symmetric under the interchange of $\mu_1,\mu_2$ and $\nu_1,\nu_2$, but we have not assumed any symmetry under the interchange of the $\mu$s with the $\nu$s. We now have to impose the conditions in~\eqref{eq:galfluidcurrentanoms}. Requiring that ${ J}_{\mu|\nu}$ is both conserved and traceless fixes the two-point function up to one free function 
\be
\begin{aligned}
\braket{{J}_{{\mu_1}|{\mu_2}}* {K}*_{{\nu_1}|{\nu_2}}} = 2c_1(p^2)\bigg[p^2\eta_{{\mu_1}({\nu_2}}\eta_{{\nu_1}){\mu_2}} &- \frac{1}{d-1}\Big(p^2 \eta_{{\mu_1}{\mu_2}}\eta_{{\nu_1}{\nu_2}} -\eta_{{\mu_1}{\mu_2}}p_{\nu_1} p_{\nu_2}-\eta_{{\nu_1}{\nu_2}}p_{\mu_1} p_{\mu_2} \Big)\\
& \hspace{-9pt}- \eta_{{\mu_2}({\nu_2}}p_{{\nu_1})}p_{\mu_1} -\eta_{{\mu_1}({\nu_2}}p_{{\nu_1})}p_{\mu_2}+ \frac{d-2}{d-1}\frac{p_{\mu_1} p_{\mu_2} p_{\nu_1} p_{\nu_2}}{p^2}\bigg] .\label{eq:CurrCurre2pt}
\end{aligned} 
\vphantom{\Bigg\}}
\ee
Even if the starting ansatz \eqref{eq:SymCurSymCur} is not symmetric under the interchange of the two currents, \eqref{eq:CurrCurre2pt} is manifestly symmetric under the interchange of the $\mu$s with the $\nu$s.
In order to fix $c_1(p^2)$, we use the magnetic conservation anomaly equation~\eqref{eq:galfluiddualconsanom}, which implies the failure of $K_{(d-1|d-1)}$ to be conserved in the two-point function is by the precise contact terms
\be 
p^{[\lambda}\braket{{J}_{{\mu_1}|{\mu_2}}*{K}*^{{\nu_1}]|{\nu_2}}} = \frac{p^2}{2}\frac{(d-1)}{d-2}p^{[\lambda}\left[\Pi^{(1)}_{({\mu_1}}{}^{{\nu_1}]}\Pi_{{\mu_2})}^{(1)}{}^{{\nu_2}}+\Pi_{({\mu_1}}^{(1)}{}^{{\nu_2}}\Pi_{{\mu_2})}^{(1)}{}^{{\nu_1}]}-\frac{2}{d-1}\Pi^{(1)}_{{\mu_1}{\mu_2}}\Pi^{(1)}{}^{{\nu_1}]{\nu_2}}\right] ,\label{eq:anomlay1}
\vphantom{\Bigg\}}
\ee
where $\Pi^{(1)}_{\mu\nu} = \eta_{\mu\nu}-\frac{p_\mu p_\nu}{p^2}$ is the transverse projector. This equation sets $c_1(p^2) = (d-1)/[2(d-2)]$ so that the two-point function is completely fixed by the conditions~\eqref{eq:galfluidcurrentanoms}
\be
\begin{aligned}
\braket{{J}_{{\mu_1}|{\mu_2}}* {K}*_{{\nu_1}|{\nu_2}}} = \frac{d-1}{d-2}\bigg[p^2\eta_{{\mu_1}({\nu_2}}\eta_{{\nu_1}){\mu_2}} &- \frac{1}{d-1}\Big(p^2 \eta_{{\mu_1}{\mu_2}}\eta_{{\nu_1}{\nu_2}} -\eta_{{\mu_1}{\mu_2}}p_{\nu_1} p_{\nu_2}-\eta_{{\nu_1}{\nu_2}}p_{\mu_1} p_{\mu_2} \Big)\\
& \hspace{-5pt}- \eta_{{\mu_2}({\nu_2}}p_{{\nu_1})}p_{\mu_1} -\eta_{{\mu_1}({\nu_2}}p_{{\nu_1})}p_{\mu_2}+ \frac{d-2}{d-1}\frac{p_{\mu_1} p_{\mu_2} p_{\nu_1} p_{\nu_2}}{p^2}\bigg] .\label{eq:CurrCurre2pt}
\end{aligned} 
\vphantom{\Bigg\}}
\ee
In the next section we verify that the spectral decomposition of this correlation function includes a massless scalar, showing that this pattern of conserved currents implies that the theory is in a gapless phase. 

Here we have focused on the most minimal presentation of the anomaly, but we can shuffle it instead into the electric conservation equation, if desired. 
In this formulation, the two-point function takes the particularly simple form
\be
\braket{{J}_{\mu_1|\mu_2}* {K}*_{\nu_1|\nu_2}} =\frac{p_{\mu_1} p_{\mu_2} p_{\nu_1} p_{\nu_2}}{p^2}\, ,\label{eq:missingcreativenames1}
\ee
which clearly obeys conservation for the magnetic current, but $J$ is neither traceless nor conserved at coincident points. Notice of course that this presentation of the correlator differs from~\eqref{eq:CurrCurre2pt} only by contact terms. We elaborate more on these other presentations of the anomaly in appendix \ref{sec:cur2ptfunctiongalileonap}.

\vspace{6pt}
\noindent
{\bf $\boldsymbol{d=2}$ dimensions:} 
In two spacetime dimensions, we cannot impose all the conditions~\eqref{eq:galfluidcurrentanoms}. In this case, there is a three-way anomaly between conservation and tracelessness of {${ J}$}, and the conservation of {${ K}$}---we can choose one of these conditions to be preserved, while the other two will be anomalous. We discuss the various possibilities in more detail in appendix~\ref{sec:cur2ptfunctiongalileonap}. Here we just report the correlator where {${ J}$} is conserved:
\be
\braket{{J}_{{\mu_1}|{\mu_2}}* {K}*_{{\nu_1}|{\nu_2}}} =  p^2 \eta_{{\mu_1}{\mu_2}}\eta_{{\nu_1}{\nu_2}} -\eta_{{\mu_1}{\mu_2}}p_{\nu_1} p_{\nu_2}-\eta_{{\nu_1}{\nu_2}}p_{\mu_1} p_{\mu_2} +\frac{p_{\mu_1} p_{\mu_2} p_{\nu_1} p_{\nu_2}}{p^2}\, .
\ee
Other choices of conditions to impose at coincident points will just change the contact terms appearing in this correlator. (For example, requiring conservation of the magnetic current yields the same answer as in general dimension~\eqref{eq:missingcreativenames1}.)

\subsubsection{\KL spectral representation}

We have seen that the structure of anomalies completely fixes the nonlocal part of the current two-point function, and the precise conditions that we decide to impose at coincident points only change the contact terms appearing in the correlator. We now want to perform a spectral decomposition to show that there is necessarily a gapless mode in the spectrum, whose presence is protected by the anomaly.

Our starting point is the \KL decomposition for a correlator of two symmetric traceless tensors (see appendix \ref{ap:KLrep} for details)
\be
\braket{{J}_{\mu_1|\mu_2}*{K}*_{\nu_1|\nu_2}} 
= \int_0^\infty\, \rd s\, \frac{s^2}{p^2 +s} \left(\rho_0(s)\tl{\Pi}^{(0)}_{\mu_1\mu_2\nu_1\nu_2}- \rho_1(s)\tl{\Pi}^{(1)}_{\mu_1\mu_2\nu_1\nu_2}+ \rho_2(s)\tl{\Pi}^{(2)}_{\mu_1\mu_2\nu_1\nu_2}\right)\, .
\label{eq:galKLdecomp}
\ee
Here $\rho_i(p^2)$ are the spin $i$ components of the spectral density (the only massless representation that can couple to a symmetric conserved current is a scalar~\cite{Weinberg:2020nsn,Distler:2020fzr}, so the spectral densities of the spin-1 and spin-2 states must go to zero as $p^2\to 0$).  We defined $\Pi^{(i)}_{\mu_1\mu_2\nu_1\nu_2}$ as the projectors onto the spin-$i$ representation that couples to the currents, given by
\begin{align}
\label{eq:galsfproj1}
\Pi^{(0)}_{\mu_1\mu_2\nu_1\nu_2}&= \frac{d}{d-1}\left(\Pi^{(0)}_{\mu_1\mu_2}-\frac{\eta_{\mu_1\mu_2}}{d}\right)\left(\Pi^{(0)}_{\nu_1\nu_2}-\frac{\eta_{\nu_1\nu_2}}{d}\right)\, ,\\
\Pi^{(1)}_{\mu_1\mu_2\nu_1\nu_2}&=\frac{1}{2}\left(\Pi^{(0)}_{{\mu_1}{\nu_1}}\Pi^{(1)}_{{\mu_2}{\nu_2}}+\Pi^{(0)}_{{\mu_1}{\nu_2}}\Pi^{(1)}_{{\mu_2}{\nu_1}}	+\Pi^{(1)}_{{\mu_1}{\nu_1}}\Pi^{(0)}_{{\mu_2}{\nu_2}}+\Pi^{(1)}_{{\mu_1}{\nu_2}}\Pi^{(0)}_{{\mu_2}{\nu_1}}\right)\, ,\\
\Pi^{(2)}_{\mu_1\mu_2\nu_1\nu_2} &= \frac{1}{2}\left(\Pi^{(1)}_{{\mu_1}{\nu_1}}\Pi^{(1)}_{{\mu_2}{\nu_2}}+\Pi^{(1)}_{{\mu_1}{\nu_2}}\Pi^{(1)}_{{\mu_2}{\nu_1}}\right)-\frac{1}{d-1}\Pi^{(1)}_{{\mu_1}{\mu_2}}\Pi^{(1)}_{{\nu_1}{\nu_2}}\, ,
\label{eq:galsfproj2}
\end{align}
where $\Pi^{(0)}_{\mu\nu} = \frac{p_\mu p_\nu}{p^2}$ and $\Pi^{(1)}_{\mu\nu} = \eta_{\mu\nu}-\frac{p_\mu p_\nu}{p^2}$ are defined as in \eqref{eq:superfluidproj}.   The projectors \eqref{eq:galsfproj2}  are orthonormal and complete on the space of traceless symmetric two-index tensors.
As in the previous case, the tensors that actually appear in the spectral decomposition of the correlator are not quite~\eqref{eq:galsfproj1}--\eqref{eq:galsfproj2}, but are instead off-shell versions of them obtained by replacing  $\Pi^{(0)}$ and $\Pi^{(1)}$ with their tilded versions defined in~\eqref{eq:superfluidproj2}. This defines the set of four-index tensors $\tl\Pi^{(i)}_{\mu_1\mu_2\nu_1\nu_2}$  appearing in \eqref{eq:galKLdecomp}.  They depend on $s$ and reduce to the projectors when $s \to -p^2$.

Given the tensors~\eqref{eq:galsfproj1}--\eqref{eq:galsfproj2}, we can write~\eqref{eq:CurrCurre2pt} quite economically as 
\be 
\braket{{J}_{\mu_1|\mu_2}* {K}*_{\nu_1|\nu_2}} =   \frac{d-1}{d-2}p^2\Pi^{(2)}_{\mu_1\mu_2\nu_1\nu_2}\, .
\label{eq:simplegal2pt}
\ee
The way that $s$ appears and the fact that $\rho_1(s)$ and $\rho_2(s)$ must go to zero as $s\to 0$ means that there is a unique way to match~\eqref{eq:simplegal2pt}, which is to set
\be 
\rho_2(s) = 0\, ,\qquad \rho_1(s) = 0\, , \qquad \rho_0(s) = \frac{d-1}{d}\delta(s)\, . 
\label{eq:galspectraldensities}
\ee
From the spectral densities~\eqref{eq:galspectraldensities}, we see that there is a gapless scalar in the spectrum. Further, since position space contact terms do not affect the spectral functions (as is discussed in appendix~\ref{ap:KLrep}), this will be the case regardless of where we choose to put the anomaly. We therefore conclude that there must always be a gapless scalar in the spectrum, whose presence is a consequence of the symmetry and anomaly structure of the currents.

\section{The graviton as a Goldstone\label{sec:LinGrav}}

What defines a theory of gravity? A common response is that gravitational theories are those that respect general coordinate invariance. However, this is a statement about the gauge redundancies in our description of the physics, and so cannot be the true essence of gravity.  A slightly better answer is that gravitational theories are those with a massless spin-2 particle in the spectrum, since powerful uniqueness results imply that the interactions of such a theory will be those of Einstein gravity, assuming it mediates a $\sim r^{-2}$ force between point-like matter sources \cite{Weinberg:1964ew,Weinberg:1965rz,Deser:1969wk}.  Despite being correct, this answer is somewhat incomplete, because gaplessness is itself something to be explained. A ubiquitous source of gapless modes is symmetry breaking. Goldstone's theorem guarantees that systems with spontaneously broken continuous symmetries will possess massless excitations. It is therefore natural to ask whether a similar explanation can underly the appearance of a massless spin-2 field in gravitational theories. The relevant symmetries necessarily belong to the family of biform symmetries discussed in section~\ref{sec:hbs}, and so we want to understand how gravity fits into this picture.

In this section, we explore  these questions. We will see that linearized gravity can be {\it defined} as a gapless phase with two conserved currents 
\be
J_{\mu_1\mu_2 | \nu_1\nu_2}\,, \qquad {\rm and} \qquad K_{\mu_1\mu_2\cdots \mu_{d-2} | \nu_1\nu_2\cdots \nu_{d-2}}\,,
\label{eq:gravcurrentdef}
\ee
where $J_{(2|2)}$ is a $(2|2)$-biform---which is traceless on-shell---and $K_{(d-2|d-2)}$ is a $(d-2|d-2)$-biform. Importantly, there is a mixed anomaly between the conservation conditions of these currents: turning on a background gauge field source for $J_{(2|2)}$ causes $K_{(d-2|d-2)}$ to no longer be conserved, and vice versa. In the deep infrared, the currents are related in a simple way as $J_{(2|2)}= a* K_{(d-2|d-2)}*$, so we can phrase the mixed anomaly in terms of $J_{(2|2)}$ alone. 

We can think of linearized gravity as being the phase defined by the following equations in the IR\footnote{These equations are valid for $d > 4$. For $d=4$ there is additionally an interesting anomaly in the trace condition for $J$, which we discuss further in the following.} 
\be
\begin{aligned}
\mathcal{J}_{~~\nu|\mu\beta}^{\mu}&= 0 \, , \qquad\quad
&  &\\ 
 \partial^\mu \mathcal{J}_{\mu\nu|\alpha\beta}&=0\,, & \qquad\quad
\partial_{[\rho}*\mathcal{K}*_{\mu\nu]|\alpha\beta} &= -\partial_{[\rho} {\cal C}_{\mu\nu]|\alpha\beta}\,,
\end{aligned}
\label{eq:gravityanomalyeqs}
\ee
where ${\cal J}_{(2|2)}$ is a gauge-invariant improvement of $J_{(2|2)}$ in the presence of a background gauge field.
The first line of~\eqref{eq:gravityanomalyeqs}
expresses the tracelessness of ${\cal J}_{(2|2)}$ .
The first equation on the second line is the conservation of ${\cal J}_{(2|2)}$ and the second equation is the manifestation of the mixed 't Hooft anomaly between the $(2|2)$-biform global symmetry generated by ${\cal J}$ and the $(d-2|d-2)$-biform symmetry related to ${\cal K}$. In the presence of a source $A_{(2|2)}$ for the current ${\cal J}_{(2|2)}$, conservation of ${\cal K}_{(d-2|d-2)}$ is lost, and the non-conservation is proportional to the field strength of this gauge field.\footnote{In detail, the gauge-invariant field strength $\rd{\cal C}_{(2|2)}$ is built from the background gauge field for ${\cal J}_{(2|2)}$, $A_{\mu\nu|\alpha\beta}$, as the exterior derivative of
\be
 {\cal C}_{\mu\nu|\alpha\beta} = \frac{d-3}{d-4}{\cal Y}_{(2|2)}\bigg(  \partial^\sigma \partial_\beta A^{(T)}_{\mu\nu\alpha\sigma} -\frac{1}{4}\square A_{\mu\nu\alpha\beta}^{(T)}+\frac{3}{2(d-3)} \eta_{\alpha\beta} \partial^\rho\partial^\sigma A_{\mu\rho\nu\sigma}^{(T)} \bigg)
\ee
where $\mathcal{Y}_{(2|2)}$ is the young projector onto the ${(2|2)}$ Young diagram corresponding to a ${(2|2)}$-biform, and $A^{(T)}$ is the traceless part of the gauge field $A_{(2|2)}$.
}
Much as we saw for superfluids, the two-point function between the operators ${\cal J}$ and ${\cal K}$ will be completely fixed by these equations. Then, upon performing a \KL decomposition, we will infer the presence of a gapless spin-2 mode (the graviton) in the spectrum of the theory in this phase.

In order to realize this physics via a quantum field theory, we have to represent the current ${\cal J}_{(2|2)}$ in terms of local fields so that the conditions~\eqref{eq:gravityanomalyeqs} follow from the equations of motion. To do this, we introduce the two curvatures
\begin{align}
J_{\mu\nu|\rho\sigma} &\equiv \frac{a}{2}\left( \partial_{\rho}\Gamma_{\mu\nu|\sigma}-\partial_{\sigma}\Gamma_{\mu\nu|\rho}\right)\, ,\label{eq:FirstR2}\\
\mathcal{Q}_{\mu\nu|\rho} &\equiv  \partial_{[\mu}h_{\nu]\rho}-\frac{1}{2}\Gamma_{\mu\nu|\rho} \, ,
\end{align}
where $\Gamma_{(2|1)}$ is a $(2|1)$-biform and $h_{\mu\nu}$ is a symmetric tensor (a $(1|1)$-biform). Note that this necessarily introduces linearized diffeomorphism invariance, as these curvatures are gauge invariant under the transformations
\be
\delta h_{\mu\nu} = 2\partial_{(\mu}\xi_{\nu)}\,, \qquad\qquad \delta \Gamma_{\mu\nu|\rho} = 2\partial_\rho \partial_{[\mu}\xi_{\nu]}\, .
\ee
The goal is then to construct an action whose equations of motion are the flatness conditions
\be
\label{eq:gravflatness1}
\mathcal{Q}_{\mu\nu|\rho} = 0\,,\qquad{\rm and}\qquad\left(\tr\, J\right)_{\mu|\nu} = 0\,.
\ee
The first of these conditions allows us to express $\Gamma_{\mu\nu|\rho}$ in terms of $h_{\mu\nu}$. Then, in terms of $h_{\mu\nu}$, $J_{(2|2)}$ is nothing other than the linearized Riemann tensor, so that the equation $\tr \,J = 0$ is precisely the linearized Einstein equation.

An action that produces~\eqref{eq:gravflatness1} as its equations of motion is
\be
 S = -a\int \rd^dx\, \left[h^{\mu\nu}\left(\partial_\alpha\Gamma\indices{^\alpha_{\mu | \nu}}-\eta_{\mu\nu}\partial_\alpha \Gamma\indices{^{\alpha\rho}_{ | \rho}}+\partial_\nu \Gamma\indices{_\mu^\rho_{ | \rho}}\right)+ \frac{1}{4}\Gamma_{\mu\nu | \rho}\Gamma^{\mu\nu | \rho}-\frac{1}{2}\Gamma\indices{^{\mu\rho}_{ | \rho}}\Gamma\indices{_\mu^\sigma_{ | \sigma}}\right]\,.
\ee
This action is precisely of the form~\eqref{eq:biformEHaction}, where the first term in brackets is the linearized Einstein tensor, written in terms of $\Gamma_{(2|1)}$. Upon integrating out this auxiliary field, we recover the ordinary Einstein action for $h_{\mu\nu}$. This action possesses a higher-biform symmetry, whose Noether current is $J_{\mu\nu|\alpha\beta}$ in~\eqref{eq:FirstR2}. Coupling this current to a background gauge field source improves the current $J$ to the gauge-invariant current ${\cal J}$, which then displays the mixed anomaly~\eqref{eq:gravityanomalyeqs} between ordinary and dual conservation.

In this section we first describe the higher form symmetries of linearized gravity and their gauging, showing how the biform current ${\cal J}_{(2|2)}$ arises from the linear Einstein action. We then invert the logic and demonstrate how one can use the higher-form currents and their anomalies as an input in order to interpret linearized gravity as a gapless phase realizing these symmetries in a particular way. This serves as a version of a Goldstone theorem for the graviton.

\subsection{Linearized gravity}
\label{sec:lingrav}

In order to ground our discussion in the familiar, we begin by reviewing the rudiments of linearized gravity. A free massless spin-2 field is described by the Fierz--Pauli action\footnote{This action also arises from linearizing the Einstein--Hilbert action as $g_{\mu\nu} = \eta_{\mu\nu}+2h_{\mu\nu}/M_{\rm Pl}^\frac{d-2}{2}$. }
\be 
S_{\rm FP} = \int \rd^dx \left[\frac{1}{2}\partial_\rho h_{\mu\nu}\partial^\rho h^{\mu\nu} -\partial_\rho h_{\mu\nu}\partial^\nu h^{\rho\mu} -\frac{1}{2}\partial_\mu h \partial^\mu h +\partial_\nu h^{\mu\nu}\partial_\mu h\right]\, .\label{eq:SFierzPauli}
\ee
In what follows we will stick to $d>3$ since this theory is topological with no propagating modes in $d\leq 3$.
It is convenient to integrate by parts to write this in terms of the linearized Einstein tensor
\be
G\left[h\right] _{\mu\nu}= 2\partial_\sigma \partial_{(\mu} h\indices{_{\nu)}^\sigma}-\partial_\mu\partial_\nu h-\square h_{\mu\nu}-\eta_{\mu\nu}\left(\partial_\alpha \partial_\beta h^{\alpha\beta}-\square h\right)\, ,
\label{eq:linGtensor}
\ee
so that~\eqref{eq:SFierzPauli} takes the form
\be 
S_{\rm FP} = \int \rd^dx \,\frac{1}{2}h^{\mu\nu}G_{\mu\nu}\, ,
\label{eq:FP2}
\ee
which makes it clear that the linearized Einstein equation is $G_{\mu\nu} = 0$.
The Fierz--Pauli action is invariant under linearized diffeomorphisms, where $h_{\mu\nu}$ shifts as
\be
\delta h_{\mu\nu} = \partial_\mu\xi_\nu+\partial_\nu\xi_\mu\,.
\label{eq:lineardiffs}
\ee
This diffeomorphism invariance is seen most simply in the form~\eqref{eq:FP2}. The Einstein tensor~\eqref{eq:linGtensor} is both identically gauge invariant and conserved, so that the action is gauge invariant after integration by parts.

The Einstein tensor is not the most general gauge-invariant local operator in linearized gravity. In fact, the full linearized Riemann tensor
\be
R_{\mu\nu\rho\sigma} = \partial_\rho \partial_\nu h_{\mu\sigma}+\partial_\sigma \partial_\mu h_{\nu\rho}-\partial_\sigma \partial_\nu h_{\mu\rho}-\partial_\rho \partial_\mu h_{\nu\sigma}\, ,\label{eq:LinRiemannT}
\ee
is gauge invariant, and has the symmetries of the GL$(d)$ Young tableau
\be 
R_{\mu\nu\rho\sigma} \,\in~ \raisebox{1.2ex}{\Yboxdimx{13pt}
\Yboxdimy{13pt}\gyoung(\mu;\rho,\nu;\sigma)}~ .\label{eq:RIemannYoung}
\ee
The Einstein tensor~\eqref{eq:linGtensor} is related to the traces of the Riemann tensor as $G_{\mu\nu} = R_{\alpha\mu~\,\nu}^{~~\,\alpha}-\tfrac{1}{2}\eta_{\mu\nu} R_{\alpha\beta}^{~~\,\,\alpha\beta}$, so the linearized Einstein equations are a partial flatness condition, setting the trace of the Riemann tensor to zero. The remaining nonzero curvature is the Weyl tensor.\footnote{Explicitly it is given in terms the Riemann tensor by
\be 
W_{\mu\nu\rho\sigma} = R_{\mu\nu\rho\sigma} -\frac{2}{d-2}\left(\eta_{\mu[\rho}R_{\sigma]|\nu}-\eta_{\nu[\rho}R_{\sigma]|\mu}\right)+ \frac{2}{(d-1)(d-2)}\eta_{\mu[\rho}\eta_{\sigma]\nu}R\, ,\label{eq:WeylGR}
\ee
where $R_{\mu\nu}\equiv R_{\alpha\mu~\,\nu}^{~~\,\alpha}$ is the linearized Ricci tensor and $R\equiv R_{\alpha\beta}^{~~\,\,\alpha\beta}$ is the linearized Ricci scalar.
}
It is straightforward to check that the symmetries of the Riemann tensor along with the linearized Einstein equation are equivalent to the following conditions on the Riemann tensor\footnote{In fact, the logic can be inverted---starting from these equations one can infer that $R_{\mu\nu\alpha\beta}$ can be written in terms of a graviton field $h_{\mu\nu}$ which solves the Einstein equations. This formulation is the one that makes electric-magnetic duality of gravity manifest in $D=4$~\cite{Hull:2001iu,Bekaert:2002dt,Bunster:2006rt}.}
\be
\begin{aligned}
R_{~~\nu\mu\beta}^{\mu}&= 0 \, , \qquad\qquad
& R_{[\mu\nu\alpha]\beta} &=0 \,,\,\,\\
 \partial^\mu R_{\mu\nu\alpha\beta}&=0\,, &
\partial_{[\rho}R_{\mu\nu]\alpha\beta} &= 0\,.
\end{aligned}
\ee
As we will see, the linearized Riemann tensor is closely related to the current ${\cal J}_{\mu\nu|\alpha\beta}$ appearing in eqs.~\eqref{eq:gravityanomalyeqs}.

\vspace{6pt}
\noindent
{\bf Symmetries:} In addition to linearized diffeomorphisms~\eqref{eq:lineardiffs}, the Fierz--Pauli action is also invariant under a global $(1|1)$-biform symmetry where 
\be 
\delta h_{\mu\nu} = b_{\mu\nu}\, ,\label{eq:hshift1}
\ee
with $b_{\mu\nu}$ a constant symmetric tensor. The Noether current associated to this continuous symmetry is a $(2|1)$-form
\be
J_{\mu\nu|\alpha} = \partial_{[\mu}h_{\nu]\alpha} +\eta_{\alpha[\mu}\partial_{\nu]}h+\frac{1}{2}\partial^\rho h_{\rho[\mu}\eta_{\nu]\alpha}\,.
\label{eq:gaugenoninvcurrent}
\ee
The divergence of this current is the linearized Einstein tensor: $\partial^\mu J_{\mu\nu|\alpha} = G_{\nu\alpha}$, so it is conserved on-shell, as expected. However, this current is {\it not} gauge invariant, and so does not really exist as an operator in the theory. This suggests that the symmetry for which the graviton is a Goldstone is a slightly different, but related, one.

The transformation~\eqref{eq:hshift1} with a constant $b_{\mu\nu}$ is not the most general $(1|1)$-form symmetry of the Fierz--Pauli action. More generally, $b_{\mu\nu}$ can have some spatial dependence,  we just need to require that the Riemann tensor built from it vanishes: $R[b]_{\mu\nu\alpha\beta} =0$, rather than $b$ being constant.\footnote{At the free level, we only have to require that the Einstein tensor built from $b_{\mu\nu}$ vanishes: $G[b]_{\mu\nu}=0$, but interactions built of the linearized Riemann tensor will only preserve shifts that have vanishing Riemann.} This condition is the analogue of the $1$-form symmetry in electromagnetism requiring that we shift by a flat connection. The relevant choice is to write $b_{\mu\nu}$ as the divergence of a three-index tensor $b_{\mu\nu} = 2\partial^\alpha\Lambda_{\alpha(\mu|\nu)}$ so that $h_{\mu\nu}$ shifts as
\be
\delta h_{\mu\nu} = 2\partial^\alpha\Lambda_{\alpha(\mu|\nu)}\,,
\label{eq:lhfsymm}
\ee
where $\Lambda_{\alpha\mu|\nu}$ is a $(2|1)$-biform
\be
\Lambda_{\alpha\mu|\nu}\, \in ~ \raisebox{1.15ex}{\Yboxdimx{13pt}
\Yboxdimy{13pt}\gyoung(\alpha;\nu,\mu)}~ .
\ee
The benefit of parametrizing $b_{\mu\nu}$ this way is that the corresponding Noether current is the Riemann tensor~\eqref{eq:LinRiemannT}, which {\it is} gauge invariant. In a precise sense, we can think of the graviton as the Goldstone mode for this nonlinearly realized higher-biform symmetry.\footnote{There is another well known interpretation of the graviton or gauge fields as the goldstone corresponding to the non-linear realization of the infinite number of broken global symmetries making up their gauge symmetries \cite{Ogievetsky:1973ik,Borisov:1974bn,Ivanov:1976zq,Ivanov:1981wn,Pashnev:1997xk,Riccioni:2009hi,Delacretaz:2014oxa,Goon:2014paa,Ivanov:2016lha}.}

We would like to study the gauging of these $(1|1)$-biform symmetries. Our desire is to source the Riemann tensor current, which is diffeomorphism invariant. But, once we promote $b_{\mu\nu}$ to be an arbitrary function, it is hard to tell the difference between~\eqref{eq:hshift1} and~\eqref{eq:lhfsymm}. We are therefore motivated to find a formulation  that decouples these two symmetries in order to simplify the gauging of the system. Precisely this happens in first-order form, as we now describe.

\subsubsection{Linearized first-order formulation}

Einstein gravity of course has a well-known metric-based first-order formulation---the so-called Palatini action. In this formalism the metric and Christoffel connection are treated as independent variables, the equation of motion for the Christoffel symbols allows us to relate them to the metric in the usual way, and integrating them out reproduces the standard Einstein--Hilbert action. Linearized gravity has an analogous Palatini-like formulation, where the action is given by~\cite{Deser:1969wk}
\be 
S = -2\int \, \rd^dx\,  \Big[\varphi^{\mu\nu}\, \left(\partial_\mu \Gamma\indices{_{\nu\alpha}^\alpha}-\partial_\alpha \Gamma\indices{_{\mu\nu}^\alpha}\right)+ \eta^{\mu\nu}\left(\Gamma\indices{_{\mu\nu}^\alpha}\Gamma\indices{_{\alpha\rho}^\rho} -\Gamma\indices{_{\rho\mu}^\alpha}\Gamma\indices{_{\alpha\nu}^\rho}\right)\Big]\, .\label{eq:Spalatini1}
\ee
Here $\varphi_{\mu\nu}$ is a symmetric tensor and $\Gamma_{\mu\nu\rho}$ is the linearized analogue of the Christoffel connection---it is symmetric in its first two indices, but the last index has no specific symmetry property. In other words, it is a tensor of symmetry type ${\gyoung(\mu;\nu)}\, \medotimes \, {\gyoung(\rho)}$. This action is invariant under the combined gauge transformations 
\begin{align}
\delta \varphi_{\mu\nu} &= 2\partial_{(\mu }\xi_{\nu)}  -\eta_{\mu\nu}\partial_\alpha \xi^\alpha\, , \\
\delta \Gamma_{\mu\nu\rho} &= \partial_\mu \partial_\nu \xi_\rho \, ,
\end{align}
with $\xi_\mu$ an arbitrary $d-$dimensional vector. 

In order to see the equivalence of~\eqref{eq:Spalatini1} with the ordinary Einstein action, we vary with respect to $\Gamma_{\mu\nu\rho}$, to obtain an algebraic equation of motion for $\Gamma_{\mu\nu\rho}$
\be 
\eta_{\mu\nu}\Gamma\indices{_{\rho\alpha}^\alpha}-2\Gamma\indices{_{(\mu\nu)\rho}}+ \eta_{\rho(\mu}\Gamma\indices{^{\alpha}_{\alpha\nu)}}+ \partial_\rho \varphi_{\mu\nu} -\eta_{\rho(\mu}\partial^\alpha \varphi\indices{_{\nu)\alpha}}=0\, . \label{eq:eom11}
\ee
This equation can be used to solve for $\Gamma_{\mu\nu\rho}$ in terms of $\varphi_{\mu\nu}$. It is simplest to express in terms of a trace-shifted field 
\be 
h_{\mu\nu}\equiv \varphi_{\mu\nu} -\frac{1}{d-2}\eta_{\mu\nu}\varphi\indices{_\alpha^\alpha}\, ,
\ee
so that the solution to~\eqref{eq:eom11} is the standard expression for the linearized Christoffel connection,
\be 
\Gamma_{\mu\nu\rho} = \frac{1}{2}\left(\partial_\mu h_{\nu\rho}+ \partial_\nu h_{\mu\rho}-\partial_\rho h_{\mu\nu}\right)\, .\label{eq:ChristoffelLG}
\ee
Substituting this relation back into~\eqref{eq:Spalatini1}, integrating by parts, and writing $\varphi_{\mu\nu}$ in terms of $h_{\mu\nu}$, we precisely recover the linearized Einstein action~\eqref{eq:SFierzPauli}.

\vspace{6pt}
\noindent
{\bf Decoupling the symmetric part of $\Gamma_{\mu\nu\alpha}$:} In the formulation of linear gravity given by~\eqref{eq:Spalatini1}, the Christoffel symbol is reducible as a representation of the symmetric group---it can be decomposed into a totally symmetric tensor, and one with the index symmetries of a hook diagram. Interestingly, in linearized gravity only the hook part contributes to the Riemann tensor. We should therefore expect that the totally symmetric part of $\Gamma_{\mu\nu\alpha}$ is unnecessary to formulate the action in first-order form. As we will now show, this is indeed the case.

We begin by splitting $\Gamma_{\mu\nu\alpha}$ into its irreducible components as
\be
\Gamma_{\mu\nu\alpha} = -\frac{1}{3}\left(\Gamma_{\alpha\nu|\mu}+\Gamma_{\alpha\mu|\nu}\right)+\Gamma^{(s)}_{\mu\nu\alpha}\,,
\ee
where $\Gamma^{(s)}$ is a symmetric tensor and $\Gamma_{\alpha\mu|\nu}$ is a $(2|1)$-biform, with the index symmetries of the Young tableau
\be
\Gamma_{\alpha\mu|\nu}\, \in ~ \raisebox{1.15ex}{\Yboxdimx{13pt}
\Yboxdimy{13pt}\gyoung(\alpha;\nu,\mu)}~ ,
\ee
and the normalization has been chosen for later convenience.
In terms of these fields, the action~\eqref{eq:Spalatini1} becomes
\begin{align}
S = \int \, \rd^dx\,  &\bigg[\,\varphi^{\mu\nu}\, \left(-2\partial_\mu \Gamma^{(s)}{}\indices{_{\nu\alpha}^\alpha}+2\partial_\alpha \Gamma^{(s)}{}\indices{_{\mu\nu}^\alpha}\right)-\frac{2}{3}
\varphi^{\mu\nu}\, \left(\partial_{(\mu} \Gamma_{\nu)}-2\partial_\alpha \Gamma^{~~\alpha}_{(\mu~\,|\nu)}\right)
\\[2pt]\nonumber
&
\!\!+ \eta^{\mu\nu}\left(2\Gamma^{(s)}{}\indices{_{\rho\mu}^\alpha}\Gamma^{(s)}{}\indices{_{\alpha\nu}^\rho}-2\Gamma^{(s)}{}^{~~\,\,\alpha}_{\mu\alpha}\Gamma^{(s)}{}^{~~\,\,\alpha}_{\nu\alpha}-\frac{1}{3}\Gamma_{\mu\alpha|\rho}\Gamma_\nu^{~\,\alpha|\rho}+\frac{4}{9}\Gamma_\mu \Gamma_\nu+\frac{2}{3}\Gamma_\mu\Gamma^{(s)}{}^{~~\,\,\alpha}_{\nu\alpha}\right)\bigg],
\end{align}
where we have defined $\Gamma_\mu \equiv \eta^{\alpha\beta}\Gamma_{\mu\alpha|\beta}$, the trace of $\Gamma_{(2|1)}$. We can then integrate out $\Gamma_{\mu\nu\alpha}^{(s)}$ using its equation of motion, which sets
\be
\Gamma_{\mu\nu\alpha}^{(s)} = \frac{1}{2(d-1)}\eta_{(\mu\nu}\Gamma_{\alpha)}+\frac{1}{2}\partial_{(\mu}\varphi_{\nu\alpha)}+\frac{1}{2(d-1)}\partial^\rho \varphi_{\rho(\alpha}\eta_{\mu\nu)}-\frac{1}{2(d-1)}\eta_{(\mu\nu}\partial_{\alpha)}\varphi\,.
\ee
Substituting this back into the action (after defining $\varphi_{\mu\nu} = h_{\mu\nu}-\tfrac{1}{2}h\eta_{\mu\nu}$ and integrating by parts), we get
\be
\begin{aligned}
S = \int \, \rd^dx\,  \bigg(&\frac{2-3d}{3(d-1)}
\left[h^{\mu\nu}\left(\partial_\alpha\Gamma\indices{^\alpha_{\mu | \nu}}-\eta_{\mu\nu}\partial_\alpha \Gamma\indices{^{\alpha\rho}_{ | \rho}}+\partial_\nu \Gamma\indices{_\mu^\rho_{ | \rho}}\right)+ \frac{1}{4}\Gamma_{\mu\nu | \rho}\Gamma^{\mu\nu | \rho}-\frac{1}{2}\Gamma\indices{^{\mu\rho}_{ | \rho}}\Gamma\indices{_\mu^\sigma_{ | \sigma}}\right]\\[2pt]
&-\frac{1}{6(d-1)}h^{\mu\nu}G[h]_{\mu\nu}-\frac{(d-2)}{12(d-1)}(\Gamma_{\mu\nu|\alpha} - 2\partial_{[\mu}h_{\nu]\alpha})(\Gamma^{\mu\nu|\alpha} - 2\partial^{[\mu}h^{\nu]\alpha})
\bigg)\,.
\end{aligned}
\label{eq:palatini2g}
\ee
The field $\Gamma_{(2|1)}$ is still auxiliary, and integrating it out using its equation of motion sets
\be
\Gamma_{\mu\nu|\alpha} = \partial_{\mu}h_{\nu\alpha}-\partial_{\nu}h_{\mu\alpha}\,.
\label{eq:linearpalatinih}
\ee
Substituting this into the action~\eqref{eq:palatini2g} reproduces the linearized Einstein--Hilbert action.

The action~\eqref{eq:palatini2g} therefore is a formulation of linearized gravity with an auxiliary $(2|1)$-biform field. However, it is not exactly of the form that we would expect from~\eqref{eq:biformEHaction}. In particular we would like to remove the quadratic dependence on $h_{\mu\nu}$. 
In reality,~\eqref{eq:palatini2g} is a representative of a two-parameter family of actions that all produce the same equations of motion, which are~\eqref{eq:linearpalatinih}, along with the linearized Einstein equation $G_{\mu\nu} = 0$. We are free to add any multiple of both $h^{\mu\nu}G_{\mu\nu}$ and $(\Gamma_{\mu\nu|\alpha} - \partial_{\mu}h_{\nu\alpha}+\partial_{\mu}h_{\nu\alpha})^2$ to the action without modifying these equations of motion. There is a unique choice, up to overall rescaling, that removes the quadratic dependence on $h_{\mu\nu}$ so that we are left with
\begin{tcolorbox}[colframe=white,arc=0pt,colback=greyish2]
\be
 S = -a\int \rd^dx \left[h^{\mu\nu}\left(\partial_\alpha\Gamma\indices{^\alpha_{\mu | \nu}}-\eta_{\mu\nu}\partial_\alpha \Gamma\indices{^{\alpha\rho}_{ | \rho}}+\partial_\nu \Gamma\indices{_\mu^\rho_{ | \rho}}\right)+ \frac{1}{4}\Gamma_{\mu\nu | \rho}\Gamma^{\mu\nu | \rho}-\frac{1}{2}\Gamma\indices{^{\mu\rho}_{ | \rho}}\Gamma\indices{_\mu^\sigma_{ | \sigma}}\right]\,.  \vphantom{\Bigg)}
 \label{eq:fractongravityaction}
\ee
\end{tcolorbox}
\vspace{-6pt}
\noindent
This action is exactly equivalent to the linearized Einstein action, after integrating out $\Gamma_{\mu\nu|\alpha}$ using~\eqref{eq:linearpalatinih}, and where $a$ parametrizes its overall scaling. Of course we could have just written down~\eqref{eq:fractongravityaction} without taking this detour through the linearized Palatini formulation~\eqref{eq:Spalatini1}, but it is conceptually useful to see how this fractonic formulation of linearized gravity arises from more familiar variables.

If we introduce the two curvatures
\begin{align}
J_{\mu\nu|\rho\sigma} &\equiv \frac{a}{2}\Big( \partial_{\rho}\Gamma_{\mu\nu|\sigma}-\partial_{\sigma}\Gamma_{\mu\nu|\rho}\Big)\, ,\label{eq:FirstR2}\\
\mathcal{Q}_{\mu\nu|\rho} &\equiv \partial_{\mu}h_{\nu|\rho}-\partial_{\nu}h_{\mu|\rho}-\Gamma_{\mu\nu|\rho}\, ,
\label{eq:FirstR3}
\end{align}
then the equations of motion of~\eqref{eq:fractongravityaction} are precisely the statements that (parts of) these curvatures vanish
\be
(\tr \,J)_{\mu|\rho} =0\,,\qquad\qquad\quad \mathcal{Q}_{\mu\nu|\rho} =0\,\,.
\ee
We see that the curvature $J_{\mu\nu|\rho\sigma}$ coincides with the linearized Riemann tensor on-shell using the vanishing of ${\cal Q}_{(2|1)}$. Consequently we find that $J_{\mu\nu|\rho\sigma}$ is also conserved and its antisymmetric derivative vanishes on shell.

The action~\eqref{eq:fractongravityaction} has the same higher-form symmetries as the original linearized Einstein--Hilbert action~\eqref{eq:SFierzPauli}, but this formulation is more convenient for their gauging.

\subsection{Gauging higher-biform symmetries}
We want to understand the higher-form symmetries of the action~\eqref{eq:fractongravityaction} and then introduce a background gauge field source for $J_{(2|2)}$. In addition to the gauge transformations
\be
\delta h_{\mu|\nu} = \partial_{\mu}\xi_{\nu}+\partial_{\nu}\xi_{\mu}\,, \qquad\qquad \delta \Gamma_{\mu\nu|\rho} = \partial_\rho \partial_{\mu}\xi_{\nu}-\partial_\rho \partial_{\nu}\xi_{\mu}\, ,
\ee
the action~\eqref{eq:fractongravityaction} is invariant under some biform symmetries. The first type is~\eqref{eq:hshift1} where we just shift $h_{\mu\nu}$:
\be
\delta h_{\mu\nu}=b_{\mu\nu}\,,\qquad\qquad\quad \delta\Gamma_{\mu\nu|\rho} = 0\,,
\ee
with $b_{\mu\nu}$ a constant tensor. The action is additionally invariant under a symmetry of the form~\eqref{eq:lhfsymm}, where now $\Gamma_{(2|1)}$ transforms as well
\be
\delta h_{\mu\nu}= \partial^\alpha\Lambda_{\alpha\mu|\nu}+\partial^\alpha\Lambda_{\alpha\nu|\mu}\,,\qquad
\delta\Gamma_{\mu\nu|\rho} = \partial_\mu\partial^\alpha \Lambda_{\alpha\nu|\rho}+\partial_\mu\partial^\alpha \Lambda_{\alpha\rho|\nu}-\partial_\nu\partial^\alpha \Lambda_{\alpha\mu|\rho}-\partial_\nu\partial^\alpha \Lambda_{\alpha\rho|\mu}\,.
\label{eq:biformlambdasym}
\ee
We now see one of the benefits of the first-order formulation, it decouples the symmetry whose Noether current is the gauge non-invariant~\eqref{eq:gaugenoninvcurrent}, from the symmetry whose current is the Riemann tensor.\footnote{Of course, we need not stop with~\eqref{eq:biformlambdasym}. We could imagine considering symmetries involving additional derivatives of tensors with more indices. Following a similar logic, we would find that their associated conserved currents are derivatives of the Riemann tensor. This is in a sense similar to the fractonic superfluids considered in section~\ref{sec:superfluid}. It is natural to consider the symmetry~\eqref{eq:biformlambdasym} because gauging it will introduce a source for the simplest gauge-invariant operator in the theory.}

We now want to gauge the symmetry~\eqref{eq:biformlambdasym}. That is, we want to promote~\eqref{eq:biformlambdasym} to a symmetry for an arbitrary function $\Lambda_{\mu\nu|\alpha}$, not just one for which the Einstein tensor built out of $\partial^\alpha\Lambda_{\alpha\mu|\nu}$ vanishes. There are two complementary viewpoints on this procedure: the first is to introduce couplings to a background gauge field into the action~\eqref{eq:fractongravityaction} so that it becomes invariant under these more general transformations. The other is to work at the level of the curvatures~\eqref{eq:FirstR2} and~\eqref{eq:FirstR3} and promote them to gauge-invariant operators. These are of course closely related and we will explore both.

The background field that gauges the symmetry~\eqref{eq:biformlambdasym} should act as a source for $J_{\mu\nu|\alpha\beta}$, so we introduce a gauge field $A_{\mu\nu|\alpha\beta}$ with the index symmetries of Riemann:
\be
A_{\mu\nu|\alpha\beta} \,\in~ \raisebox{1.2ex}{\Yboxdimx{13pt}
\Yboxdimy{13pt}\gyoung(\mu;\alpha,\nu;\beta)}~.
\label{eq:atableau}
\ee
Under the gauged version of~\eqref{eq:biformlambdasym}, the fields in the theory transform as
\begin{align}
\label{eq:Lgauge1}
\delta A_{\mu\nu|\alpha\beta} &= 12{\cal Y}_{(2|2)} \,\partial_\mu \Lambda_{\nu\alpha|\beta} = \partial_\mu \Lambda_{\nu\alpha|\beta}+\cdots\,,\\[2pt]
\delta h_{\mu\nu} &= 3\left(1+g\right)\left(\partial^\alpha\Lambda_{\alpha\mu|\nu}+\partial^\alpha\Lambda_{\alpha\nu|\mu}\right)\,,\\[2pt]
\delta\Gamma_{\mu\nu|\rho} &=3\left(1+g\right)\left( \partial_\mu\partial^\alpha \Lambda_{\alpha\nu|\rho}+\partial_\mu\partial^\alpha \Lambda_{\alpha\rho|\nu}-\partial_\nu\partial^\alpha \Lambda_{\alpha\mu|\rho}-\partial_\nu\partial^\alpha \Lambda_{\alpha\rho|\mu}\right)\,,
\label{eq:Lgauge2}
\end{align}
where ${\cal Y}_{(2|2)}$ is the projector onto the tableau~\eqref{eq:atableau}, and where we have introduced the free parameter, $g$, which captures the relative normalization between the gauge transformations of the dynamical fields $h_{\mu\nu}, \Gamma_{\mu\nu|\rho}$ and the background field $A_{\mu\nu|\alpha\beta}$. The slightly strange parameterization is chosen for later convenience.

At this point, we can just directly construct combinations of $A_{\mu\nu|\alpha\beta}$ that promote~\eqref{eq:FirstR3} to a gauge-invariant current (note that ${\cal Q}_{(2|1)}$ is already gauge invariant). It is mechanically simpler, however, to first construct a gauge invariant action and then derive the gauge-improved current ${\cal J}_{\mu\nu|\alpha\beta}$ from it. There is a two-parameter family of actions invariant under the gauge symmetry, which can be parameterized as
\begin{align}
S = \int\rd^dx \bigg[& \hspace{-0.2em}-h^{\mu\nu}\hspace{-0.1em}\Big(\partial_\alpha \Gamma_{~\,\mu|\nu}^{\alpha}\hspace{-0.1em}-\eta_{\mu\nu}\partial_\alpha \Gamma^{\alpha\rho}_{~~\,|\rho}+\partial_\nu \Gamma_{\mu~|\rho}^{~\rho}+\partial^\alpha\partial^\beta A_{\mu\alpha|\nu\beta}\Big) - \frac{1}{4}\Gamma_{\mu\nu|\rho}\Gamma^{\mu\nu|\rho}+\frac{1}{2}\Gamma^{\mu\rho}_{~~\,|\rho}\Gamma_{\mu~|\sigma}^{~\sigma}\nonumber\\[2pt]
&+g\,A^{\mu|\nu}R_{\mu\nu}-\frac{1}{4}(1+2g)AR+\kappa\, (\partial_\alpha A_{\mu\nu|\rho\sigma})^2-4\kappa\,(\partial^\alpha A_{\mu\nu|\rho\alpha})^2+c_1(\partial_\alpha A_{\mu|\nu})^2\nonumber\\[2pt]
&-2c_1(\partial^\alpha A_{\alpha|\nu})^2+c_2 A_{\mu\nu|\alpha\beta}\partial^\mu\partial^\alpha A^{\nu|\beta}+c_3(\partial A)^2+c_4A^{\mu|\nu}\partial_\mu\partial_\nu A
\bigg]\,,
\label{eq:gaugeinvaction}
\end{align}
where the coefficients $c_1,c_2,c_3,c_4$ can be written in terms of the free parameters $g, \kappa$ as:
\begin{align}
\label{eq:cvars1}
c_1 &\equiv \frac{1}{2}\left(g^2-8\kappa-1\right)\,, \quad \quad& c_2 &\equiv -g-8\kappa-1\,,\\
c_3 &\equiv -\frac{1}{2}g(g+1)+\kappa\,  , \quad\quad& c_4&\equiv -\frac{1}{2}\Big(g(2g+1)-8\kappa-1\Big)\,.
\label{eq:cvars2}
\end{align}
In writing the action~\eqref{eq:gaugeinvaction} it was convenient to introduce both the trace of $A_{\mu\nu|\alpha\beta}$, defined as $A_{\mu|\nu} \equiv A_{\mu \phantom{\nu}|\nu\alpha}^{\phantom{\mu}\alpha}$, and the double trace $A\equiv \eta^{\mu\nu}A_{\mu|\nu}$. 

Using the action~\eqref{eq:gaugeinvaction} we can extract gauge-invariant versions of the Einstein tensor and the current $J_{(2|2)}$ as
\begin{align}
{\cal G}_{\mu\nu} &\equiv \frac{\delta S}{\delta h^{\mu\nu}}\,,\\
{\cal J}_{\mu\nu|\alpha\beta} &\equiv \frac{\delta S}{\delta A^{\mu\nu|\alpha\beta}}\,.
\end{align}
Note that the $\Gamma_{(2|1)}$ equation of motion is unchanged in the presence of $A_{(2|2)}$ and still sets $\Gamma_{\mu\nu|\alpha} = \partial_{\mu}h_{\nu\alpha}-\partial_{\nu}h_{\mu\alpha}$. Explicitly, the gauge-invariant improvement of the Einstein tensor is given by
\begin{tcolorbox}[colframe=white,arc=0pt,colback=greyish2]
\be
\begin{aligned}
{\cal G}_{\mu\nu} = G[\Gamma]_{\mu\nu}
&-\partial^\alpha\partial^\beta A_{\mu\alpha|\nu\beta}-g\left(
\square A_{\mu|\nu}-2\partial^\alpha\partial_{(\mu}A_{\nu)|\alpha}+\eta_{\mu\nu}\partial^\alpha\partial^\beta A_{\alpha|\beta}
\right)\\
&-\frac{1}{2}\left(1+2g\right)\Big(\partial_\mu\partial_\nu A-\eta_{\mu\nu}\square A\Big)\,,
\label{eq:gaugeinveins}
\end{aligned}
\ee
\end{tcolorbox}
\vspace{-6pt}
\noindent
where $G[\Gamma]_{\mu\nu}$ is the ordinary Einstein tensor written in terms of $\Gamma$,
\be
G[\Gamma]_{\mu\nu}\equiv -\partial_\alpha\Gamma\indices{^\alpha_{(\mu | \nu)}}+\eta_{\mu\nu}\partial_\alpha \Gamma\indices{^{\alpha\rho}_{ | \rho}}-\partial_{(\mu} \Gamma\indices{_{\nu)}^\rho_{ | \rho}}\,.
\ee

Similarly, we can derive the gauge-invariant current:
\begin{tcolorbox}[colframe=white,arc=0pt,colback=greyish2]
\be
{\cal J}_{\mu\nu|\alpha\beta} = \frac{1}{4}R_{\mu\nu\alpha\beta} +  \frac{3}{4}{\cal Y}_{(2|2)}\Big( g\, \eta_{\nu\beta}R_{\mu\alpha}-\frac{1}{4} (1+2g)\eta_{\mu\alpha}\eta_{\nu\beta}R\Big)-C_{\mu\nu|\alpha\beta}\,,
\label{eq:Jcurrentgauge}
\ee
\end{tcolorbox}
\vspace{-6pt}
\noindent
where $R_{\mu\nu\alpha\beta}$ is the linearized Riemann tensor, $R_{\mu\alpha}$ is the linearized Ricci tensor, $R$ is the linearized Ricci scalar, and we have defined here the tensor
\be
\begin{aligned}
C_{\mu\nu|\alpha\beta} \equiv -{\cal Y}_{(2|2)} \bigg(
&2\kappa\, \left(4\partial^\rho \partial_\mu A_{\nu\alpha|\beta \rho}- \,\square A_{\mu\nu|\alpha\beta}\right)+\frac{3c_2}{4}\Big[\eta_{\mu\alpha}\partial^\rho\partial^\sigma A_{\rho\nu|\sigma\beta}+ \partial_\mu\partial_\alpha A_{\nu|\beta}\Big]\\
&+3c_1\eta_{\mu\alpha}\left[\partial_\nu\partial^\rho A_{\beta|\rho}-\frac{1}{2}\square A_{\nu|\beta}\right]+\frac{3c_4}{4}\eta_{\mu\alpha}\Big[\eta_{\nu\beta}\partial^\rho\partial^\sigma A_{\rho|\sigma}+\partial_\nu\partial_\beta A\Big]\\
&-\frac{3c_3}{2}\eta_{\mu\alpha}\eta_{\nu\beta}\square A
\bigg)\,.
\end{aligned}
\ee
where $\mathcal{Y}_{(2|2)}$ is the projection on the tableau of equation \eqref{eq:atableau}. It is relatively straightforward to check that both~\eqref{eq:gaugeinveins} and~\eqref{eq:Jcurrentgauge} are invariant under the gauge transformations~\eqref{eq:Lgauge1}--\eqref{eq:Lgauge2}. Note that as a consequence of the index symmetries of $A_{\mu\nu|\alpha\beta}$, the current satisfies the condition
$
{\cal J}_{[\mu\nu|\alpha]\beta} = 0\,,
$
even off-shell. In the next section we will explore the on-shell properties of this current.

Though we have been proceeding in first-order form, there is an elegant simplification of the action~\eqref{eq:gaugeinvaction} that occurs in second-order form. If we use the equation~\eqref{eq:linearpalatinih} to integrate out $\Gamma_{(2|1)}$, we can write the action as
\be 
S = \frac{1}{2}\int \rd^dx \bigg(h^{\mu\nu}\mathcal{G}_{\mu\nu} + A^{\mu\nu|\rho\sigma}\mathcal{J}_{\mu\nu|\rho\sigma}\bigg)\, .\label{eq:NiceActGeom}
\ee
Varying this action with respect to $h_{\mu\nu}$ produces~\eqref{eq:gaugeinveins} (with $\Gamma_{(2|1)}$ written in terms of $h_{\mu\nu}$), and varying with respect to $A_{\mu\nu|\rho\sigma}$ produces~\eqref{eq:Jcurrentgauge}. {Note that the existence of this formulation of the action is entirely nontrivial because both terms contribute to each variation, so there must be a conspiracy between the two terms to produce the correct variations.}

\subsection{Anomalies}
\label{sec:actionanom}
We want to understand how the introduction of the background gauge field $A_{(2|2)}$ changes the properties of the current $J_{(2|2)}$. Recall that this current satisfies all of the equations
\be
\begin{aligned}
J_{~~\nu|\mu\beta}^{\mu}&= 0 \, , \qquad\qquad
&  & \,\,\\
 \partial^\mu J_{\mu\nu|\alpha\beta}&=0\,, &
\partial_{[\rho}J_{\mu\nu]|\alpha\beta} &= 0\,,
\end{aligned}
\label{eq:eqsthatJsatsfies}
\ee
on shell. We have coupled the theory to $A_{(2|2)}$ in a way that preserves the gauge transformation~\eqref{eq:Lgauge1}, so ${\cal J}_{(2|2)}$ will be conserved on-shell.
However, the other two conditions in~\eqref{eq:eqsthatJsatsfies} are not necessarily satisfied because $A_{\mu\nu|\alpha\beta}$ is not traceless, and does not have the correct gauge symmetry to guarantee dual conservation. So, we want to see how we can use the freedom to select the parameters $\kappa$ and $g$ in~\eqref{eq:gaugeinvaction} to enforce as many of the conditions~\eqref{eq:eqsthatJsatsfies} as possible for ${\cal J}_{(2|2)}$ on shell. We will find two interesting features: the first is that not all the conservation conditions~\eqref{eq:eqsthatJsatsfies} can be satisfied---which implies a nontrivial mixed 't Hooft anomaly. Additionally, we will find that the $d=4$ dimensional case is special, where we can impose even fewer conditions.

\vspace{6pt}
\noindent
{\bf Conservation conditions on-shell:} In order to explore the anomaly structure, we need to compute the on-shell conditions~\eqref{eq:eqsthatJsatsfies} for ${\cal J}_{(2|2)}$. This means that we can use the conditions ${\cal G}_{\mu\nu} = 0$, where ${\cal G}$ is defined in~\eqref{eq:gaugeinveins} and ${\cal Q} = 0$, where ${\cal Q}$ is defined in~\eqref{eq:FirstR3}. We can then compute
\begin{itemize}

\item {\bf Trace:} We first compute the trace ${\cal J}_{\mu|\nu} =\eta^{\alpha\beta}{\cal J}_{\mu\alpha|\nu\beta}$:
\be
\begin{aligned}
{\cal J}_{\mu|\nu} =  \frac{\big(3-d-8(d-4)\kappa\big)}{4}\bigg[\partial^\rho\partial^\sigma A_{\mu\rho|\nu\sigma} &-\square A_{\mu|\nu}+2\partial^\alpha \partial_{(\mu}A_{\nu)|\alpha}-\eta_{\mu\nu}\partial^\rho\partial^\sigma A_{\rho|\sigma} \\
&-\frac{1}{2}\Big(\partial_\mu\partial_\nu A-\eta_{\mu\nu}\square A\Big)
\bigg]\,,\vphantom{\Bigg)}
\end{aligned}
\label{eq:trace1}
\ee
which we see in general does not necessarily vanish. We can also check that the right hand side is gauge invariant, as it must be.

\item {\bf Conservation:} Next we can compute the divergence, which vanishes on shell
\be
\partial^\mu {\cal J}_{\mu\nu|\alpha\beta} = 0\,.
\label{eq:cons2}
\ee
This is a consequence of the way that we have introduced the gauge field $A_{\mu\nu|\alpha\beta}$. Since its gauge transformation involves a derivative of the gauge parameter, gauge invariance guarantees that it couples to an on-shell conserved current. If we had chosen to gauge the theory in a different way this condition would not necessarily be satisfied.

\item {\bf Dual conservation:} Finally, we can compute the antisymmetric derivative
\be
\begin{aligned}
\partial_{[\rho} {\cal J}_{\mu\nu]|\alpha\beta} = -\partial_{[\rho} {\cal C}_{\mu\nu]|\alpha\beta} \,,
\end{aligned}
\label{eq:dualcons3}
\ee
where we have defined the tensor
\be
\begin{aligned}
{\cal C}_{\mu\nu|\alpha\beta} \equiv- {\cal Y}_{(2|2)} \bigg[2&\kappa \left(
4 \partial^\rho \partial_\mu A_{\nu\alpha|\beta \rho}-\square A_{\mu\nu|\alpha\beta}\right)
\\
&-\frac{3(1+8\kappa)}{4}\eta_{\mu\alpha}\Big(\partial^\rho\partial^\sigma A_{\rho\nu|\sigma\beta}+ 2\partial_\nu\partial^\rho A_{\beta|\rho}-\square A_{\nu|\beta}\Big)\\
&+\frac{3}{8}\left(8\kappa+1+\frac{1}{d-2}\right)\eta_{\mu\alpha}\eta_{\nu\beta}\left(\partial^\rho\partial^\sigma A_{\rho|\sigma}-\frac{1}{2} \square A\right)
\bigg]\,.
\end{aligned}
\ee
It is easy to check that the field strength $\partial_{[\rho} {\cal C}_{\mu\nu]|\alpha\beta} $ is gauge invariant, as expected.

\end{itemize}
Now that we have the three on-shell equations~\eqref{eq:trace1},~\eqref{eq:cons2}, and~\eqref{eq:dualcons3} we want to see how many of them we can set to zero simultaneously. It is clear from~\eqref{eq:trace1} and~\eqref{eq:dualcons3} that it is not possible to make them all vanish. In particular, we cannot make~\eqref{eq:dualcons3} vanish for any choice of parameters, because we have required~\eqref{eq:cons2} to be satisfied,  
which is the expression of a mixed anomaly between the electric and magnetic biform symmetries. However, we see from~\eqref{eq:trace1} that we can set the trace to zero in $d \neq 4$, so it is
convenient to split the discussion into two cases, $d>4$ and $d=4$. We will consider the generic case first.

\vspace{6pt}
\noindent
{\bf Generic dimension $\boldsymbol{d>4}$:} In the general case it is possible to make ${\cal J}_{(2|2)}$ both traceless and conserved, in addition to being dual traceless. We accomplish this by setting
\be
\kappa = -\frac{(d-3)}{8(d-4)}\,.
\label{eq:kingendim}
\ee
In addition, it is convenient to further set $g = 1/(2-d)$. With this choice, we can partially fix the gauge invariance to set $A = 0$, after which the trace $A_{\mu|\nu}$ completely decouples and the action~\eqref{eq:gaugeinvaction} becomes
\be
\begin{aligned}
S = \int\rd^dx \bigg[& \!-h^{\mu\nu}\Big(\partial_\alpha \Gamma_{~\,\mu|\nu}^{\alpha}\hspace{-0.1em}-\eta_{\mu\nu}\partial_\alpha \Gamma^{\alpha\rho}_{~~\,\,|\rho}+\partial_\nu \Gamma_{\mu~|\rho}^{~\rho}+\partial^\alpha\partial^\beta A^{(T)}_{\mu\alpha|\nu\beta}\Big) - \frac{1}{4}\Gamma_{\mu\nu|\rho}\Gamma^{\mu\nu|\rho}+\frac{1}{2}\Gamma^{\mu\rho}_{~~|\rho}\Gamma_{\mu~|\rho}^{~\rho}\\[2pt]
&\!-\frac{(d-3)}{8(d-4)} (\partial_\alpha A^{(T)}_{\mu\nu|\rho\sigma})^2+\frac{(d-3)}{2(d-4)}(\partial^\alpha A^{(T)}_{\mu\nu|\rho\alpha})^2
\bigg]\,,
\end{aligned}
\label{eq:gaugeinvactiongenD}
\ee
where we have introduced the traceless part of $A_{\mu\nu|\alpha\beta}$ defined as
\be
 A^{(T)}_{\mu\nu|\alpha\beta} \equiv  A_{\mu\nu|\alpha\beta} - \frac{3}{d-2} {\cal Y}_{(2|2)} A_{\mu|\alpha}\eta_{\nu\beta}\,,
\ee
and where $A_{\mu|\nu}$ itself is traceless because of the partial gauge fixing we have done. The action~\eqref{eq:gaugeinvactiongenD} is invariant under the gauge transformations~\eqref{eq:Lgauge1}--\eqref{eq:Lgauge2} with a traceless gauge parameter. The fact that only the traceless part of $A_{(2|2)}$ couples to the dynamical fields will imply that the corresponding current ${\cal J}_{\mu\nu|\alpha\beta}$ is now traceless {\it off shell}.
We can write the relevant current more explicitly as
\be
{\cal J}_{\mu\nu|\alpha\beta} = \frac{1}{4}W_{\mu\nu\alpha\beta} -C^{(T)}_{\mu\nu|\alpha\beta}\,,
\label{eq:Jingend}
\ee
where the $C$ tensor built from the traceless part of $A$ is:
\be
C^{(T)}_{\mu\nu|\alpha\beta} =\frac{d-3}{d-4}{\cal Y}_{(2|2)}\left(
\partial^\rho \partial_\mu A^{(T)}_{\nu\alpha|\beta \rho}- \frac{1}{4} \square A^{(T)}_{\mu\nu|\alpha\beta}-\frac{3}{2(d-2)}\eta_{\mu\alpha}\partial^\rho\partial^\sigma A^{(T)}_{\rho\nu|\sigma\beta}\right)\,.
\label{eq:genctens}
\ee
Note that it is not obvious in the way \eqref{eq:genctens} is written, but $C^{(T)}_{\mu\nu|\alpha\beta}$ is traceless as it should. In fact,~\eqref{eq:genctens} is the unique tensor that is traceless, has the correct index symmetries, and transforms oppositely to the Weyl tensor, so that the current~\eqref{eq:Jingend} is gauge invariant. Indeed, we could have worked directly at the level of the current and introduced~\eqref{eq:genctens} in order to gauge the relevant symmetries, and we would have ended up with the same result.

The current~\eqref{eq:Jingend} has many desired properties; 
however the right hand side of~\eqref{eq:dualcons3} continues to be nonzero, so that all together we have
\begin{tcolorbox}[colframe=white,arc=0pt,colback=greyish2]
\be
\begin{aligned}
{\cal J}_{~~\nu|\mu\beta}^{\mu}&= 0 \, , \qquad\qquad
&&\,\,\\
 \partial^\mu {\cal J}_{\mu\nu|\alpha\beta}&=0\,, &
\partial_{[\rho}{\cal J}_{\mu\nu]|\alpha\beta} &= -\partial_{[\rho} {\cal C}^{(d)}_{\mu\nu]|\alpha\beta}\,,
\end{aligned}
\label{eq:dimeqsgravity}
\ee
\end{tcolorbox}
\vspace{-6pt}
\noindent
where the tensor appearing in the magnetic conservation equation is
\be
 {\cal C}_{\mu\nu|\alpha\beta}^{(d)} = \frac{d-3}{d-4}{\cal Y}_{(2|2)}\bigg( \partial^\sigma \partial_\beta A^{(T)}_{\mu\nu|\alpha\sigma} -\frac{1}{4}\square A_{\mu\nu|\alpha\beta}^{(T)}-\frac{3}{4(d-3)} \eta_{\mu\alpha} \partial^\rho\partial^\sigma A_{\rho\nu|\sigma\beta}^{(T)} \bigg)\,.
\ee
Note that this differs slightly from~\eqref{eq:genctens} because there are contributions from the equations of motion to the right hand side of the dual conservation equation.
Our inability to satisfy both the electric and magnetic conservation laws at the same time is a consequence of a mixed 't Hooft anomaly between these global biform symmetries. We could, of course, have instead chosen $J_{(2|2)}$ to be dual conserved (preserving the magnetic symmetry), but we would then have found both electric conservation and tracelessness would fail to hold.
We can understand the presence of the graviton as being a consequence of this mixed anomaly, as we explore in section~\ref{sec:grav2pt}.

\vspace{6pt}
\noindent
{\bf Four dimensions:} As we can see from the appearance of $(d-4)$ factors in~\eqref{eq:kingendim} and~\eqref{eq:genctens}, something is special about $d=4$. Indeed, we can no longer choose parameters to make ${\cal J}_{(2|2)}$ both traceless and conserved.\footnote{As in the galileon superfluid case, the tensor~\eqref{eq:genctens} (after multiplying through by $d-4$) is gauge invariant in $d=4$, which prevents us from using it to gauge  $J_{(2|2)}$.} 
In particular, setting $\kappa$ as in~\eqref{eq:kingendim} is not possible, because it will diverge in $d=4$. In this case, the conservation equations take the form
\begin{tcolorbox}[colframe=white,arc=0pt,colback=greyish2]
\be
\begin{aligned}
{\cal J}_{~~\nu|\mu\beta}^{\mu}&= {\cal C}_{\mu|\nu}^{(4)} \, , \qquad\qquad
&\,\,\\
 \partial^\mu {\cal J}_{\mu\nu|\alpha\beta}&=0\,, &
\partial_{[\rho}{\cal J}_{\mu\nu]|\alpha\beta} &= -\partial_{[\rho} {\cal C}^{(4)}_{\mu\nu]|\alpha\beta}\,,
\end{aligned}
\ee
\end{tcolorbox}
\vspace{-6pt}
\noindent
where the field strengths appearing  on the right hand sides are  
\be
{\cal C}_{\mu|\nu}^{(4)} =  -\frac{1}{4}\bigg[\partial^\rho\partial^\sigma A_{\mu\rho|\nu\sigma} -\square A_{\mu|\nu}+2\partial^\alpha \partial_{(\mu}A_{\nu)|\alpha}-\eta_{\mu\nu}\partial^\rho\partial^\sigma A_{\rho|\sigma}-\frac{1}{2}\partial_\mu \partial_\nu A+\frac{1}{2}\eta_{\mu\nu}\square A\bigg]\,,
\label{eq:ctracetens}
\ee
along with
\be 
\begin{aligned}
{\cal C}^{(4)}_{\mu\nu|\alpha\beta} \hspace{-0.1em}=\hspace{-0.1em}{\cal Y}_{(2|2)} \! \bigg[\vspace{-0.1em}&2\kappa \! \left(\square A_{\mu\nu|\alpha\beta}-
4 \partial^\rho \partial_\mu A_{\nu\alpha|\beta \rho}\right)\hspace{-0.1em}+\hspace{-0.1em}\left(\! \frac{3}{4}+6\kappa\! \right)\! \eta_{\mu\alpha}\vspace{-0.1em}\Big(\partial^\rho\partial^\sigma A_{\rho\nu|\sigma\beta}+ 2\partial_\nu\partial^\rho A_{\beta|\rho}-\square A_{\nu|\beta}\Big)\\
& - \frac{3}{16}\left(3+16\kappa\right)\eta_{\mu\alpha}\eta_{\nu\beta}\Big(\partial^\rho\partial^\sigma A_{\rho|\sigma} -\frac{1}{2}\square A  \Big)
\bigg]\,,
\end{aligned}
\label{eq:cdualdivtens}
\ee
where $A$ is the trace of $A_{\mu|\nu}$.

It is easy to check that ${\cal C}_{(1|1)}^{(4)}$ and $\rd{\cal C}_{(2|2)}^{(4)}$ are gauge invariant under the transformation~\eqref{eq:Lgauge1}. Interestingly, we see that the minimal anomaly in $d=4$ also involves failure of tracelessness, in contrast to the $d>4$ case.

Here we have given only one presentation of the mixed anomaly between the various conservation conditions~\eqref{eq:eqsthatJsatsfies}. By including different contact terms (corresponding to terms quadratic in the gauge field) or gauging the theory in different variables one can shuffle around the anomaly into failures of different conservations conditions. However, the incompatibility between electric and magnetic conservation cannot be changed. Indeed, one can understand gravity as a gapless phase mandated by the presence of this anomaly. In the following section, we explore this viewpoint, and show how the anomalies uncovered here can be used to prove that there is a massless spin-2 field in the spectrum of the theory, without making any reference to an underlying Lagrangian that realizes the physics. {This approach also has the benefit of making manifest the origin of the anomaly and showing that it cannot be removed by clever choice of field variables or gauging of the theory.}

\subsection{Gravity as a phase of matter}
\label{sec:grav2pt}

The preceding discussion has centered on a particular realization of the physics described by the current conservation conditions~\eqref{eq:gravityanomalyeqs}, but we now wish to show that this effective description is actually universal. To do so, we will prove a version of a Goldstone theorem, showing that any theory that has two currents of the type~\eqref{eq:gravcurrentdef} with a mixed anomaly {\it necessarily} has a gapless spin-2 mode in the spectrum. In the deep infrared, this graviton is of course described by the linearized Einstein action discussed in section~\ref{sec:lingrav}. The philosophy is then that we can define a (linear) theory of gravity as the gapless phase with a particular structure of conserved currents associated to biform global symmetries.

Concretely, we consider a theory that has two currents of the form~\eqref{eq:gravcurrentdef} 
\be
J_{\mu_1\mu_2 | \nu_1\nu_2}\,, \qquad {\rm and} \qquad K_{\mu_1\mu_2\cdots \mu_{d-2} | \nu_1\nu_2\cdots \nu_{d-2}}\,.
\ee
The two natural conditions that we can impose on these currents {\it at separated points} are 
\be
\begin{aligned}
(\tr J)_{\mu|\nu}&= 0 \, , \qquad\quad
& & \\
 \partial^{\mu_1} J_{\mu_1\mu_2|\nu_1\nu_2}&=0\,, &\qquad\quad
\partial^{\mu_1}K_{\mu_1\mu_2\cdots \mu_{d-2} | \nu_1\nu_2\cdots \nu_{d-2}} &= 0\,.
\end{aligned}
\label{eq:consconditionsgpm}
\ee
These relations hold as operator equations in the quantum theory, so long as operators never collide. However, the conditions~\eqref{eq:consconditionsgpm} {\it cannot} all be made to hold at coincident points as well. Instead the failure of these conditions is universal, and is dictated by the anomaly.

We now wish to show how the combination of the conservation conditions~\eqref{eq:consconditionsgpm} at separated points, along with the equation for the anomaly completely fixes the $\braket{J*\hspace{-2pt}K\hspace{-1pt}*}$ two point function. The spectral decomposition of this two point function will then include a massless spin-2 state, establishing a Goldstone theorem for the graviton.

\subsubsection{The current two-point function}

We now use~\eqref{eq:consconditionsgpm} to fix the two point function between the currents $J_{(2|2)}$ and $K_{(d-2|d-2)}$. First, it is convenient to dualize the current $K_{(d-2|d-2)}$ into $*\hspace{-.5pt}K\hspace{-1pt}*_{(2|2)}$, so that the two currents have the same index symmetries.  We can then construct the most general ansatz for the Fourier-space correlator $\braket{J_{\mu_1\mu_2|\nu_1\nu_2}*\hspace{-2pt}K\hspace{-1pt}*_{\alpha_1\alpha_2|\beta_1\beta_2}}$, where the two currents have the following index symmetries
\be
J_{\mu_1\mu_2|\nu_1\nu_2} \,\in~ \raisebox{1.25ex}{\Yboxdimx{13.5pt}
\Yboxdimy{13.5pt}\gyoung({{\hspace{.1em}\mu_1}};{{\hspace{.1em}\nu_1}},{{\hspace{.1em}\mu_2}};{{\hspace{.1em}\nu_2}})}~,\qquad\quad
*\hspace{-.5pt}K\hspace{-1pt}*_{\alpha_1\alpha_2|\beta_1\beta_2} \,\in~ \raisebox{1.25ex}{\Yboxdimx{13.5pt}
\Yboxdimy{13.5pt}\gyoung({{\hspace{.1em}\alpha_1}};{{\hspace{.1em}\beta_1}},{{\hspace{.1em}\alpha_2}};{{\hspace{.1em}\beta_2}})}~,
\label{eq:gravcurrents}
\ee
which is built solely out of the Lorentz-invariant metric $\eta_{\mu\nu}$, and the single momentum that the correlator depends on, $p_\mu$. This general ansatz has eleven different independent tensor structures, which can each be multiplied by an arbitrary function of $p^2$. 

Next, we want to see how many of the conditions~\eqref{eq:consconditionsgpm} we can simultaneously impose everywhere (meaning both separated and coincident points). Unsurprisingly, it is not possible to satisfy all of these conditions simultaneously, and depending on which equations we require different possibilities are allowed. We enumerate all the possibilities in appendix~\ref{app:lingravanom} and here just focus on the maximal case, where we impose as many conditions as possible. There is again a difference between what happens in $d=4$ and in $d>4$, so we treat them separately.

\vspace{6pt}
\noindent
{\bf Generic dimension $\boldsymbol{d>4}$:} In general dimensions $\geq 4$, it is possible to require that $J_{(2|2)}$ is both conserved and traceless even at coincident points. We can write these conditions as
\be
\begin{aligned}
(\tr J)_{\mu|\nu}&= 0 \,, \\
 \partial^{\mu_1} J_{\mu_1\mu_2|\nu_1\nu_2}&=0\,.
\end{aligned}
\label{eq:dg4conditions}
\ee
These conditions  on $J$
completely fix the current two-point function, up to a  function of $p^2$:
\be
\braket{J_{\mu_1\mu_2|\nu_1\nu_2}*\hspace{-1.5pt}K\hspace{-1pt}*_{\alpha_1\alpha_2|\beta_1\beta_2}}
  = -\frac{1}{4} (d -4) (d -3) f(p^2)\Big[ p^2\Pi^{(1|1)}_{\mu_1\mu_2\nu_1\nu_2\alpha_1\alpha_2\beta_1\beta_2} +{\rm contact~terms}\Big]\,,
\label{eq:current2ptbeforeanom}
\ee
where ``contact terms" indicates terms that are purely analytic in $p^2$, and 
in this expression we have defined the tensor structure (for more details see appendix~\ref{app:GravProj})
\be 
\Pi^{(1|1)}_{\mu_1\mu_2\nu_1\nu_2\alpha_1\alpha_2\beta_1\beta_2} \equiv  {\cal P}\,\frac{9(d-2)}{8(d-3)} \frac{p_{\mu_1}p_{\nu_1}p_{\alpha_1}p_{\beta_1}}{p^4}\left(\eta_{\mu_2\alpha_2}\eta_{\nu_2\beta_2} + \eta_{\nu_2\alpha_2}\eta_{\mu_2\beta_2}-\frac{2}{d-2}\eta_{\mu_2\nu_2}\eta_{\alpha_2\beta_2}\right)\, ,
\label{eq:Pi11def}
\ee
with  ${\cal P} \equiv {\cal Y}_{(2|2)^T} {\cal Y}_{(2|2)^T} $ a Young projector onto the tableau
\be
 \raisebox{1.25ex}{\Yboxdimx{13.5pt}
\Yboxdimy{13.5pt}\gyoung({{\hspace{.1em}\mu_1}};{{\hspace{.1em}\nu_1}},{{\hspace{.09em}\mu_2}};{{\hspace{.09em}\nu_2}})}^T~\medotimes~\, \raisebox{1.25ex}{\Yboxdimx{13.5pt}
\Yboxdimy{13.5pt}\gyoung({{\hspace{.075em}\alpha_1}};{{\hspace{.05em}\beta_1}},{{\hspace{.075em}\alpha_2}};{{\hspace{.05em}\beta_2}})}^T\, ,
\label{eq:weylweylprojectortraceless}
\ee
{where the superscript $T$ means the tableaux are traceless.}

The expression~\eqref{eq:current2ptbeforeanom} satisfies all the conditions~\eqref{eq:dg4conditions} exactly, but there is no choice of $f(p^2)$ for which $K_{(d-2|d-2)}$ is identically conserved (corresponding to $\partial_{[\alpha}\hspace{-1.25pt}*\hspace{-1pt}K\hspace{-1pt}*_{\mu_1\mu_2]|\nu_1\nu_2}  = 0$), aside from the trivial $f=0$. Instead, the anomaly equation for ${\cal K}$ (which is the gauge-improvement of $K$ in a background field that sources $J$)\footnote{This equation can be obtained from~\eqref{eq:dimeqsgravity}, but it can alternatively be thought of as a definition of the theory because ${\cal K}$ is uniquely determined in the presence of a traceless source for $J$, as was discussed below equation~\eqref{eq:genctens}.
}
\be
 p_{[\lambda}\hspace{-1.5pt}*\hspace{-1.4pt}{\cal K}\hspace{-1pt}*_{\mu\nu]|}^{\hphantom{\mu\nu]}\,\,\alpha\beta}  = -\frac{d-3}{d-4}p_{[\lambda}{\cal Y}_{(2|2)}\bigg( p^\rho p^\beta A^{(T)\,\alpha}_{\mu\nu]|\phantom{\alpha}\rho} -\frac{1}{4}p^2 A_{\mu\nu]|}^{(T)\, \alpha\beta}-\frac{3}{4(d-3)} \delta_{\mu}^{\phantom{\mu}\alpha} p^\rho p^\sigma A_{\nu]\rho|\phantom{\beta}\sigma}^{(T)\,\,\beta} \bigg)\,.
 \label{eq:DualCons1}
\ee
expresses the fact that the non-conservation of $K$ only fails at coincident points. Taking a functional derivative of~\eqref{eq:DualCons1} with respect to $A^{(T)}_{\alpha_1\alpha_2|\beta_1\beta_2}$ yields an explicit expression for the terms analytic in $p$ that appear in the non-conservation of $K$. Applying this to~\eqref{eq:current2ptbeforeanom} sets
$f(p^2) = -\frac{1}{(d-2)(d-4)}$, and so the current-current two-point function is completely fixed:
\begin{tcolorbox}[colframe=white,arc=0pt,colback=greyish2]
\vspace{-7pt}
\be
\braket{J_{\mu_1\mu_2|\nu_1\nu_2}*\hspace{-1pt}K\hspace{-1pt}*_{\alpha_1\alpha_2|\beta_1\beta_2}}
  = \frac{d-3 }{4 (d -2)}\bigg[ p^2\Pi^{(1|1)}_{\mu_1\mu_2\nu_1\nu_2\alpha_1\alpha_2\beta_1\beta_2} +{\rm contact~terms}\bigg]\,.
\label{eq:current2ptgenD}
\ee
\vspace{-10pt}
\end{tcolorbox}
\vspace{-6pt}
\noindent
Looking at the definition~\eqref{eq:Pi11def}, we see that~\eqref{eq:current2ptgenD} scales like $p^{-2}$, so we can already anticipate that there will be a massless particle in the spectrum of the theory, and indeed we will see that this is the two-point function of a massless spin-2 particle.

The conditions~\eqref{eq:dg4conditions} are the maximal set that we can impose identically, shuffling the anomaly into the failure of conservation of the magnetic current at coincident points. However, it is possible to require other conditions to be satisfied identically, making the anomaly appear elsewhere. All this does is change the precise contact terms that appear in~\eqref{eq:current2ptgenD}, but does not alter the nonlocal part of the correlator. Since only the nonlocal part is relevant to extract the spectrum of the theory, there is no loss of generality in making the choices of conservation conditions to impose that we have made. Nevertheless, for completeness in appendix~\ref{app:lingravanom}, we discuss the other possible conditions that one could impose.

\vspace{6pt}
\noindent
{\bf Four dimensions:} In four dimensions there are several subtle differences compared with the general dimensional case. First, we have to take into account dimension-dependent identities, which cause some linear combinations of the general tensor structures in the current two-point function ansatz to vanish. This removes one free parameter, leading to an ansatz with 10 independent structures. 

In four dimensions,
it is not possible to require all of~\eqref{eq:dg4conditions} at coincident points, consistent with what we saw in section~\ref{sec:actionanom}.
Instead, the maximal condition we can impose everywhere is
\be
\begin{aligned}
 \partial^{\mu_1} J_{\mu_1\mu_2|\nu_1\nu_2}&=0\,.
\end{aligned}
\label{eq:d4conditions}
\ee
This, completely fixes the correlator, up to overall normalization, which can be fixed by the anomaly equations
\be
\begin{aligned}
{\cal J}_{~~\nu|\mu\beta}^{\mu}= {\cal C}_{\mu|\nu}^{(4)} \, , \qquad\qquad
\partial_{[\rho}*\hspace{-.5pt}{\cal K}\hspace{-.5pt}*_{\mu\nu]|\alpha\beta} = -\partial_{[\rho} {\cal C}^{(4)}_{\mu\nu]|\alpha\beta}\,,
\end{aligned}
\ee
where the field strengths ${\cal C}^{(4)}_{\mu|\nu}$ and $\rd{\cal C}^{(4)}_{(2|2)}$ are given by~\eqref{eq:ctracetens} and~\eqref{eq:cdualdivtens}, respectively. Accounting for all these constraints, the current two-point function is again completely fixed to be~\eqref{eq:current2ptgenD} with $d=4$. Note that the contact terms are different between the $d=4$ case and the $d>4$ case. In particular, in four dimensions, we need traceful contact terms to be able to write the full two-point function. The $d=4$ case is written explicitly in equation \eqref{eq:curcur2ptd4} in appendix \ref{sec:Ansatz}.
As before, we see that this correlator has a $1/p^{2}$ pole, indicating the presence of a massless spin-2 field  in the spectrum.

\subsubsection{\KL decomposition}

We have seen that the two-point function of the currents $J_{(2|2)}$ and $K_{(d-2|d-2)}$ is completely fixed by their conservation conditions and the mixed 't Hooft anomaly. Much like we did for the superfluid, we would now like to prove a Goldstone-like theorem and show that any theory with these two conserved currents must necessarily have a gapless spin-2 excitation in the spectrum. The universal effective action for this gapless phase is then the linearized Einstein action (plus irrelevant corrections), whose properties we explored in sections~\ref{sec:lingrav}--\ref{sec:actionanom}.

The strategy is as before: we decompose the current two-point function~\eqref{eq:current2ptgenD} by means of the \KL spectral representation. We can then read off the (gapless) spectrum of the theory by matching the spectral density to the correlator, which is entirely fixed by the structure of symmetries and anomalies. 

Since contact terms do not affect the spectral decomposition, we only are interested in the \KL representation of the nonlocal part of the correlator, which is identically traceless. The spectral decomposition for traceless currents of Riemann symmetry type is given by (see appendix \ref{ap:KLrep} for details) 
\be
\begin{aligned}
\braket{J_{\mu_1\mu_2|\nu_1\nu_2}*\hspace{-1.25pt}K\hspace{-.8pt}*_{\alpha_1\alpha_2|\beta_1\beta_2}}
= \int_0^\infty\frac{s^2\,\rd s}{p^2+s} \bigg[&\rho_{(2|2)}(s)\,\tl{\Pi}_{\mu_1\mu_2\nu_1\nu_2\alpha_1\alpha_2\beta_1\beta_2}^{(2|2)}-\rho_{(2|1)}(s)\,\tl{\Pi}_{\mu_1\mu_2\nu_1\nu_2\alpha_1\alpha_2\beta_1\beta_2}^{(2|1)}\\[2pt]
&\hspace{.25cm} + \rho_{(1|1)}(s)\tl{\Pi}_{\mu_1\mu_2\nu_1\nu_2\alpha_1\alpha_2\beta_1\beta_2}^{(1|1)}\bigg]\, ,
\end{aligned}
\ee
where the on-shell projectors $\tl{\Pi}_{\mu_1\mu_2\nu_1\nu_2\alpha_1\alpha_2\beta_1\beta_2}^{(i|j)}$ are defined in appendix \ref{app:GravProj}.  The spectral densities $\rho^{(i|j)}$ are associated with the internal propagation of states with $(i|j)$-biform Lorentz representations.  { Since a massless $(2|2)$ or $(2|1)$ state cannot couple to a current of Riemann type~\cite{Weinberg:2020nsn,Distler:2020fzr}, the densities $\rho_{(2|2)}$, $\rho_{(2|1)}$ must go to zero as $s\rightarrow 0$.}

In order to match the two-point function~\eqref{eq:current2ptgenD}, the spectral densities have to be given by
\be 
\rho_{(2|2)}(s) = 0\, ,\qquad\quad \rho_{(2|1)}(s) = 0\, , \qquad\quad \rho_{(1|1)}(s) = \frac{d-3}{4(d-2)}\delta(s)\, ,
\ee
which shows that there is a massless spin-2 particle in the spectrum---the graviton. Since the spectral decomposition is insensitive to contact terms, the presence of the graviton is completely robust, and does not depend on where we choose to put the anomaly. This establishes a Goldstone-like theorem: any theory with conserved currents of the form~\eqref{eq:gravcurrents} with a mixed anomaly will be in a gapless phase where the massless degree of freedom has spin two.\footnote{Note that, much as in the previous case, there is a more efficient route to the same conclusion. The fact that the non-local part of the correlator has a pole as $p\to0$ already indicates that there is a massless particle in the spectrum, and this massless degree of freedom cannot be gapped by local interactions in the effective field theory, so we can restrict the spectral decomposition to massless states. We can then apply the results of~\cite{Weinberg:2020nsn,Distler:2020fzr}, which show that only massless particle that can appear in matrix elements with a conserved current with the symmetries of the Weyl tensor is a spin-2.}

\subsection{Charged solutions}\label{sec:chargedsol}

We now want to study the conserved charges associated to the electric and magnetic currents $J_{(2|2)}$ and $K_{(d-2|d-2)}$ and the defects charged under these symmetries. In linearized gravity, on-shell these currents are just the Weyl tensor and its double dual, so we will take the Weyl tensor as our starting point. Recall that it is given by\footnote{The normalization convention is that the linearization of the usual fully non-linear Weyl tensor is given by ${M_{\rm Pl}^{-{1 \over 2}(d-2)}} W_{\mu_1\mu_2\mu_3\mu_4}$ when $h_{\mu\nu}$ is the canonically normalized fluctuation of the graviton around flat space, $h_{\mu\nu}=\frac{1}{2}M_{\rm Pl}^{{1\over 2}(d-2)}(g_{\mu\nu}-\eta_{\mu\nu})$.}
\be
W_{\mu\nu\rho\sigma}=-3{\cal Y}^T_{(2|2)}\,\partial_{\mu}\partial_\rho h_{\nu\sigma} \, ,\label{eq:Weyl23}
\ee
where ${\cal Y}^T_{(2|2)}$ is a Young projector onto the traceless Young tableau with the symmetries of the Riemann tensor~\eqref{eq:RIemannYoung}. The Weyl tensor is conserved on-shell
\be 
\partial^{\mu} W_{\mu\nu\rho\sigma}=0,\label{eq:WeylCons1}
\ee
where we used $R_{\mu\nu}= R=0$ on shell. In order to get a conserved 2-form current that we can integrate from the Weyl tensor (which is a $(2|2)$-biform), we proceed as in section~\ref{sec:charges}
and contract two of its indices with a 2-form Killing tensor $\zeta^{\mu\nu}$ to obtain
\be 
J^{(\zeta)}_{\mu\nu}=W_{\mu\nu\, \alpha\beta}\zeta^{\alpha\beta}\, .\label{eq:SmallCurrent}
\ee
Using~\eqref{eq:WeylCons1}, we can write the divergence of this current as
\be  
\partial^{\nu} J_{\mu\nu}^{(\zeta)}=W_{\mu\nu\, \alpha\beta} \partial^{\nu} \zeta^{\alpha\beta}.\label{conscondweq}
\ee
In order for this to vanish, the mixed symmetry traceless part of the derivative of the Killing vector must vanish, the special case $p=1$ of \eqref{eq:zetaconstraint},
\be \partial^{\nu} \zeta^{\alpha\beta}-\partial^{[\alpha} \zeta^{\beta]\nu}+{3\over d-1} \eta^{\nu[\alpha}\partial_\rho \zeta^{\beta]\rho} =0 \, .
\ee
From \eqref{eq:chargedsolntensors}, the solutions to this equation are parametrized by four constant fully antisymmetric tensors
$c^\mu$, $c^{\mu_1\mu_2}$, $\bar{c}^{\mu_1\mu_2}$, $c^{\mu_1\mu_2\mu_3}$ as
\be  
\zeta^{\alpha\beta}=c^{\alpha\beta}+c^{\alpha\beta\rho}x_\rho +c^{[\alpha}x^{\beta]}+\bar{c}^{[\alpha}_{\ \ \ \rho} x^{\beta]}x^\rho-{1\over 4} \bar{c}^{\alpha\beta}x^2  .\label{xi2eqatopmece}
\ee

The conserved charges are then constructed by integrating the current~\eqref{eq:SmallCurrent} over a codimension two surface:
\be 
Q^{(\zeta)}(\Sigma_{d-2})=\int_{\Sigma_{d-2}} \ast J^{(\zeta)}\, .\label{chargeswe}
\ee
There is an independent charge for each choice of the constant $c$ tensors.
For $d=4$, there are a total of 20 such charges.  These are the same charges discussed recently in non-relativistic language in \cite{Benedetti:2021lxj} and covariantly in \cite{Benedetti:2022zbb}. Earlier references include \cite{Penrose:1986ca,Jezierski:2002mn,Kastor:2004jk,Jezierski:2014gka,Jezierski:2019xxl}.  

As we pointed out before, we expect only $c^\mu$ and $\bar{c}^{\mu\nu}$ to yield non-trivial charges for topologically trivial field configurations that are regular on $\Sigma_{d-2}$. Still, we will consider solutions that turn on all charges and we will comment on their regularity and significance.

Let us then describe solutions of the equations of motion of linearized gravity that carry these charges.  In electromagnetism the Coulomb and Dirac monopole solutions carry respectively the 1-form electric and magnetic charges. Analogous solutions exist in this case, carrying \eqref{chargeswe}. In $d=4$ we will see explicitly these include their magnetic counterparts.  These will be solutions which are singular at the spatial origin, and can be considered as defect line operators that are charged under the 1-form symmetries corresponding to the topological operators~\eqref{chargeswe}. Furthermore, there might be Dirac strings coming out of these singularities for topologically non-trivial solutions.
The Lorentzian $d=4$ solutions are given in spherical coordinates by
\begin{subequations}
\begin{align}
h_{tt} &={M\over r} +\frac{L \cos (\theta ) \left(r^2+t^2\right)}{2 r^2}  \,, & h_{rr} & = {M\over r}-\frac{L \cos (\theta ) \left(\cos (2 \theta ) \left(t^2-r^2\right)+2 t^2\right)}{2 r^2} \,,\\
h_{tr} &=  -\frac{L t \sin ^2(\theta ) \cos (\theta )}{r} \,, & h_{r\theta} &= -\frac{L \sin (\theta ) \left(\cos (2 \theta ) \left(3 r^2+t^2\right)-r^2-3 t^2\right)}{4 r}  \,,\\
h_{t\theta} &=  -L t \sin (\theta ) \cos ^2(\theta ) \,, & h_{\theta\theta} & =L \sin (\theta ) \sin (2 \theta ) \left(r^2+t^2\right)  \,,\\
h_{t\phi} & =- N\left(c+\cos\theta\right)+ {J\over r}\sin^2\theta \,, & h_{\phi\phi} & =  L r^2 \sin ^2(\theta ) (b-2 \cos (\theta ))\,,
\end{align}
\label{genlinemetricse}
\end{subequations}
with the other components zero.\footnote{This solution with $c=-1$ and $J=L=b=0$ was presented in \cite{Bunster:2006rt}.}

This general solution depends on six parameters: $M$, $J$, $N$, $L$, $c$ and $b$. It solves the vacuum linearized Einstein equations for any choice of these parameters.  
The interpretation of the parameters appearing in the solution is as follows: $M$ is a mass, the solution for this parameter is the linearized Schwarzschild solution \cite{Schwarzschild:1916uq}. The parameter $J$ is an angular momentum along the $z$ direction, the solution for this parameter is the linearized Kerr solution \cite{Kerr:1963ud}.\footnote{Note that in the linearized theory there is no extremality or positive energy constraint and we can take any values for $M$, $J$.} The parameter $N$ is a NUT charge, which is the magnetic dual of the mass, and the solution for this parameter is the linearized Taub--NUT solution \cite{Taub:1950ez,Newman:1963yy}.  The parameter $L$ is a magnetic dual version of angular momentum called acceleration, and the solution associated to this parameter is a linearization of the GR solution known as the $C$-metric \cite{Weyl:1917gp,Kinnersley:1970zw,Hong:2003gx,Griffiths:2006tk,Gregory:2017ogk,Scoins:2022gim}.  Finally, $c,b$ are proportional to pure gauge solutions and will be important in our discussion of singularities.

The solutions for $M$ and $J$ are manifestly regular away from the spatial origin.  The solutions with $N$ and $L$ have Dirac string type singularities away from the origin along the $z$ axis. Nevertheless, the Weyl tensor calculated from these solutions is regular away from the origin, showing that these singularities are gauge artifacts.  This is analogous to electromagnetism, were the electric solution can be described with a single gauge field, but the magnetic solution requires patching together gauge fields with the singularities in various places, related on overlaps by a gauge transformation.  Here, solutions with different singularity placements than those shown in~\eqref{genlinemetricse} will be related by linearized diffeomorphisms.  For instance, consider the $N$ part of the solution,
\be h_{t\phi} = -N\left(c + \cos\theta\right),\label{Cnutsole}
\ee
with other components zero.  Here the $c$ part is pure gauge, and this solution has the same Weyl tensor independent of the value of $c$.
In cartesian coordinates, the non-zero metric components of the solution \eqref{Cnutsole} are $h_{t,{\bf x}^\perp}={\hat{\bf x}^\perp\over \rho}\left(c+{z\over \sqrt{z^2+\rho^2}}\right)$, where 
${\bf x}^\perp$ are the coordinates perpendicular to $z$, and $\rho$ the distance from the $z$ axis.  Expanding for small $\rho$ we see that there is a singularity $\sim 1/\rho$  extending along the positive $z$ axis unless $c=-1$, and along the negative $z$ axis unless $c=1$.   Given that the Weyl tensor is independent of $c$, these solutions must be related by a gauge transformation on the overlap where they are both non-singular,
\be\label{gt1} h^{c=1}_{\mu\nu}-h^{c=-1}_{\mu\nu}=\nabla_\mu\xi_\nu+\nabla_\nu\xi_\mu,\ \ {\rm with} \ \ \xi^\mu=(-2\phi,0,0,0).\ee
In the above equation, the covariant derivative of the background flat metric in spherical coordinates has been used.

A similar situation occurs for the $L$ solution.  
We can see the structure of the singularity at the origin by expanding near the $z$ axis in cartesian coordinates, where the upper sign applies for $z>0$ and the lower sign applies for $z<0$,
\be h_{\mu\nu}\xrightarrow{\rho\rightarrow0}\left(
\begin{array}{cccc}
\pm \frac{t^2+z^2}{2 z^2} & 0 & 0 & 0 \\
 0 & \frac{(b\mp 2) y^2}{\rho ^2} & -\frac{(b\mp2) x y}{\rho ^2} & 0 \\
 0 & -\frac{(b\mp2) x y}{\rho ^2} & \frac{(b\mp2) x^2}{\rho ^2} & 0 \\
 0 & 0 & 0 &\pm  \frac{1}{2}\left(1-\frac{3 t^2}{ z^2}\right) \\
\end{array}
\right)\,,
\ee
where $\rho$ is the distance to the $z$ axis.
Note that there is no $1/\rho$ type singularity as there was for the Taub--NUT solution, everything is finite at the $z$ axis, but for generic $b$ the metric depends on the direction of approach to the $z$ axis, so there is still a discontinuity.  We can remove this discontinuity for $z>0$ by choosing $b=2$, or we can remove this discontinuity for $z<0$ by choosing $b=-2$.  Given that the Weyl tensor is independent of $b$, these solutions must be related by a gauge transformation on the overlap where they are both non-singular.  In spherical coordinates we have $h^{b=2}_{\mu\nu}-h^{b=-2}_{\mu\nu}={\rm diag}(0,0,0,4r^2\sin^2\theta)$ which is pure gauge with the gauge parameter
\be\label{gt2} h^{b=2}_{\mu\nu}-h^{b=-2}_{\mu\nu}=\nabla_\mu\xi_\nu+\nabla_\nu\xi_\mu,\ \ {\rm with}\ \ \xi^\mu=(0,0,0,2\phi).\ee

Let us now comment on the meaning of equations~\eqref{gt1} and~\eqref{gt2}. As can easily be seen, these gauge transformations are not globally well defined. This is completely analogous to the situation in electromagnetism for the Dirac monopole. In that case, the solution is only considered acceptable when the associated higher form symmetry is compact. When that happens, only the imaginary exponential of the gauge parameter is observable and the gauge transformation is allowed to have a discrete winding number. The same considerations apply here. If the biform symmetries associated with $c^{\mu\nu}$ and $c^{\mu\nu\rho}$ are compact, we expect the gauge transformations~\eqref{gt1} and~\eqref{gt2} to be allowed. Otherwise, we must conclude that these charges are trivial. Notice that, in the compact case when these charges are allowed, they will furthermore introduce quantization conditions both for these charges and for their magnetic duals, as expected from Dirac quantization.

We now evaluate the charges \eqref{chargeswe} for the metric~\eqref{genlinemetricse}. This can be done on any surface surrounding the worldline of the spatial origin, due to the topological nature of the charge.
We will choose a spherical surface of radius $R$ at time $T$ surrounding the origin:
\be 
Q^{(\zeta)}(\Sigma_{2})= \int_{\Sigma_{2}} \ast J^{(\zeta)}=\int \rd\theta \rd\phi\  {1\over 2}\epsilon_{\theta\phi}^{\ \ \ \mu_1\mu_2}W_{\mu_1\mu_2\, \nu_1\nu_2} \zeta^{\nu_1\nu_2}\bigg|_{r=R,t=T},
\label{eq:chargeexp}
\ee
 and the fact that the result  is independent of $R$ and $T$ is a check on the topological invariance of the charge.
In $d=4$, the tensors $c^{\mu_1\mu_2}$ and $\bar{c}^{\mu_1\mu_2}$ each have 6 parameters, and $c^\mu$ and $c^{\mu\nu\rho}$ each have 4 parameters, for a total of 20 parameters.  Evaluating~\eqref{eq:chargeexp} with all these parameters turned on we get (switching to cartesian coordinates) 
\be 
 Q^{(\zeta)}(\Sigma_{2})=16\pi\Big(c^0\, M-\bar{c}^{12} \, J+ c^{123} \, N-c^{03}L\Big).
 \label{eq:electriccharges}
\ee

Since the dual of the Weyl tensor is also conserved and traceless (or equivalently, the Weyl tensor is dual conserved), 
we can also consider the dual charges,
\be 
\tl Q^{(\zeta)}(\Sigma_2)=\int_{\Sigma_2}  J^{(\zeta)}.
\ee
Evaluating these explicitly, we find a different packaging of the same charges:
\be
\tl Q^{(\zeta)}(\Sigma_2)=\int \rd\theta \rd\phi\ W_{\theta\phi\, \mu_1\mu_2} \zeta^{\mu_1\mu_2}\bigg|_{r=R,t=T}=16\pi\left( c^{123} \, M- \bar{c}^{03} \, J-c^0\, N -c^{12}L\right)\,,
\ee
consistent with the electric-magnetic duality invariance of linear gravity in $d=4$ \cite{Hull:2001iu,Bekaert:2002dt}. Notice that this is only the case in $d=4$ at leading order in the EFT expansion. Higher derivative terms are expected to change the above result. In particular, we do not expect the magnetic dual current to be traceless, which changes the construction of charges above. The same comment applies to gravity in $d>4$.

We see from~\eqref{eq:electriccharges} that the conformal Killing tensor parameters $c^\mu$  extract the standard energy-momentum, while $c^{\mu\nu\rho}$---dual to a vector in $d=4$---picks up the dual energy momentum (in general $d$, the dual energy-momentum is a $3$-form \cite{Bunster:2006rt}).  The $2$-form parameters $c^{\mu\nu}$, $\bar{c}^{\mu\nu}$ pick up the angular momentum and dual angular momentum. 
Note that these charges are constructed as integrals of gauge-invariant currents.  This is unlike the standard Abbott--Deser charges \cite{Abbott:1981ff}, which are constructed as gauge-invariant integrals of non-gauge-invariant integrands. Furthermore, the integrals above can be computed on any surface that surrounds the defect and not only in asymptotic regions, as is typically done in the ADM formalism.

In quantum field theory, the energy-momentum and angular momentum usually arise from 0-form symmetries, whose 1-form currents are given by contracting the stress tensor with a Killing vector.  Linearized gravity, however, has no stress tensor \cite{Dorigoni:2009ra,Farnsworth:2021zgj} (a direct consequence of the Weinberg--Witten theorem \cite{Weinberg:1980kq}) and we see here that energy-momentum and angular momentum instead arise from 1-form symmetries with 2-form currents.

\newpage
\section{Discussion \label{sec:conclusion}}

In this work we have taken a modest first step toward answering a deep and important question: what is gravity? It is quite surprising that more than three hundred years after Newton's---and more than one hundred after Einstein's---ideas started this research field, we do not have a precise and simple definition. Much of the effort in the last fifty years has revolved around the study of the ultraviolet properties of theories of gravity. The non-renormalizability of Einstein gravity catalyzed the search for a consistent theory of nature incorporating both gravity and quantum mechanics to arbitrarily high energy.
While string theory provides an answer to this search, away from perturbation theory it does not really answer the posed question.
What property or structure in string theory fundamentally makes it a theory of gravity? 
The issue becomes more acute in the context of the Swampland program. 
Since the space of actual solutions to string theory might correspond to isolated points in its landscape, we are left with a lack of adjustable parameters. From the ultraviolet perspective this is a great advantage, as it helps singling out a unique theory. From the infrared however, it points to a strange coincidence that the theory gives rise to a gapless phase we associate with Einstein gravity.

What then, is the organizing principle? In condensed matter systems we have learned a great deal from the Landau paradigm, which states that phases of matter and phase transitions are associated to physical symmetries and the emergence of order parameters associated to them. This organizing principle for infrared physics has proved useful even at strong coupling, where perturbative techniques are unavailable.
How does gravity fit into this paradigm? An important challenge is that we do not expect any exactly conserved global charges in quantum gravity. While this presents a real difficulty in synthesizing gravity into this framework, it is not necessarily fatal.
In the last few years, the concept of symmetry has been greatly enlarged to include more exotic concepts (e.g., higher-form, 2-group, non-invertible symmetries, and beyond). There may still be a structure, compatible with the constraints of quantum gravity, that defines this theory in terms of relations between observables.

In this note we have approached the problem from the infrared by considering EFTs with a novel symmetry, biform symmetry. We have explored how theories with these symmetries can give rise to a gapless phase that is protected by a mixed anomaly between electric and magnetic biform currents. This structure of symmetries and anomalies is sufficient to completely fix the non-local part of the two-point correlators of the conserved currents in the theory. For theories with a maximal electric $(1|1)$-biform symmetry, the spectral decomposition of this correlator reveals the presence of a massless spin-2 excitation: the graviton. Using these global symmetries as a guiding principle, we can construct the universal EFT that reproduces this anomaly structure, which ends up being nothing but linearized gravity plus its (linear gauge invariant) irrelevant corrections.

A similar story should go through for the higher biforms: for theories with a maximal electric $(p|p)$-biform symmetry, the spectral decomposition of the correlator determined from the anomaly should reveal the presence of a massless particle with the Lorentz symmetries of a $(p-1|p-1)$-biform.  Note that this does not include the traditional higher-spin particles, which transform in completely symmetric Lorentz representations.  (A further generalization to multiforms would be needed to account for these and other representations~\cite{deMedeiros:2002qpr,Francia:2004lbf}.)

Importantly, the EFTs so constructed are not restricted to free theories; they can include an infinite number of interaction terms. For example, corrections to the linearized Einstein action of the form
 $W^3$---where $W$ is the linearized Weyl tensor---are included in this EFT, and provide a cubic interaction for the graviton. So, these theories are far from trivial.
 Nevertheless, they do not include the usual graviton vertices we would obtain by expanding the fully nonlinear Einstein action, as these terms break the biform symmetries. However, we can get some guidance from the traditional gauge theory viewpoint. Starting from the linearized Einstein action, there are two paths that one can take \cite{Wald:1986bj}: the first is to include interactions that are exactly invariant under the linearized gauge transformations of the free theory \cite{Bonifacio:2018van}, this reproduces the EFTs discussed here that have exact nonlinearly realized biform symmetries.  Alternatively, we can introduce the Einstein-like irrelevant interactions that are not linearized gauge invariant. These interactions do not destroy the gauge invariance of the system, but rather deform it into full diffeomorphism invariance. This strongly suggests that there is a deformation of the biform symmetries that survive in the nonlinear Einstein theory, but it remains an important open challenge to understand precisely how this works, and to frame this physics from the perspective of global symmetries. Some further evidence that this is not a futile endeavor is 
provided by the fact that the charged solutions presented in section~\ref{sec:chargedsol} persist in nonlinear Einstein gravity~\cite{Podolsky:2021zwr}.

An interesting ingredient in all of these field theories is the presence of an anomaly. This anomaly is encoded by the Einstein term, which is the most relevant one in the EFT expansion. Interestingly, some of the physics associated to this term is quite similar to that of Chern--Simons (or BF) theories. This explains why the Einstein action produces equations of motion that set parts of curvatures to zero, instead of just implying their conservation, like what happens for Maxwell actions. A crucial difference with topological field theories, is, of course, that the associated biform symmetry allows for a gapless phase in our case---essentially because the equations of motion are only a partial flatness condition. Nevertheless, many of the lessons of the physics of anomalies in Chern--Simons theories can be imported into in these new cases. 
In this context, the presence of boundaries typically play an important role. The anomaly structure we have discussed can be reproduced from a higher-dimensional topological theory via inflow, which raises interesting questions: Can we realize the linearized graviton as a boundary mode for a dynamical gapped bulk theory? Or, can we construct an Einstein-type action on boundaries of theories which have only Maxwell-type terms?\footnote{Relatedly, $(p|p)$-biform theories with $d < 2p+2$ have no local propagating degrees of freedom. It would be interesting to understand their features as topological field theories.}

An important detail in the study of anomalies and their connection to different phases for these systems is whether the biform symmetry of interest is compact or non-compact. We have been a bit cavalier about this important aspect of the problem. It is the compactness of the symmetry that might lead to the quantization of anomalies and might prove crucial in defining a UV complete model with these symmetries. We saw one small hint of the role of compactness in the study of charged solutions in linearized gravity where NUT and acceleration charges become available only in that case. The consequences of this fact must be further explored.

A deeper understanding of this general structure could also be useful for more systematically studying other phases of quantum gravity that could be gapped or, alternatively, present a large number of massless excitations (as in higher-spin theories, for example). An intriguing notion is that string theory could display other, more symmetric, phases (for example, phases with tensionless strings), whose understanding may be crucial to decipher the full non-perturbative structure of the theory. One might hope that biform (or more extended multiform) symmetries could present an avenue of approach to these questions.

Along the road, we have also remarked on the intriguing connection between this approach and the physics of fractons. Indeed, gravity can be understood as a gauging of a fractonic global symmetry, and its massless phase corresponds to the gapping of the degrees of freedom charged under the fractonic symmetry, not unlike the massless phase of electromagnetism. Curiously, some stringy physics appeared in this connection, for example one of the structures in the theory naturally couples to a worldsheet (see~\eqref{eq:dipoleact}, for example). This is rather surprising and mysterious, and it would be interesting to deepen this connection.

Another interesting fact we stumbled upon when coupling linearized gravity to background sources associated to the biform symmetries is that this theory is special in $d=4$. Only in this dimension does the theory present a trace anomaly for the background fields. The scalar case shows a similar feature in $d=2$, which is known to be associated to conformality and the appearance of a Kac--Moody algebra in the extreme infrared. In~\cite{Hofman:2018lfz}, a similar structure was uncovered for electromagnetism. It is therefore natural to suspect that a similar Kac--Moody enlargement will occur for linearized gravity in $d=4$.
This possibility is particularly interesting in connection with both the double copy formalism and with soft graviton theorems. 

One arena where these ideas may find practical application is cosmology. An understanding of the fundamental organizing principles of EFTs for gravitational physics is surely of importance in developing cosmological models. Concretely, understanding the symmetries of gravity will help shed further light on the early universe, where it is expected that spacetime experienced a period of inflationary expansion. This can be interpreted as a partial higgsing of gravity~\cite{Arkani-Hamed:2003pdi,Creminelli:2006xe,Cheung:2007st}, where matter degrees of freedom mix with the graviton in the infrared.
In much the same way that the Higgs phase of electromagnetism can be viewed as a symmetry-restored phase of the magnetic 1-form symmetry~\cite{Kovner:1990pz,Rosenstein:1990py}, one could imagine that inflation can be understood as a different phase of gravity. In any case, having an EFT definition of the inflationary era based purely on global symmetries is clearly of importance for the further understanding of the microphysics of inflation, and we expect that the (nonlinear extension of) the symmetries explored here will play a role. A step in this direction would be the formulation of these symmetries and anomalies on (anti) de Sitter backgrounds.  It would also be interesting to see how the new types of representations such as partially massless fields that are possible in de Sitter space fit into this picture. Beyond this, it is tempting to speculate that gapped phases of gravity could themselves play a role in the early universe, possibly along the lines of~\cite{Agrawal:2020xek}.

Having discussed at length the infrared properties of gravity, let us come full circle and return to the ultraviolet. One obvious question we can ask is: Are there nontrivial UV-complete theories that have these symmetries, say only in their magnetic form. What would such a theory look like?\footnote{Such a theory is unlikely to look anything like a local QFT. We certainly know that it cannot have a stress tensor by Weinberg--Witten~\cite{Weinberg:1980kq}. Likely it would be some sort of theory of extended objects, along the lines of~\cite{Polyakov:1980ca,Iqbal:2021rkn}.} Below the mass scale of whatever electric matter breaks the electric biform symmetry, we would expect this electric symmetry to be restored in an emergent way, giving rise to linearized gravity in the infrared. While this is not the theory of gravity that describes our universe (as it would have different nonlinearities from Einstein gravity), it would nevertheless amount to a version of emergent gravity at low energies. More ambitiously, we might hope that these symmetries can deformed in some way so that the nonlinearities of Einstein gravity also emerge at low energies. This would amount to nothing short of an understanding of gravity in the infrared within the framework of the Landau paradigm, which is clearly a worthy goal. We hope to return to these interesting problems in the near future.

\vspace{-15pt}
\paragraph{Acknowledgments:}

We would like to thank Horacio Casini, Luca Delacr\'etaz, Tom Hartman, Jonathan Heckman, Nabil Iqbal, Javier Magan, David Meltzer, and Dam Son for interesting and helpful discussions. We would also like to thank Valentin Benedetti for pointing out an incorrect equation in the first version of this draft. KH acknowledges support from DOE grant DE-SC0009946 and from Simons Foundation Award Number 658908.  DH and GM are supported in part by the ERC starting grant {\scriptsize{GENGEOHOL}} (grant agreement No 715656). GM is supported in part by NSF grant PHY-2014071.

\appendix

\section{Spectral representation\label{ap:KLrep}}

In this appendix, we provide some technical details about the \KL spectral representation of correlators. The goal is to derive the form of the spectral function that we used throughout the main text. Ultimately we are interested in euclidean correlation functions, but it will prove to be convenient to first construct the spectral representation in Lorentzian signature and then analytically continue. Our discussion somewhat follows~\cite{Karateev:2020axc,Karateev:2019ymz}.

\subsection{Spectral density}

We would like to decompose the two-point function between currents with the same number of Lorentz indices, $q$. (This assumption that both currents have an equal number of Lorentz indices is not generally necessary, but it will simplify our discussion, and is sufficiently general for our purposes.) We do not put any constraints on the index symmetries of the two currents.

The basic object of interest is the Fourier transform of the position space correlator
\be 
\braket{J_{\mu_1\cdots\mu_q}(p)K_{\nu_1\cdots\nu_q}(-p)} = \int \rd^dx\, 	 e^{ixp}\, \braket{J_{\mu_1\cdots\mu_q}(x) K_{\nu_1\cdots\nu_q}(0)}\, .
\ee
The Wightman function in position space is defined as an ordered vacuum expectation value, where the operators are ordered from left to right in Lorentzian time. This is achieved via the usual $i\epsilon$ prescription, such that 
\be 
\braket{0|J_{\mu_1\cdots\mu_q}(x)K_{\nu_1\cdots\nu_1}(0) |0}_W \equiv \lim_{\epsilon\rightarrow 0^+}\braket{0|J_{\mu_1\cdots\mu_q}(\hat{x})K_{\nu_1\cdots\nu_1}(0) |0}\, ,
\ee
where we have defined $\hat{x}^\mu = (x^0-i\epsilon,\vec{x})$, with $\epsilon>0$.
As usual, we perform all the computations at finite $\epsilon$ before taking the limit where $\epsilon$ approaches $0$ from above at the end of the computation.

We define the spectral density $\rho(-p^2)$ as the Fourier transform 
\be 
(2\pi) \theta(p^0) \rho^{\mu_1\cdots\mu_p\nu_1\cdots\nu_p}(-p^2)\equiv \int \rd^d x\, e^{-ip\cdot x}\braket{0|J^{\mu_1\cdots\mu_p}(x)K^{\nu_1\cdots\nu_p}(0)|0}_W\, .
\ee
The appearance of the Heavyside function $\theta(p^0)$ ensures that we are working with positive energies~\cite{Karateev:2020axc}.
From this definition, it is clear that 
\be 
\braket{0|J^{\mu_1\cdots\mu_p}(x)K^{\nu_1\cdots\nu_p}(0)|0}_W = \lim_{\epsilon\rightarrow 0^+}\int \frac{\rd^dp}{(2\pi)^d}e^{ip\cdot x}(2\pi)\theta(p^0)\rho^{\mu_1\cdots\mu_p\nu_1\cdots\nu_p}(-p^2)\, .\label{eq:rhoKL}
\ee
It is now helpful to define a non-negative variable as $s \equiv -p^2$.
We can then rewrite \eqref{eq:rhoKL} by inserting a delta function enforcing the condition $s= -p^2$
\be 
\braket{0|J^{\mu_1\cdots\mu_p}(x)K^{\nu_1\cdots\nu_p}(0)|0}_W =\lim_{\epsilon\rightarrow 0^+} \int_0^\infty \rd s\int \frac{\rd^dp}{(2\pi)^d}e^{ip\cdot x}(2\pi)\theta(p^0)\rho^{\mu_1\cdots\mu_p\nu_1\cdots\nu_p}(s)\delta (s+ p^2)\, .\label{eq:A4}
\ee

In any unitary quantum field theory, we can assemble a complete basis of states, that we collectively denote as  $\ket{n}$. These states must transform in unitary representations of the Poincar\'e group, and in particular, we require that they diagonalize the generators of translations $P^\mu$ as $P^\mu\ket{n}= p_n^\mu\ket{n}$. The completeness relation is then 
\be 
\mathds{1} = \sum_n \lvert n\rangle\langle n\rvert\, .\label{eq:comp1}
\ee
In this case, the summation over $n$ stands for a sum over any required quantum numbers needed to characterize these states, as well as Lorentz indices. We can insert \eqref{eq:comp1} within the Wightman function and  then perform the following manipulations
\begin{align}
\braket{0|J^{\mu_1\cdots\mu_p}(x)K^{\nu_1\cdots\nu_p}(0)|0}_W &=\lim_{\epsilon\rightarrow 0^+}  \sum_n \langle0|J^{\mu_1\cdots\mu_p}(x)|n\rangle\langle n|K^{\nu_1\cdots\nu_p}(0)|0\rangle\, ,\\
&=\lim_{\epsilon\rightarrow 0^+} \sum_n \langle0|e^{-iP\cdot x}J^{\mu_1\cdots\mu_p}(0)e^{iP\cdot x}|n\rangle\langle n|K^{\nu_1\cdots\nu_p}(0)|0\rangle\, ,\\
&=\lim_{\epsilon\rightarrow 0^+} \sum_n e^{ip_n\cdot x}\langle0|J^{\mu_1\cdots\mu_p}(0)|n\rangle\langle n|K^{\nu_1\cdots\nu_p}(0)|0\rangle\, ,\\
&=\lim_{\epsilon\rightarrow 0^+} \int\frac{\rd^dp}{(2\pi)^d}e^{ip\cdot x}\sum_n (2\pi)^d\delta^{(d)}(p-p_n)\times\nonumber\\
&\phantom{=}\qquad \qquad \qquad \langle0|J^{\mu_1\cdots\mu_p}(0)|n\rangle\langle n|K^{\nu_1\cdots\nu_p}(0)|0\rangle\, .\label{eq:A12}
\end{align}
By comparing \eqref{eq:rhoKL} with \eqref{eq:A12}, we conclude that the spectral density is given by
\be 
(2\pi)\theta(p^0)\rho^{\mu_1\cdots\mu_p\nu_1\cdots\nu_p}(-p^2)= \sum_n (2\pi)^d\delta^{(d)}(p-p_n)\braket{0|J^{\mu_1\cdots\mu_p}(0)|n}\braket{n|K^{\nu_1\cdots\nu_p}(0)|0}\, .\label{eq:Apspectral1}
\ee
This is the general form of the spectral density that we consider. (Note that it is not necessarily positive-definite, since it involves two different operators, though it is positive when the operators are identical.)

The spectral density $\rho^{\mu_1\cdots\mu_p\nu_1\cdots\nu_p}(-p^2)$ can be decomposed using Lorentz invariance. The strategy is to construct dimensionless projectors $\Pi_i^{\mu_1\cdots\mu_p\nu_1\cdots\nu_p}$ that depend on the $d$-momenta $p^\mu$ and the metric $\eta^{\mu\nu}$ which are orthonormal and complete:
\be 
\Pi\indices{_i^{\mu_1\cdots\mu_p}_{\rho_1\cdots\rho_p}}\Pi\indices{_j^{\rho_1\cdots\rho_p\nu_1\cdots\nu_p}} = \delta_{ij}\Pi\indices{_j^{\mu_1\cdots\mu_p\nu_1\cdots\nu_p}}\, ,\qquad \quad \sum_i \Pi\indices{_i^{\mu_1\cdots\mu_p\nu_1\cdots\nu_p}} =\mathds{1}^{\mu_1\cdots\mu_p\nu_1\cdots\nu_p}\, ,\label{eq:ProjAp}
\ee
where $\mathds{1}^{\mu_1\cdots\mu_p\nu_1\cdots\nu_p}$ is the identity in the relevant space of tensors. In many cases we will be interested in decomposing traceless tensors, so the projectors will be traceless.

We can thus write the spectral density as  
\be 
\rho^{\mu_1\cdots\mu_p\nu_1\cdots\nu_p}(-p^2) = \sum_i (-p^2)^{\Delta-d/2+1}\rho_i(-p^2)\Pi_i^{\mu_1\cdots\mu_p\nu_1\cdots\nu_p}\, ,\label{eq:Decomp1}
\ee
where $\rho_i(-p^2)$ are scalar functions that can depend only on $p^2$, and $\Delta$ is the mass dimension of the currents appearing in the \KL decomposition. Note that we sometimes have to extract a minus sign from the definition of the spectral function to ensure that they are positive-definite when the two operators are identical~\cite{Karateev:2020axc}.

Using \eqref{eq:Decomp1} within \eqref{eq:A4}, we can rewrite the Wightman function as
\be 
\braket{0|J^{\mu_1\cdots\mu_p}(x)K^{\nu_1\cdots\nu_p}(0)|0}_W = \sum_i\int_0^\infty \rd s\, (s)^{\Delta-d/2 + 1}\rho_i(s)\Delta^{\mu_1\cdots\mu_p\nu_1\cdots\nu_p}_i(x;s)\, .\label{eq:A16}
\ee
where 
\be 
\Delta^{\mu_1\cdots\mu_p\nu_1\cdots\nu_p}_{W,i}(x;s)= \lim_{\epsilon\rightarrow 0^+}\int \frac{\rd^dp}{(2\pi)^d}e^{ip\cdot x}(2\pi)\theta(p^0)\delta (s+ p^2)\Pi_i^{\mu_1\cdots\mu_p\nu_1\cdots\nu_p}\, .\label{eq:A14}
\ee
In general, we can use the explicit form of the projectors that appear in \eqref{eq:Decomp1} to rewrite the propagators \eqref{eq:A14} in terms of the scalar Wightman two-point function, which is given as: 
\be 
\Delta_W(x;s)  = \lim_{\epsilon\rightarrow 0^+}\int \frac{\rd^d p}{(2\pi)^d}e^{ip\cdot x}(2\pi)\theta(p^0)\delta(s+p^2)\, .\label{eq:A18}
\ee
For example, in the simple case of two spin 1 currents of dimension $\Delta = d-1$, the projectors are 
\be 
\Pi_0^{\mu\nu} = \frac{p^\mu p^\nu}{p^2}\, , \qquad \qquad \Pi_1^{\mu\nu} = \eta^{\mu\nu}-\frac{p^\mu p^\nu}{p^2}\, .
\ee
We can thus obtain the building blocks that appear in \eqref{eq:A14} from \eqref{eq:A18} as  
\be 
\Delta_{W,0}^{\mu\nu}(x;s) = \frac{\partial^\mu \partial^\nu}{\square}\Delta_W(x;s)\, , \qquad \qquad \Delta_{W,1}^{\mu\nu}(x;s) = \left(\eta^{\mu\nu}-\frac{\partial^\mu \partial^\nu}{\square}\right)\Delta_W(x;s)\, .
\ee
Now that we understand how to decompose Wightman functions, we would like to consider the case of time-ordered (or Feynman) propagators. 

\noindent
{\bf Time-ordered two-point functions:}
To obtain the time-ordered two-point function from the Wightman function, we use the following relation:
\begin{align}
&\braket{0|J^{\mu_1\cdots\mu_p}(x)K^{\nu_1\cdots\nu_p}(0)|0}_T =\\
&\qquad \qquad  \theta(x^0)\braket{0|J^{\mu_1\cdots\mu_p}(x)K^{\nu_1\cdots\nu_p}(0)|0}_W + \theta(-x^0)\braket{0|K^{\nu_1\cdots\nu_p}(0)J^{\mu_1\cdots\mu_p}(x)|0}_W\nonumber\, ,
\end{align}
where the subscript $T$ indicates that this is a time-ordered correlator.  Using~\eqref{eq:A16}, we can decompose the time-ordered correlator as 
\be 
\braket{0|J^{\mu_1\cdots\mu_p}(x)K^{\nu_1\cdots\nu_p}(0)|0}_T  = -i\sum_j\int_0^\infty \rd s\, s^{\Delta-d/2+1}\rho_j(s)\Delta^{\mu_1\cdots\mu_p\nu_1\cdots\nu_p}_{F,j}(x;s)\, ,
\ee
where the Feynman propagators are defined as
\begin{align}
-i\Delta^{\mu_1\cdots\mu_p\nu_1\cdots\nu_p}_{F,j}(x;s)& = \theta(x^0)\Delta^{\mu_1\cdots\mu_p\nu_1\cdots\nu_p}_{W,j}(x;s)+\theta(-x^0)\Delta^{\mu_1\cdots\mu_p\nu_1\cdots\nu_p}_{W,j}(-x;s)\, ,\\
&= \lim_{\epsilon\rightarrow 0^+}\frac{\rd^d p}{(2\pi)^d}e^{ip\cdot x}\frac{-i \Pi^{\mu_1\cdots\mu_p\nu_1\cdots\nu_p}_j}{p^2 + s -i\epsilon}\, .
\end{align}

\noindent
{\bf Euclidean space:} We next want to analytically continue to euclidean signature.
This amounts to the following change of variables:
\be 
x_L^0\rightarrow -ix_E^0\, ,\qquad \qquad p_L^0\rightarrow -ip_E^0\, .
\ee
The  $i\epsilon$ prescription is no longer needed since there are no poles to be regularized for real $p^2$. The general decomposition is now\footnote{The overall signs of the various spectral densities, $\rho_i$, have been left implicit here, and can be fixed by requiring that the spectral density is positive-definite for identical operators.}
\begin{tcolorbox}[colframe=white,arc=0pt,colback=greyish2]
\be 
\braket{0|J^{\mu_1\cdots\mu_p}(x)K^{\nu_1\cdots\nu_p}(0)|0}_E = \sum_i \int_0^{\infty}\rd s \, s^{\Delta-d/2+1}\rho_i^E(s)\Delta_{E,i}^{\mu_1\cdots\mu_p\nu_1\cdots\nu_p}\, ,
\label{eq:spectraldensityapp}
\ee
\end{tcolorbox}
\vspace{-6pt}
\noindent
where the subscript $E$ denotes that these are Euclidean objects, and we have used
\be 
\Delta_{E,i}^{\mu_1\cdots\mu_p\nu_1\cdots\nu_p} = \int \frac{\rd^d p_E}{(2\pi)^d} e^{iP_E\cdot x_E}\frac{\Pi_i^{\mu_1\cdots\mu_p\nu_1\cdots\nu_p}}{p_E^2+s}\, .
\ee
The spectral decomposition~\eqref{eq:spectraldensityapp} is the form that we use in the main text.

\subsection{Contact terms}
In this section, we show that we can neglect contact terms when we perform a \KL decomposition, as they never contribute to the spectral function. For simplicity we consider the spectral function of scalars, but the arguments can be readily generalized to cases with spin.

We begin by considering a Euclidean momentum space two-point function for real scalars of mass dimension $\Delta$ in $d$ dimensions, denoted $G(p^2)$. The two-point function has mass dimension $2\Delta-d$, and is well-defined and real for all $p^2>0$. If we let $z\equiv p^2$, we can analytically continue to the complex $z$-plane, and the result should be analytic everywhere except for the negative $z$ axis, i.e., $z\leq 0$, where it can have isolated simple poles and a branch cut extending to $z = -\infty$. Because we continued a real function, the correlator should obey  $\bar{G}(z) = G(\bar{z})$.
In particular, if we cross the branch cut, the discontinuity is
\be 
\text{Disc}\, G(z) \equiv \lim_{\epsilon\rightarrow 0}\left[G(z+i\epsilon)-G(z-i\epsilon)\right] = 2i\,  \text{Im}\,G(z)\, ,\label{eq:DiscAp}
\ee
where the imaginary part is taken above the branch cut. 

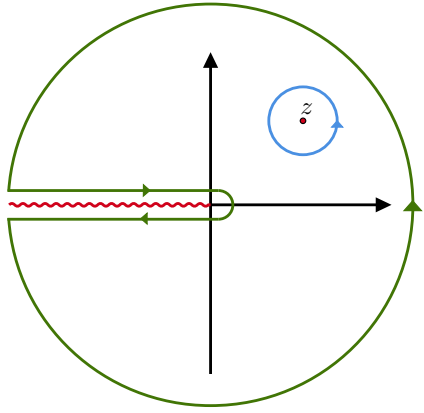
\begin{figure}
\begin{center}
 \tikzset{every picture/.style={line width=0.5pt}} 

\scalebox{.85}{
\begin{tikzpicture}[x=0.75pt,y=0.75pt,yscale=-0.65,xscale=0.65]

\draw [line width=1.25]    (313,304) -- (471.5,304) ;
\draw [shift={(476.5,304)}, rotate = 180] [fill={rgb, 255:red, 0; green, 0; blue, 0 }  ][line width=0.08]  [draw opacity=0] (14.29,-6.86) -- (0,0) -- (14.29,6.86) -- cycle    ;
\draw [line width=1.25]    (313,458) -- (313,170) ;
\draw [shift={(313,165)}, rotate = 90] [fill={rgb, 255:red, 0; green, 0; blue, 0 }  ][line width=0.08]  [draw opacity=0] (14.29,-6.86) -- (0,0) -- (14.29,6.86) -- cycle    ;
\draw [color={rgb, 255:red, 208; green, 2; blue, 27 }  ,draw opacity=1 ][line width=1.25]    (131,304) .. controls (132.67,302.33) and (134.33,302.33) .. (136,304) .. controls (137.67,305.67) and (139.33,305.67) .. (141,304) .. controls (142.67,302.33) and (144.33,302.33) .. (146,304) .. controls (147.67,305.67) and (149.33,305.67) .. (151,304) .. controls (152.67,302.33) and (154.33,302.33) .. (156,304) .. controls (157.67,305.67) and (159.33,305.67) .. (161,304) .. controls (162.67,302.33) and (164.33,302.33) .. (166,304) .. controls (167.67,305.67) and (169.33,305.67) .. (171,304) .. controls (172.67,302.33) and (174.33,302.33) .. (176,304) .. controls (177.67,305.67) and (179.33,305.67) .. (181,304) .. controls (182.67,302.33) and (184.33,302.33) .. (186,304) .. controls (187.67,305.67) and (189.33,305.67) .. (191,304) .. controls (192.67,302.33) and (194.33,302.33) .. (196,304) .. controls (197.67,305.67) and (199.33,305.67) .. (201,304) .. controls (202.67,302.33) and (204.33,302.33) .. (206,304) .. controls (207.67,305.67) and (209.33,305.67) .. (211,304) .. controls (212.67,302.33) and (214.33,302.33) .. (216,304) .. controls (217.67,305.67) and (219.33,305.67) .. (221,304) .. controls (222.67,302.33) and (224.33,302.33) .. (226,304) .. controls (227.67,305.67) and (229.33,305.67) .. (231,304) .. controls (232.67,302.33) and (234.33,302.33) .. (236,304) .. controls (237.67,305.67) and (239.33,305.67) .. (241,304) .. controls (242.67,302.33) and (244.33,302.33) .. (246,304) .. controls (247.67,305.67) and (249.33,305.67) .. (251,304) .. controls (252.67,302.33) and (254.33,302.33) .. (256,304) .. controls (257.67,305.67) and (259.33,305.67) .. (261,304) .. controls (262.67,302.33) and (264.33,302.33) .. (266,304) .. controls (267.67,305.67) and (269.33,305.67) .. (271,304) .. controls (272.67,302.33) and (274.33,302.33) .. (276,304) .. controls (277.67,305.67) and (279.33,305.67) .. (281,304) .. controls (282.67,302.33) and (284.33,302.33) .. (286,304) .. controls (287.67,305.67) and (289.33,305.67) .. (291,304) .. controls (292.67,302.33) and (294.33,302.33) .. (296,304) .. controls (297.67,305.67) and (299.33,305.67) .. (301,304) .. controls (302.67,302.33) and (304.33,302.33) .. (306,304) .. controls (307.67,305.67) and (309.33,305.67) .. (311,304) -- (313,304) -- (313,304) ;
\draw  [fill={rgb, 255:red, 208; green, 2; blue, 27 }  ,fill opacity=1 ] (394,227.5) .. controls (394,226.12) and (395.12,225) .. (396.5,225) .. controls (397.88,225) and (399,226.12) .. (399,227.5) .. controls (399,228.88) and (397.88,230) .. (396.5,230) .. controls (395.12,230) and (394,228.88) .. (394,227.5) -- cycle ;
\draw  [color={rgb, 255:red, 74; green, 144; blue, 226 }  ,draw opacity=1 ][line width=1.25]  (365.61,227.5) .. controls (365.61,210.44) and (379.44,196.61) .. (396.5,196.61) .. controls (413.56,196.61) and (427.39,210.44) .. (427.39,227.5) .. controls (427.39,244.56) and (413.56,258.39) .. (396.5,258.39) .. controls (379.44,258.39) and (365.61,244.56) .. (365.61,227.5) -- cycle ;
\draw  [color={rgb, 255:red, 74; green, 144; blue, 226 }  ,draw opacity=1 ][fill={rgb, 255:red, 74; green, 144; blue, 226 }  ,fill opacity=1 ] (427.39,227.5) -- (432.43,233.26) -- (422.36,233.26) -- cycle ;
\draw [color={rgb, 255:red, 65; green, 117; blue, 5 }  ,draw opacity=1 ][line width=1.25]    (129.7,290.99) -- (320.2,290.99) ;
\draw [color={rgb, 255:red, 65; green, 117; blue, 5 }  ,draw opacity=1 ][line width=1.25]    (129.44,317.11) -- (319.94,317.11) ;
\draw  [draw opacity=0][line width=1.25]  (320.2,290.99) .. controls (320.2,290.99) and (320.2,290.99) .. (320.2,290.99) .. controls (327.41,291.06) and (333.2,296.97) .. (333.13,304.18) .. controls (333.06,311.4) and (327.15,317.19) .. (319.94,317.11) -- (320.07,304.05) -- cycle ; \draw  [color={rgb, 255:red, 65; green, 117; blue, 5 }  ,draw opacity=1 ][line width=1.25]  (320.2,290.99) .. controls (320.2,290.99) and (320.2,290.99) .. (320.2,290.99) .. controls (327.41,291.06) and (333.2,296.97) .. (333.13,304.18) .. controls (333.06,311.4) and (327.15,317.19) .. (319.94,317.11) ;  
\draw  [color={rgb, 255:red, 65; green, 117; blue, 5 }  ,draw opacity=1 ][fill={rgb, 255:red, 65; green, 117; blue, 5 }  ,fill opacity=1 ] (258.2,290.93) -- (252.26,296.04) -- (252.35,285.72) -- cycle ;
\draw  [color={rgb, 255:red, 65; green, 117; blue, 5 }  ,draw opacity=1 ][fill={rgb, 255:red, 65; green, 117; blue, 5 }  ,fill opacity=1 ] (250.01,316.53) -- (255.65,311.61) -- (255.63,321.47) -- cycle ;
\draw  [draw opacity=0][line width=1.25]  (130.58,292.01) .. controls (136.76,196.6) and (216.13,121.14) .. (313.1,121.2) .. controls (414.06,121.25) and (495.86,203.14) .. (495.8,304.1) .. controls (495.75,405.06) and (413.86,486.86) .. (312.9,486.8) .. controls (216.03,486.75) and (136.79,411.35) .. (130.59,316.05) -- (313,304) -- cycle ; \draw  [color={rgb, 255:red, 65; green, 117; blue, 5 }  ,draw opacity=1 ][line width=1.25]  (130.58,292.01) .. controls (136.76,196.6) and (216.13,121.14) .. (313.1,121.2) .. controls (414.06,121.25) and (495.86,203.14) .. (495.8,304.1) .. controls (495.75,405.06) and (413.86,486.86) .. (312.9,486.8) .. controls (216.03,486.75) and (136.79,411.35) .. (130.59,316.05) ;  
\draw  [color={rgb, 255:red, 65; green, 117; blue, 5 }  ,draw opacity=1 ][fill={rgb, 255:red, 65; green, 117; blue, 5 }  ,fill opacity=1 ] (495.63,300.3) -- (503.51,309.3) -- (487.76,309.3) -- cycle ;
\draw (392.29,209.4) node [anchor=north west][inner sep=0.75pt]    {$z$};
\end{tikzpicture}
}
 \caption{The contour deformation used to derive the spectral function. We deform the blue contour centered around $z$ into the green contour that runs along the branch cut with an arc at infinity.}
  \label{fig:boat1}
  \end{center}
\end{figure}

\noindent
{\bf Unsubtracted dispersion relation:}
We can now apply the Cauchy integral formula to a general point $z$ away from the negative axis to write
\be 
G(z) = \frac{1}{2\pi i}\oint \rd\omega \frac{G(\omega)}{\omega-z}\, ,
\ee
where the contour is a small circle encircling the point $z$. We can then deform the contour as shown in Figure~\ref{fig:boat1}.
As long as $G(z)$ is bounded by a constant as $z\to \infty$,
we can ignore the circle at infinity and we pick up integrals along the singular points. This implies that 
\be 
G(z) = \frac{1}{2\pi i}\lim_{\epsilon\rightarrow 0}\left[\int_{-\infty}^0\, \rd s \frac{G(s + i\epsilon)}{s-z} + \int_0^\infty\,\rd s (-1)\frac{G(-s-i\epsilon)}{-s-z}\right]\, .
\ee
The first integral runs to the right over the branch cut with $z = s + i\epsilon,\, s = (-\infty,0)$ while the second integral runs to the left under the branch cut with $z = -s-i\epsilon,\, s = (0,\infty)$. We can then change variables as $s\rightarrow -s$ in the first integral, so that we can combine the integrals as 
\be 
G(z) = -\frac{1}{2\pi i}\lim_{\epsilon\rightarrow 0}\int_{0}^\infty \, \rd s \frac{G(-s+i\epsilon)-G(-s-i\epsilon)}{s+z}\, . 
\ee
We can now use \eqref{eq:DiscAp} to rewrite this last expression as 
\be 
G(z) = -\frac{1}{2\pi i}\int_0^\infty\, \rd s \frac{\text{Disc}\left(G(-s)\right)}{s+z} = -\frac{1}{\pi}\int_0^\infty\, \rd s \frac{\text{Im}\,G(-s)}{s + z}\, . 
\ee
Defining the spectral function 
\be 
s^{\Delta-\frac{d}{2}+1}\rho(s)\equiv -\frac{1}{\pi}\text{Im}\,G(-s)\, ,\label{eq:spectralap1}
\ee
so that $\rho(s)$ has mass dimension $[\rho(s)] =-2$---as required for a density---we can finally write the spectral representation as 	
\be 
G(z) = \int_0^\infty \rd s\, \frac{\rho(s) s^{\Delta-\frac{d}{2}+1}}{s + z}\, . \label{eq:spectralap2}
\ee
Note that the condition that $G(z)$ is bounded by a constant together with \eqref{eq:spectralap1} ensures that $\rho(s) s^{\Delta-\frac{d}{2}+1}<\text{const.}$ as $s\rightarrow \infty$, so the integral \eqref{eq:spectralap2} converges at infinity.

\noindent
{\bf Subtracted dispersion relation:}
In general, $G(z)$ is only  bounded by some power at infinity: $G(z) < \mathcal{O}\left(z^n\right)\, \quad \text{as}\quad z\rightarrow \infty$.
In this case, we cannot ignore the contribution from the contour at infinity. Nevertheless, taking the $n$-th derivative as $G^{(n)}(z)\equiv \frac{\rd^n}{\rd z^n}G(z)$, we can obtain something that is bounded by a constant so that we can apply the Cauchy theorem for $G^{(n)}(z)$, 
\be 
G^{(n)}(z) = \frac{n!}{2\pi i}\oint \rd\omega \frac{G(\omega)}{(\omega-z)^{n+1}}\, ,
\ee
and then deform the contour as in Figure~\ref{fig:boat1}. We obtain 
\begin{align}
G^{(n)}(z) & = \frac{n!}{2\pi i}\left[\int_{-\infty}^0\, \rd s\frac{G(s + i\epsilon)}{(s-z)^{n+1}} + \int_0^\infty\, \rd s (-1)\frac{G(-s-i\epsilon)}{(-s-z)^{n+1}}\right]\, ,\\
&= \frac{n!}{\pi}(-1)^{n+1}\int_0^\infty \,\rd s \frac{\text{Im}\,G(-s)}{s+z}\, .
\end{align}
Using the same definition \eqref{eq:spectralap1}, we can write
\be 
G^{(n)}(z) = n!(-1)^n\int_0^\infty\, \rd s\frac{\rho(s)s^{\Delta-\frac{d}{2}+1}}{(s + z)^{n+1}}\, .\label{eq:Gnap}
\ee
Note that \eqref{eq:Gnap} is formally the derivative of the expression \eqref{eq:spectralap2}, but \eqref{eq:Gnap} converges at infinity given the behavior of $G(z)$ at infinity, whereas \eqref{eq:spectralap2} diverges.

In order to recover $G(z)$ from $G^{(n)}(z)$, we take the $n$-th indefinite integral, which introduces $n$ integration constants $c_0,\cdots,c_{n-1}$ so that
\be 
G(z) = c_0 + c_1 z + c_2 z^2 +\cdots + c_{n-1}z^{n-1} + \int_z \rd z_1 \int_{z_1}\rd z_2 \cdots \int_{z_n-1} \rd z_n G^{(n)}(z_n)\, .
\ee
This is what we call an $n$-subtracted dispersion relation. The constants $c_1,\cdots, c_n$ are subtraction constants and are not determined. Nevertheless, these integration constants multiply polynomials in $z=p^2$, so although they are undetermined, they are just contact terms, contributing only at coincident points in position space.  Note that contact terms in correlators are everywhere analytic in $z$, so they do not affect the discontinuity along the branch cut and hence do not enter the spectral density $\rho(s)$. 

\newpage
\section{More on the Galileon superfluid\label{ap:GalSup}}

In section~\ref{sec:GalSup}, we analyzed the EFT that describes a Galileon superfluid. There we defined the theory in terms of a current $J_{\mu\nu}$ that was conserved on shell:  $\partial^\mu J_{\mu\nu}=0$. This current failed to be antisymmetrically conserved, which was the manifestation of the mixed anomaly in that context. In this appendix, we wish to show how one can move the anomaly around into other conservation conditions by an appropriate definition of the gauge-invariant current.

\subsection{Currents from the action}
\label{app:galfluidaction}
Here we want to utilize the Galileon superfluid action to construct currents that manifest the mixed anomaly in a different way from the main text. Our starting point is again~\eqref{eq:superfluid1order1}:
 \be 
S = -a\int \rd^d x\left(\phi\, \partial^\mu s_\mu + \frac{1}{2}s^\mu s_\mu\right)\, .\label{eq:superfluid1order1ap}
\ee
In order to gauge the symmetries of this action, we proceed slightly differently. 
We introduce a background two-index symmetric gauge field $\tilde{A}_{\mu|\nu}$ (or a $(1|1)$-biform), that we will separate into a traceless part $A_{\mu|\nu}$ and its trace $B$. As in section~\ref{sec:GalSup}, in $d\neq 2$, the trace $B$ completely decouples so that we can take the background gauge field to be traceless, while in $d=2$, we must keep the trace $B$ to obtain a gauge-invariant action.

The gauge transformations are slightly different since we have explicitly separated the trace $B$ (they agree with the ones from section~\ref{sec:GalSup} if we reabsorb $B$ into $A_{\mu|\nu}$):
\begin{align}
\delta\phi &= \partial_\alpha \Lambda^\alpha\, ,  &
\delta s_\mu &= \partial_\mu \partial_\alpha \Lambda^\alpha\, ,\\
\delta A_{\mu|\nu} &= \frac{1}{(1-g)}\Big(\partial_{(\mu} \Lambda_{\nu)}-\frac{1}{d}\eta_{\mu\nu}\partial_\alpha \Lambda^\alpha\Big)\, ,& \delta B& = \frac{1}{(1-g)}\partial_\alpha\Lambda^\alpha\, .
\end{align}
The most general gauge-invariant improvement of~\eqref{eq:superfluid1order1ap} is
\be
\begin{aligned}
S &= -a\int \rd^dx\Big[\phi\left(\partial^\mu s_\mu - \partial^\mu\partial^\nu A_{\mu|\nu} +  \left(g-\frac{1}{d}\right)\square B\right) + \frac{1}{2}s_\mu s^\mu\\
&\phantom{=}\qquad \qquad +\frac{d\,\kappa}{2(d-2)}\left( \kappa\partial_\mu A_{\nu|\alpha}\partial^\mu A^{\nu|\alpha} -2\partial^\mu A_{\mu|\alpha}\partial_\nu A^{\nu|\alpha}\right)\\
&\phantom{=}\qquad \qquad  +(\kappa + g-1)\partial^\mu B \partial^\nu A_{\mu|\nu} +\frac{-d(\kappa + g^2-1)+\kappa + 2g -2 }{2d}\partial_\mu B \partial^\mu B\Big]\, , \label{eq:ActionSuperfluidGauged}
\end{aligned}
\ee
with $a$ a free normalization. The action \eqref{eq:ActionSuperfluidGauged} has two free parameters ($g$ and $\kappa$), and is slightly different than \eqref{eq:ActionSuperfluidGauged2} because of the way we separated the trace. Looking at \eqref{eq:ActionSuperfluidGauged}, it is already clear that $d=2$ is special.

When $d\neq 2$, we can choose 
\be 
\kappa = \frac{d-1}{d}\, ,\qquad \qquad g = \frac{1}{d}\, ,
\ee
in which case the trace $B$ totally decouples from the rest of the action and we obtain the same action as in section~\ref{sec:GalSup}. When $d=2$, the choice of parameters is impossible because the terms that are quadratic in $A_{\mu|\nu}$ diverge, and we must keep the trace $B$ explicitly.

The equation of motion for $s_\mu$ is unchanged compared to the ungauged version of the action
\be 
\partial_\mu \phi -s_\mu =0\, ,\label{eq:ScalarEOM1}
\ee
which allows us to solve for $s_\mu$ in terms of $\phi$,
while varying the action \eqref{eq:ActionSuperfluidGauged} with respect to $\phi$, we obtain
\be 
\square \phi - \partial^\mu \partial^\nu A_{\mu|\nu} + \left(g-\frac{1}{d}\right)\, \square  B = 0\, , \label{eq:ScalarEOM2}
\ee
where we have used \eqref{eq:ScalarEOM1}. 

\subsubsection*{Conserved current}

To derive the current, we need to vary the action \eqref{eq:ActionSuperfluidGauged} with respect to the background gauge fields that we introduced. Since $A^{\mu|\nu}$ is identically traceless, we need to remove the traces when we perform the variation, this produces
\be
\begin{aligned} 
J_{\mu|\nu} \equiv \frac{\delta S}{\delta A^{\mu|\nu}} &= a\Big[\partial_{(\mu}\partial_{\nu)}\phi  -\frac{\kappa\, d}{d-2}\left[2\partial^\alpha \partial_{(\mu}A_{\nu)|\alpha} -\square A_{\mu|\nu}\right]+(\kappa+g-1 )\partial_{(\mu}\partial_{\nu)}B\\
&\phantom{=}\quad\quad  -\frac{\eta_{\mu\nu}}{d}\left(\square  \phi-\frac{2\kappa\,d}{d-2}\partial^\alpha \partial^\mu A_{\mu|\alpha}+(\kappa + g -1)\square  B\right)\Big]\, .
\end{aligned}
\ee
In order to be able to 
impose dual conservation on the current instead of standard conservation, we note that the variation of the action with respect to $B$ is also gauge invariant
\be 
J \equiv \frac{\delta S}{\delta B} = a\Big[-\left(g-\frac{1}{d}\right)\, \square  \phi +(\kappa + g-1) \partial^\mu \partial^\nu A_{\mu|\nu} + \frac{-d(\kappa + g^2-1)+\kappa + 2g -2 }{d}\square  B\Big]\, ,
\ee
so that
we can consider the following general current
\be 
\mathcal{J}_{\mu|\nu}\equiv  J_{\mu|\nu} + \beta\,  \eta_{\mu\nu}\, J\, ,\label{eq:UpgradedCurrentSF}
\ee
where $\beta$ is a free parameter. For different choice of parameters $g,\, \kappa,$ and $\beta$, the current \eqref{eq:UpgradedCurrentSF} will obey different conditions. In order to express them simply, it is useful to define the gauge-invariant combination
\be 
\mathcal{F}_d \equiv -\partial^\alpha\partial^\beta A_{\alpha|\beta}+\frac{d-1}{d}\square  B\, .
\ee
We now consider the various equations that ${\cal J}_{\mu|\nu}$ could satisfy:

\begin{itemize}
\item \textbf{Trace:} Computing the trace of \eqref{eq:UpgradedCurrentSF} yields 
\be 
\mathcal{J}\indices{_\mu^\mu} =-a\beta(1 + d(\kappa-1))\mathcal{F}_d\, .
\ee
\item \textbf{Divergence:} Computing the divergence of \eqref{eq:UpgradedCurrentSF} yields 
\be 
\partial^\mu \mathcal{J}_{\mu|\nu}= -a(\beta-1)\frac{(1 + d(\kappa-1))}{d}	\partial_\nu\mathcal{F}_d\, .
\ee
\item \textbf{Dual divergence:} Computing the dual divergence of \eqref{eq:UpgradedCurrentSF} yields 
\begin{align}
\partial_{[\alpha}\mathcal{J}_{\mu]|\nu}=a\Big[ &-\frac{d\, \kappa}{d-2}\left(\partial^\beta \partial_\nu \partial_{[\alpha}A_{\mu]|\beta} -\square \partial_{[\alpha}A_{\mu]|\nu}\right) \\
\phantom{=}\quad \quad& +\Big(\beta\left(1-\kappa -\frac{1}{d}\right)+\frac{1}{d}-\frac{2\kappa}{d-2}\Big)\eta_{\nu[\alpha}\partial_{\mu]}\partial^\rho \partial^\sigma A_{\rho|\sigma}\nonumber\\
\phantom{=}\quad \quad &-\frac{(1 + d(\kappa-1))(1 + \beta (d-1)}{d^2}\eta_{\nu[\alpha}\partial_{\mu]}\square  B\Big]\, . \nonumber
\end{align}
\end{itemize}
We now want to explore how many of these conditions we can impose simultaneously. As before, it will be necessary to consider $d=$ and $d\neq2$ separately.

\begin{itemize}

\item $\boldsymbol{d >2:}$
In this case, we can 
choose 
\be 
\kappa  = \frac{d-1}{d} \, ,\label{eq:Choiceb2}
\ee
so that the current is both traceless and conserved for any value of $g$ and $\beta$. This is the case that we considered in detail in the main text. In this case,  we cannot choose $\beta$ to ensure dual conservation, and we obtain 
\be
\partial_{[\alpha}\mathcal{J}_{\mu]|\nu} = -\frac{a}{d-2}\Big[(d-1)\left(\partial^\beta \partial_\nu \partial_{[\alpha}A_{\mu]|\beta} -\square \partial_{[\alpha}A_{\mu]|\nu}\right)+\eta_{\nu[\alpha}\partial_{\mu]}\partial^\rho \partial^\sigma A_{\rho|\sigma}\Big]\, .
\ee

Alternatively, we can choose 
\be 
\kappa =0\, , \qquad \qquad \beta  = -\frac{1}{d-1}\, ,
\ee
which ensures dual conservation. In this case, the current is neither traceless nor conserved:
\be
\mathcal{J}\indices{_\mu^\mu} =-a\mathcal{F}_d\, ,\qquad\qquad 
\partial^\mu \mathcal{J}_{\mu|\nu} =-a \partial_\nu \mathcal{F}_d\, .
\ee

\item $\boldsymbol{d =2:}$
In two dimensions, we cannot choose $\kappa $ as in~\eqref{eq:Choiceb2} because then the action \eqref{eq:ActionSuperfluidGauged} has some divergent contact terms. We then choose  $\kappa = 0$
which implies
\begin{align}
\mathcal{J}\indices{_\mu^\mu} &= a\, \beta\, \mathcal{F}_2\, ,\label{eq:2dtrace}\\
\partial^\mu \mathcal{J}_{\mu|\nu} &= \frac{1}{2}a(\beta-1) \partial_\nu \mathcal{F}_2, \\
\partial_{[\alpha}\mathcal{J}_{\mu]|\nu}&=a \frac{1+\beta}{2}\eta_{\nu[\alpha}\partial_{\mu]}\mathcal{F}_2\, .\label{eq:2ddualcons}
\end{align}
Looking at \eqref{eq:2dtrace}--\eqref{eq:2ddualcons}, it is clear that in two dimensions, we have a three-way anomaly, where we can choose the current $\mathcal{J}_{\mu\nu}$ to be one of traceless ($\beta=0$), conserved ($\beta=1$) or dual conserved ($\beta=-1$).

\end{itemize}

\noindent This structure of anomalies exactly matches the possible conditions that we can impose on the current two-point function, as we now explore.

\subsection{Current two-point function \label{sec:cur2ptfunctiongalileonap}}

Next, we enumerate the possible conditions that one can impose on the currents appearing in the two-point function.
In momentum space, the most general form for this correlator with the appropriate symmetries is~\eqref{eq:SymCurSymCur}:
\be
\begin{aligned}
\braket{{J}_{{\mu_1}|{\mu_2}}* {K}*_{{\nu_1}|{\nu_2}}} &=\, 2c_1(p^2)p^2\eta_{{\mu_1}({\nu_2}}\eta_{{\nu_1}){\mu_2}} + c_2(p^2)p^2 \eta_{{\mu_1}{\mu_2}}\eta_{{\nu_1}{\nu_2}} + c_3(p^2)\eta_{{\mu_1}{\mu_2}}p_{\nu_1} p_{\nu_2}
\\
&\phantom{=}~ + c_4(p^2)\eta_{{\nu_1}{\nu_2}}p_{\mu_1} p_{\mu_2}+ 2c_5(p^2)\Big(\eta_{{\mu_2}({\nu_2}}p_{{\nu_1})}p_{\mu_1} +\eta_{{\mu_1}({\nu_2}}p_{{\nu_1})}p_{\mu_2}\Big)\\
&\phantom{=}~ + c_6(p^2) \frac{p_{\mu_1} p_{\mu_2} p_{\nu_1} p_{\nu_2}}{p^2}\, ,
\end{aligned} 
\label{eq:SymCurSymCurap}
\ee
where $c_{1}(p^2),\cdots,c_6(p^2)$ are all arbitrary functions of $p^2$ and we have not assumed any symmetry under $({\mu_1}{\mu_2})\leftrightarrow({\nu_1}{\nu_2})$ exchange. We now impose various conditions on the currents. As before we see that we have to separate the $d=2$ case from the generic case.

\vspace{6pt}
\noindent
{\bf In $\boldsymbol{d \neq 2}$ dimensions:}
When $d\neq 2$, we can impose the following:
\begin{itemize}
\item
\noindent \textbf{Tracelessness and conservation:}  Requiring  the current $J$ to be both traceless and conserved at coincident points implies
\begin{align}
c_1(p^2) &= -\frac{1}{2}(d-1)c_2(p^2)\, ,  & c_3(p^2) &=c_4(p^2)= -c_2(p^2)\, ,\\
c_5(p^2) &= \frac{1}{2}(d-1)c_2(p^2)\, , & c_6(p^2) &= -(d-2)c_2(p^2)\, .
\end{align}
Using the anomaly equation as in the main text, which is given in \eqref{eq:anomlay1}, we can fix $c_2(p^2)$---which cannot be chosen so that $J$ is also dual conserved:
\be 
c_2(p^2) = -\frac{1}{d-2}\, ,
\ee
such that the correlator is the one given in \eqref{eq:CurrCurre2pt}.
\item 
\noindent \textbf{Dual conservation:} In this case, we require that the current is dual conserved , which means $p_{[\alpha}\braket{{J}_{{\mu_1}|{\mu_2}}(p)* {K}*_{{\nu_1]}|{\nu_2}}(-p)}=0$ everywhere. This implies
\be 
c_1(p^2) = c_2(p^2) = c_4(p^2) = c_5(p^2) =0\, .
\ee
We cannot choose $c_{3,6}(p^2)$ to either ensure tracelessness nor conservation. Using the trace and conservation anomaly equations 
\begin{equation}
\braket{{J}\indices{_{{\mu_1}|}^{{\mu_1}}}(p)*{K}*\indices{_{{\nu_1}|}_{\nu_2}}(-p)} =  p_{\nu_1} p_{\nu_2} \, ,\qquad \quad 
p^{\mu_1}\braket{{J}\indices{_{{\mu_1}|{\mu_2}}}(p)*{K}*_{{\nu_1}|{\nu_2}}(-p)} =  p_{\mu_2} p_{\nu_1} p_{\nu_2} \, ,
\end{equation}
we can fix $c_3(p^2) = 0,\, c_6(p^2) = 1$
so that the correlator is
\be
\braket{{J}_{{\mu_1}|{\mu_2}}(p)* {K}*_{{\nu_1}|{\nu_2}}(-p)} =\frac{p_{\mu_1}p_{\mu_2}p_{\nu_1}p_{\nu_2}}{p^2}\, .\label{eq:missingcreativenames1ap}
\ee

\end{itemize}

\vspace{6pt}
\noindent
{\bf In $\boldsymbol{d = 2}$ dimensions:}
When $d=2$, the tensor structure multiplied by $c_1(p^2)$ in the ansatz~\eqref{eq:SymCurSymCurap} is not linearly independent from the other structures, so we can set $c_1(p^2)=0$. We now see that we can impose fewer conditions:
\begin{itemize}

\item \textbf{Conserved: } Requiring conservation, and using the  trace anomaly equation 
\be 
\braket{{J}\indices{_{{\mu_1}|}^{{\mu_1}}}(p)*{K}*\indices{_{{\nu_1}|}_{{\nu_2}}}(-p)}  = p^2 \eta_{{\nu_1}{\nu_2}} - p_{\nu_1} p_{\nu_2}\, ,
\ee
we can uniquely fix the two-point function to be 
\begin{align}
\braket{{J}_{{\mu_1}|{\mu_2}}(p)* \! {K}\! *_{{\nu_1}|{\nu_2}}(-p)}_{d=2} &=  p^2 \eta_{{\mu_1}{\mu_2}}\eta_{{\nu_1}{\nu_2}} -\eta_{{\mu_1}{\mu_2}}p_{\nu_1} p_{\nu_2}-\eta_{{\nu_1}{\nu_2}}p_{\mu_1} p_{\mu_2} +\frac{p_{\mu_1} p_{\mu_2} p_{\nu_1} p_{\nu_2}}{p^2}\, .
\end{align}
We can check that this also agrees with the dual conservation anomaly equation.

\item \textbf{Dual conserved: }
For this case, there is actually nothing that changes compared to the $d\neq 2$ computation that we performed above. The two-point function thus has the form shown in \eqref{eq:missingcreativenames1}
\end{itemize}
We see that the possible conditions and anomalous conservation equations agree perfectly with the analysis of section~\ref{app:galfluidaction} based on the IR EFT action.

\newpage
\section{Galileonic electromagnetism \label{sec:EM}}

In this appendix, we work out the details of the EFT that has a $(1|0)$-biform symmetry, which is the simplest example of the theories discussed in section~\ref{sec;bms}. The relevant Goldstone mode is a vector gauge field, so this EFT coincides with electromagnetism in the deep infrared, but has different (more restricted) allowed interactions. Much as in the other examples, the presence of the gapless photon in this theory is a consequence of a mixed anomaly between electric and magnetic symmetries.

\subsection{Summary of Maxwell theory \label{sec:symEM}}

We first briefly review Maxwell electromagnetism in its usual formulation, in order to contrast with the presentation in the subsequent section.
More details can be found for example in~\cite{Gaiotto:2014kfa,Hofman:2017vwr,Cordova:2018cvg}.  We work in general dimension $d$ and denote the gauge coupling as $g$. 
This theory is usually described using a $1$-form (electric) gauge field, $a_{(1)}$, that is associated to the usual (electric) field strength $F_{(2)} = \rd a_{(1)}$. This theory has two different conserved currents
\be 
J_{(2)} \equiv \frac{1}{g^2}F_{(2)}\, ,\qquad \qquad K_{(d-2)} \equiv  * F_{(2)}\, .\label{eq:EMCurrents1}
\ee
These currents are associated to two global higher-form symmetries: An electric $1$-form symmetry with current $J_{(2)}$ and a $(d-3)$-form magnetic symmetry with current $K_{(d-2)}$. 
The objects charged under these symmetries are Wilson (electric) and 't Hooft (magnetic) lines, and one can define conserved charges as:
\be 
Q_e = \oint_{\Sigma_{d-2}} * F_{(2)} \, ,\qquad \qquad Q_m  = \oint_{\Sigma_2} F_{(2)}\, .
\ee

We can gauge the electric symmetry by including a background source $B_{(2)}$ for the electric current so that the theory is invariant under arbitrary shifts $a_{(1)} \mapsto a_{(1)}+ \xi_{(1)}$.  The currents~\eqref{eq:EMCurrents1} can be improved to preserve gauge invariance as
\be 
J_{(2)} \rightarrow {\cal J}_{(2)} = \frac{1}{g^2}\left(F_{(2)}- B_{(2)}\right)\, , \qquad \qquad K_{(d-2)}\rightarrow {\cal K}_{(d-2)}=* (F_{(2)}-B_{(2)})\, , \label{eq:EMCurrent1}
\ee
where under a gauge transformation ${B}_{(2)} \mapsto {B}_{(2)} + \rd \xi_{(1)}$. The equations above imply a mixed anomaly between the conserved currents in the form
\begin{align} 
\rd * {\cal J}_{(2)} &= 0\label{eq:EMEOM1p}\, ,\\
\rd *{\cal K}_{(d-2)}&= -\,\rd B_{(2)}\label{eq:EMEOM2p}\, .
\end{align}
These equations continue to hold even in the presence of interactions, so long as they preserve the higher-form symmetries. Interactions would change the form of the electric symmetry current, but it would have the same anomaly with the magnetic symmetry.

\subsubsection{Two-point function}

This anomaly structure is enough to  guarantee that the theory is in a gapless phase, with the photon appearing as a Goldstone mode.
With the above input, we can compute the current-current two-point function in electromagnetism. We are interested in the correlator
\be 
\braket{{J}_{\mu_1\mu_2}(p)*K_{\nu_1\nu_2}(-p)} \equiv \int \rd^dx\, e^{ix\cdot p}\braket{{J}_{\mu_1\mu_2}(x)*{K}_{\nu_1\nu_2}(0)}\, .
\ee 
The most general form that this can take consistent with Lorentz invariance is
\be
\braket{{J}_{\mu_1\mu_2}(p)*K_{\nu_1\nu_2}(-p)} =-2c_1(p^2)p^2\left(\eta_{{\mu_1}[{\nu_1}}\eta_{{\nu_2]}{\mu_2}}\right)+2c_2(p^2)\left(\eta_{{\nu_1}[{\mu_1}}p_{{\mu_1}]} p_{\nu_2} - \eta_{{\nu_2}[{\mu_1}}p_{{\mu_2}]}p_{\nu_1}\right) \, ,
\ee
which depends on two unknown functions $c_1(p^2)$ and $c_2(p^2)$ that are fixed by the conservation equations.
Requiring~\eqref{eq:EMEOM1p} and~\eqref{eq:EMEOM2p} to be satisfied, we have 
\begin{align} 
p^{\mu_1} \braket{J_{{\mu_1}{\mu_2}}(p)*K_{{\nu_1}{\nu_2}}(-p)}& = 0\, ,  &\implies&&  c_1(p^2) &= c_2(p^2)\, ,\\
p_{[\alpha}\braket{J^{{\mu_1}{\mu_2}}(p)*K_{{\nu_1}{\nu_2}]}(-p)}& = p_{[\alpha}\delta\indices{^\mu_\sigma}\delta\indices{^\nu_{\rho]}}\, , & \implies && c_2(p^2)& = \frac{1}{2p^2}\, .
\end{align}
The mixed correlator is therefore entirely fixed and takes the form 
\be 
\braket{J_{{\mu_1}{\mu_2}}(p)*K_{{\nu_1}{\nu_2}}(-p)}=\frac{\eta_{{\mu_2}{\nu_2}}p_{\mu_1} p_{\nu_1}+ \eta_{{\mu_1}{\nu_1}}p_{\mu_2} p_{\nu_2} -\eta_{{\mu_1}{\nu_2}}p_{\mu_2} p_{\nu_1} -\eta_{{\mu_2}{\nu_1}}p_{\mu_2} p_{\nu_2}}{2p^2}-\eta_{{\mu_1}{[\nu_1}}\eta_{{\nu_2]}{\mu_2}} .\label{eq:EMcorr1}
\ee
The $1/p^2$ pole is a robust consequence of the anomaly, as in the superfluid case. Moreover, the contact terms can be redefined at will to choose which of the two currents should be conserved even at coincident points. If we instead require the magnetic current to be identically conserved, we obtain
\be 
\braket{J_{{\mu_1}{\mu_2}}(p)*K_{{\nu_1}{\nu_2}}(-p)} =\frac{1}{2}\frac{\eta_{{\mu_2}{\nu_2}}p_{\mu_1} p_{\nu_1}+ \eta_{{\mu_1}{\nu_1}}p_{\mu_2} p_{\nu_2}-\eta_{{\mu_1}{\nu_2}}p_{\mu_2} p_{\nu_1} -\eta_{{\mu_2}{\nu_1}}p_{\mu_2} p_{\nu_2} }{p^2}\, .\label{eq:EMcorr2}
\ee
The difference between the two correlators \eqref{eq:EMcorr1} and \eqref{eq:EMcorr2} is simply a local contact term. 

\vspace{6pt}
\noindent
{\bf \KL decomposition:}
We can now perform the spectral decomposition of this correlator to verify that there is a massless spin-1 particle in the spectrum: the photon. Using the results from appendix~\ref{ap:KLrep}, we can write the spectral function for an antisymmetric current,
\be
\braket{J^{{\mu_1}{\mu_2}}(p)*K^{{\nu_1}{\nu_2}}(-p)}
= \int_0^\infty \rd s  {s\over p^2+s} \left(-\rho_1(s)\tl{\Pi}_{1,{\rm em}}^{{\mu_1}{\mu_2}{\nu_1}{\nu_2}} + \rho_{(2|0)}(s)\tl{\Pi}_{(2|0),{\rm em}}^{{\mu_1}{\mu_2}{\nu_1}{\nu_2}}\right) \, . \label{emspectede}
\ee
The spectral densities $\rho_1$ and $\rho_{(2|0)}$ capture the coupling of spin-1 states and antisymmetric tensor states respectively (Massless antisymmetric tensor states cannot couple to the anti-symmetric current~\cite{Weinberg:2020nsn,Distler:2020fzr} so the density $\rho_{(2|0)}$ must go to zero as $s\rightarrow 0$).

The projector $\Pi_{1, {\rm em}}^{{\mu_1}{\mu_2}{\nu_1}{\nu_2}}$ is that of a spin-$1$ particle while $\Pi_{(2|0),{\rm em}}^{{\mu_1}{\mu_2}{\nu_1}{\nu_2}}$ is the projector for a rank $2$ antisymmetric tensor particle (a $2$-form). They are uniquely defined by the requirements that they be traceless, transverse, and complete on the space of traceless antisymmetric tensors. Explicitly,
\begin{align}
\Pi_{1,{\rm em}}^{{\mu_1}{\mu_2}{\nu_1}{\nu_2}} &= \frac{1}{p^2}\left(\eta^{{\mu_1}{[\nu_1}}p^{\nu_2]}p^{\mu_2}-\eta^{{\mu_2}[{\nu_1}}p^{{\nu_2}]}p^{{\mu_1}}\right)\, ,\\
\Pi_{(2|0),{\rm em}}^{{\mu_1}{\mu_2}{\nu_1}{\nu_2}}
&= \left(\eta^{{\mu_1}{[\nu_1}}\eta^{{\nu_2}]{\mu_2}}\right)-\frac{1}{p^2}\left(\eta^{{\mu_1}{[\nu_1}}p^{\nu_2]}p^{\mu_2}-\eta^{{\mu_2}[{\nu_1}}p^{{\nu_2}]}p^{{\mu_1}}\right)\, .
\end{align}
The tensors appearing in \eqref{emspectede} are projectors only on-shell, given by replacing $p^2\rightarrow -s$, 
\begin{align}
\tl{\Pi}_{1,{\rm em}}^{{\mu_1}{\mu_2}{\nu_1}{\nu_2}} &= -\frac{1}{s}\left(\eta^{{\mu_1}{[\nu_1}}p^{\nu_2]}p^{\mu_2}-\eta^{{\mu_2}[{\nu_1}}p^{{\nu_2}]}p^{{\mu_1}}\right)\, ,\\
\tl{\Pi}_{(2|0),{\rm em}}^{{\mu_1}{\mu_2}{\nu_1}{\nu_2}}
&= \left(\eta^{\mu[\rho}\eta^{\sigma]\nu}\right)+\frac{1}{s}\left(\eta^{{\mu_1}{[\nu_1}}p^{\nu_2]}p^{\mu_2}-\eta^{{\mu_2}[{\nu_1}}p^{{\nu_2}]}p^{{\mu_1}}\right)\, .
\end{align}
Using these two projectors, we can write respectively \eqref{eq:EMcorr1} and \eqref{eq:EMcorr2} as 
\begin{align}
\braket{J^{{\mu_1}{\mu_2}}(p)*K^{{\nu_1}{\nu_2}}(-p)} &= -\frac{1}{2}\Pi_{1,{\rm em}}^{{\mu_1}{\mu_2}{\nu_1}{\nu_2}}\,,\\
\braket{J^{{\mu_1}{\mu_2}}(p)*K^{{\nu_1}{\nu_2}}(-p)} &=\frac{1}{2}\Pi_{(2|0),{\rm em}}^{{\mu_1}{\mu_2}{\nu_1}{\nu_2}} \, .
\end{align}
Writing \eqref{emspectede} as 
\begin{align}
\braket{J^{{\mu_1}{\mu_2}}(p)*K^{{\nu_1}{\nu_2}}(-p)}
&= \left(\int_0^\infty \rd s\, \frac{s\rho_{(2|0)}(s)}{p^2+s}\right)\,\eta^{\mu[\rho}\eta^{\sigma]\nu}\\
&\phantom{=}\quad +\left(\int_0^\infty \rd s\,\frac{\rho_{(2|0)}(s) +\rho_1(s)}{p^2+s}\right)\left(\eta^{{\mu_1}{[\nu_1}}p^{\nu_2]}p^{\mu_2}-\eta^{{\mu_2}[{\nu_1}}p^{{\nu_2}]}p^{{\mu_1}}\right)\, , \nonumber
\end{align}
we see that the two-point function of pure Maxwell theory, \eqref{eq:EMcorr2}, is recovered by taking 
\be 
\rho_{(2|0)}(s) = 0\, ,\qquad \qquad \rho_{1}(s)  = \delta(s)\, .
\ee
This shows that there is a massless spin-1 particle (the photon) in the spectrum.

\subsection{Galileon electromagnetism\label{sec:FirstOrderEM}}

We now turn to the study of the galileonic version of electromagnetism that has biform symmetries.
The approach is philosophically similar to the discussion of the Galileon superfluid in section~\ref{sec:GalSup}. The EFT that we construct will have a larger set of global symmetries than ordinary electromagnetism, and so will have different irrelevant corrections.

We begin by writing the free Maxwell action in a first-order form that manifests more of its symmetries.
Consider a vector field $a_\mu$ together with a symmetric two-index tensor (or a $(1|1)$-biform) $s_{\mu|\nu}$. Electromagnetism can be written in a Palatini-like form as
\be 
S = -\frac{1}{g^2}\int \rd^dx\,\left[a^\nu\left(\partial^\mu s_{\mu | \nu}-\partial_\nu s\right) + \frac{1}{2}\left(s_{\mu | \nu}s^{\mu | \nu} -s^2\right)\right]\, , \label{eq:EMpalatini}
\ee
where we defined the trace $s=s\indices{_{\mu|}^\mu}$.
The action \eqref{eq:EMpalatini} is invariant under the gauge transformations
\be
\delta a_\mu = \partial_\mu \Lambda\, ,
\qquad \qquad \delta s_{\mu\nu}= \partial_\mu \partial_\nu \Lambda\, ,
\ee
where $\Lambda$ is an arbitrary scalar function. The equations of motion of~\eqref{eq:EMpalatini} are 
\be 
s_{\mu|\nu} = \partial_{(\mu}a_{\nu)}\, , \quad\quad\quad \partial^\alpha s_{\alpha|\nu}-\partial_\nu s= 0\, .\label{eq:SintermsofA}
\ee
Combining these two equations, we find the ordinary Maxwell equation in vacuum
\be 
\partial^\mu \left(\partial_\mu a_\nu - \partial_\nu a_\mu \right) = \partial^\mu F_{\mu\nu} = 0\, . 
\ee
It is interesting to note that we obtained an equation for the antisymmetric Maxwell tensor despite the fact that the fundamental object $s_{\mu\nu}$ that we used to construct the action~\eqref{eq:EMpalatini} is symmetric. Integrating out $s_{\mu\nu}$ via its equation of motion recovers the usual Maxwell action.

From the first-order action~\eqref{eq:EMpalatini}, we can now see that it has a second set of $1$-form symmetries, which can be written as 
\begin{align}
\delta a_\nu &=\left(\partial^\alpha\Lambda_{\mu|\alpha}+ \partial^\alpha \xi_{\mu\alpha}\right)+ \partial_\mu \Lambda\indices{^\alpha_{|\alpha}}\, ,\label{eq:gaugeEM4}\\
\delta s_{\mu| \nu} &=\left(\partial^\alpha \partial_{(\mu} \Lambda_{\nu)\alpha}+  \partial^\alpha \partial_{(\mu}\xi_{\nu)\alpha}\right)+\partial_\mu \partial_\nu \Lambda\indices{^\alpha_{|\alpha}}\label{eq:gaugeEM5}\, ,
\end{align}
where $\Lambda_{\mu|\nu}$ is a symmetric two tensor, while $\xi_{\mu\nu}$ is a two-form Killing tensor that solves \eqref{eq:3derivativesvector}. This is a symmetry of the action \eqref{eq:EMpalatini}~provided we impose the flatness condition
\be 
\partial^\alpha \partial_{[\mu}\Lambda_{\nu]|\alpha}=0\, .
\ee
We now want to gauge these symmetries so that $\Lambda$ can be an arbitrary function.

\subsubsection{Coupling to sources }

We gauge the higher-form symmetries \eqref{eq:gaugeEM4} and \eqref{eq:gaugeEM5} by introducing a background gauge field that couples to the conserved current. Because we anticipate a trace anomaly in special dimension, we will consider a traceful background gauge field $B_{\mu\nu|\rho}$ expecting  that we can decouple its trace generally. The background gauge field $B_{\mu \nu | \rho}$ is a $(2|1)$-biform. The current associated to this symmetry, following the discussion in \ref{sec;bms}  is
\be
H_{\mu \nu | \rho} = \partial_\mu s_{\nu | \rho} - \partial_\nu s_{\mu |\rho}\,,
\ee
and is traceless on-shell using the equations of motion~\eqref{eq:SintermsofA}. 

We consider the following gauge transformations
\begin{align}
\delta a_\nu &=6(1-\ell)\left(\frac{1}{3}\partial^\alpha\Lambda_{\mu|\alpha}+ \partial^\alpha \xi_{\mu\alpha}\right)+ \partial_\mu \Lambda\indices{^\alpha_{|\alpha}}\label{eq:gaugeEM1}\, ,\\
\delta s_{\mu| \nu} &=6(1-\ell)\left(\frac{1}{3}\partial^\alpha \partial_{(\mu} \Lambda_{\nu)\alpha}+ \partial^\alpha \partial_{(\mu}\xi_{\nu)\alpha}\right)+\partial_\mu \partial_\nu \Lambda\indices{^\alpha_{|\alpha}}\, ,\label{eq:gaugeEM2}\\
\delta B_{\mu\nu|\rho} &= 3\mathcal{Y}_{(2|1)}\left(\partial_{\mu }\Lambda_{\nu\rho}+\partial_\mu \xi_{\nu\rho}\right)\, ,\label{eq:gaugeEM3}
\end{align}
where we have chosen the normalization for later convenience, and $\ell$ is a free coefficient. In \eqref{eq:gaugeEM1}-\eqref{eq:gaugeEM3}, the gauge parameter $\Lambda_{\mu|\nu}$ is a symmetric 2-tensor (or a $(1|1)$-biform) while $\xi_{\mu\nu}$ is an antisymmetric tensor (or a $2$-form). 
The most general action that we can write that is gauge-invariant (for any choice of $\kappa, \ell$) under these transformations is 
\begin{align} 
S&= -\frac{1}{g^2}\int \rd^dx\,\bigg[a^\nu\left(\partial^\mu s_{\mu|\nu}-\partial_\nu s - \partial^\mu \partial^\rho B_{\mu\nu | \rho}-\ell\left(\partial^\beta \partial_\nu B_\beta-\square B\indices{_\nu}\right)\right) \label{eq:EMpalatini3}\\
&\qquad \qquad \phantom{=}+ \frac{1}{2}\left(s_{\mu| \nu}s^{\mu| \nu} -s^2\right)-\kappa\left(\partial _{\mu  }B^{\mu  \nu | \rho  } \partial^{\sigma 
}B_{\rho\sigma| \nu  }-\partial _{\mu  }B^{\nu \rho | \mu  } \partial^{\sigma  }B_{\nu \rho |\sigma  } +\frac{1}{2}\partial_{\sigma  }B_{\mu  \nu | \rho  } \partial ^{\sigma  }B^{\mu \nu|  \rho  }\right) \nonumber\\
&\qquad \qquad \phantom{=} -(\ell-1-\kappa)\partial^\mu B\indices{^\nu}\partial^\sigma B\indices{_{\mu\nu|\sigma}} -\left(\ell^2-1-\kappa\right)\left(\partial^\mu B\indices{_{\nu}}\partial_\mu B\indices{_{\nu}}-\partial_\mu B^\mu \partial_\nu B^\nu\right)\bigg]\nonumber\, .  
\end{align}
where we have defined $B_\mu \equiv B\indices{_{\rho\mu|}^\rho}$, which is the trace of $B_{\mu\nu|\alpha}$.
The equation of motion for $s_{\mu|\nu}$  is unchanged from~\eqref{eq:SintermsofA}---and sets $s_{\mu\nu}$ in terms of $a_\mu$---while the equation of motion for $a_\mu$ is
\be 
\square a_\mu - \partial_\mu \partial^\rho a_{\rho}-2 \partial^\rho \partial^\sigma B_{\rho \mu | \sigma}-2\ell\left(\partial_\mu \partial_\alpha B^{\alpha}-\square B_{\mu}\right) =0\, .\label{eq:EOM15}
\ee
To obtain the current of interest, we vary the action \eqref{eq:EMpalatini3} with respect to $B_{\mu\nu|\rho}$:
\be
\mathcal{H}_{\mu\nu | \rho} = \frac{1}{g^2}\left(\partial_\rho \partial_{[\mu}a_{\nu]}-\mathcal{B}_{\mu\nu|\rho} \right)\, , \label{eq:femcur}
\ee
where we have defined
\begin{align}
\mathcal{B}_{\mu\nu|\rho}=-\mathcal{Y}_{(2|1)}\Big[&\kappa\left(3 \partial^\alpha\partial_\mu  B_{\rho\alpha|\nu}+3 \partial^\alpha\partial_\nu  B_{\rho\alpha|\mu}-\square B_{\mu\nu|\rho}\right)\label{eq:CurlyB2}\\
\phantom{=}&-3\left(1 + \kappa\right)\left(\eta_{\mu\nu}\partial^\alpha \partial^\beta B_{\rho\alpha|\beta}+ \eta_{\mu\nu}\left(\square B_{\rho}-\partial_\rho \partial^\alpha B_{\alpha}\right)\right)+3(1+\kappa -\ell)\partial_\mu \partial_\nu B_{\rho}\Big]\, ,\nonumber
\end{align}
and have used the equation of motion~\eqref{eq:EOM15} to replace $\square a_\mu$. It is straightforward to check that the current \eqref{eq:femcur} is gauge invariant. Moreover, it is conserved for both divergences
\be
\partial^\mu \mathcal{H}_{\mu\nu|\rho} = 0\, ,\qquad \qquad \partial^\rho \mathcal{H}_{\mu\nu|\rho}=0\, ,
\ee
for any $\kappa$. This is a consequence of the way we have introduced the gauge field $B_{\mu\nu|\rho}$, which is designed to couple to a conserved current. 

\subsubsection{Anomalies}

We now want to understand how the introduction of the background gauge field $B_{\mu\nu|\rho}$ changes the properties of the current \eqref{eq:femcur}. The current is conserved on-shell, but the other conditions (tracelessness and dual conservation) do not need to be satisfied, and we have some freedom to tune $\kappa$ and $\ell$ to enforce different conditions.
In order to explore the anomaly structure, we need to compute the on-shell conditions for \eqref{eq:femcur}.
\begin{itemize}

\item \textbf{Trace: } The on-shell trace of the current \eqref{eq:femcur} is
\be  
\mathcal{H}\indices{^\rho_{\mu|\rho}} = -\frac{(d-2+\kappa(d-3))}{g^2}\left(\partial^\alpha \partial^\beta B_{\alpha \mu|\beta}+\partial_\mu \partial_\alpha B^{\alpha}-\square B_\mu \right)\, .\label{eq:TraceEMOS}
\ee
It is straightforward to check that this is gauge invariant.

\item \textbf{Dual conservation:} The on-shell dual conservation is 
\be 
\partial_{[\alpha}\mathcal{H}_{\mu\nu]|\rho} =-\frac{1}{g^2}\partial_{[\alpha}\mathcal{B}_{\mu\nu]|\rho}\, ,\label{eq:EMDCons}
\ee
which is also gauge invariant.
\end{itemize}

\noindent 
We now want to explore what conditions we can require that the current satisfy. It is already clear that the current cannot be made to be both traceless and dual conserved. In fact, 
it is impossible to enforce dual conservation~\eqref{eq:EMDCons} for any choice of parameters, because we have required the current to be conserved instead. This is the expression of a mixed anomaly between the electric and magnetic biform symmetries. However, we see from \eqref{eq:TraceEMOS} that we can set the trace to zero in $d \neq  3$, so it is convenient to split the discussion into two cases, $d\neq  3$ and $d = 3$. We will consider the generic case first.

\vspace{6pt}
\noindent
{\bf In $\boldsymbol{d \neq 3}$ dimensions:}
In general dimension, it is possible to make the current $\mathcal{H}_{\mu\nu|\rho}$ simultaneously conserved and traceless by choosing
\be 
\kappa = -\frac{d-2}{d-3}\, .\label{eq:kappad3}
\ee
With the further choice
\be 
\ell =  \frac{1}{d-1}\, ,
\ee
the trace $B_{\mu}$ decouples from the action~\eqref{eq:EMpalatini3} so that we obtain 
\begin{align} 
S&= -\frac{1}{g^2}\int \rd^dx\,\left[a^\nu\left(\partial^\mu s_{\mu|\nu}-\partial_\nu s - \partial^\mu \partial^\rho B_{\mu\nu | \rho}^{(T)}\right)+ \frac{1}{2}\left(s_{\mu| \nu}s^{\mu| \nu} -s^2\right) \right.\label{eq:EMpalatini4}\\
&\qquad \qquad \phantom{=}\left.-\frac{d-2}{d-3}\left(\partial _{\mu  }B_{(T)}^{\mu  \nu | \rho  } \partial^{\sigma 
}B_{\rho\sigma| \nu  }^{(T)}-\partial _{\mu  }B^{\nu \rho | \mu  }_{(T)} \partial^{\sigma  }B_{\nu \rho |\sigma  }^{(T)} +\frac{1}{2}\partial_{\sigma  }B_{\mu  \nu | \rho  }^{(T)} \partial ^{\sigma  }B^{\mu \nu|  \rho  }_{(T)}\right)\right] \nonumber\, ,
\end{align}
where we have introduced the traceless part of $B_{\mu\nu|\rho}$ defined as 
\be 
B_{\mu\nu|\rho}^{(T)} = B_{\mu\nu|\rho}- \frac{2}{d-1}\eta^{\rho[\mu}B^{\nu]}\, .
\ee
The action \eqref{eq:EMpalatini4} is invariant under the gauge transformations \eqref{eq:gaugeEM1}-\eqref{eq:gaugeEM3} with traceless gauge parameters. The fact that only the traceless part of $B_{\mu\nu|\rho}$ couples to the dynamical fields will
imply that the corresponding current $\mathcal{H}_{\mu\nu|\rho}$ is now traceless off-shell. This current can be written as 
\begin{align}
\mathcal{H}_{\mu\nu | \rho} &= \frac{1}{g^2}\left(\partial_\rho \partial_{[\mu}a_{\nu]}-\mathcal{B}_{\mu\nu|\rho} ^{(T)}\right)\, , \label{eq:femcur2}
\end{align} 
where the traceless version of ${\cal B}$ is
\be  
\begin{aligned}
\mathcal{B}_{\mu\nu|\rho}^{(T)}=\frac{d-2}{d-3}\mathcal{Y}_{(2|1)}\Big[&3 \partial^\alpha\partial_\mu  B_{\rho\alpha|\nu}^{(T)}+3 \partial^\alpha\partial_\nu  B_{\rho\alpha|\mu}^{(T)}-\square B_{\mu\nu|\rho}^{(T)}+\frac{3}{d-2}\eta_{\mu\nu}\partial^\alpha \partial^\beta B_{\alpha\rho|\beta}^{(T)}\Big]\, .\label{eq:CurlyB5}
\end{aligned}
\ee
In fact, \eqref{eq:CurlyB5} is the unique tensor that is traceless, has the correct index symmetries, and transforms appropriately under the gauge transformations, so that the current \eqref{eq:femcur2} is gauge invariant. Indeed, we could have worked directly at the level of the current and introduced \eqref{eq:CurlyB5} in order to gauge the relevant symmetries. This current satisfies 
\begin{tcolorbox}[colframe=white,arc=0pt,colback=greyish2]
\be
\begin{aligned}
\mathcal{H}\indices{^\mu_{\nu|\mu}}&= 0 \, , \qquad\qquad
& \partial^\rho \mathcal{H}_{\mu\nu|\rho} &=0 \,,\,\,\\
 \partial^\mu {\cal H}_{\mu\nu|\rho}&=0\,, &
\partial_{[\sigma}{\cal H}_{\mu\nu]|\rho} &= \partial_{[\sigma} \tl{\cal B}^{(d)}_{\mu\nu]|\rho}\, ,
\end{aligned}
\label{eq:dimeqsgravity2}
\ee
\end{tcolorbox}
\vspace{-6pt}
\noindent where the tensor appearing in the magnetic conservation equation is 
\be 
\tl{\cal B}^{(d)}_{\mu\nu|\rho} = \frac{d-2}{g^2(d-3)}\mathcal{Y}_{(2|1)}\left(-3\partial^\alpha \partial_\mu B^{(T)}_{\rho\alpha|\nu}+\square B^{(T)}_{\mu\nu|\rho}-\frac{3}{d-2}\eta_{\mu\nu}\partial^\alpha \partial^\beta B^{(T)}_{\alpha\rho|\beta}\right)\, .
\ee
Note that this differs slightly from \eqref{eq:CurlyB5} because there are contributions from the equations of motion to the right hand side of the dual conservation equation. Our inability to satisfy both the electric and magnetic conservation laws at the same time is a consequence of a mixed 't Hooft anomaly between these global biform symmetries.

\vspace{6pt}
\noindent
{\bf In $\boldsymbol{d = 3}$ dimensions:}
As we can see from this appearance of $(d-3)$ factors in the previous discussion (for example in \eqref{eq:kappad3}), something is special in $d=3$. In this case, it is impossible to choose the two free parameters $\ell$ and $\kappa$ to make the current simultaneously traceless and conserved. The  best that we can do is to satisfy the equations
\begin{tcolorbox}[colframe=white,arc=0pt,colback=greyish2]
\be
\begin{aligned}
\mathcal{H}\indices{^\mu_{\nu|\mu}}&= \tl{\cal B}_{\nu}^{(3)}\, , \qquad\qquad
& \partial^\rho \mathcal{H}_{\mu\nu|\rho} &=0 \,,\,\,\\
 \partial^\mu {\cal H}_{\mu\nu|\rho}&=0\,, &
\partial_{[\sigma}{\cal H}_{\mu\nu]|\rho} &= \partial_{[\sigma} \tl{\cal B}^{(3)}_{\mu\nu]|\rho}\, ,
\end{aligned}
\label{eq:dimeqsgravity22}
\ee
\end{tcolorbox}
\vspace{-6pt}
\noindent where the field strengths appearing in \eqref{eq:dimeqsgravity22} are 
\be 
\tl{\cal B}_{\nu}^{(3)} = -\frac{1}{g^2}\left(\partial^\alpha \partial^\beta B_{\alpha \nu|\beta}+(\partial_\nu \partial_\alpha B^{\alpha}-\square B_\nu \right)\, ,
\ee
together with 
\be  
\begin{aligned}
\tl{\cal B}_{\mu\nu|\rho}^{(3)}=\frac{1}{g^2}\mathcal{Y}_{(2|1)}\Big[&\kappa\left(3 \partial^\alpha\partial_\mu  B_{\rho\alpha|\nu}-\square B_{\mu\nu|\rho}\right)-3\left(1 + \kappa\right)\eta_{\mu\nu}\left(\partial^\alpha \partial^\beta B_{\rho\alpha|\beta}+\square B_{\rho}\right)\Big]\, .
\end{aligned}
\ee
Interestingly, we see that the minimal anomaly in $d = 3$ also involves a trace anomaly.

Here we have given a particular presentation of the mixed anomaly between the various conservation conditions. By including different contact terms (corresponding to terms quadratic in the gauge field) or gauging the theory in different variables one can shuffle around the anomaly into failures of different conservations conditions. However, the incompatibility between electric and magnetic conservations cannot be changed. This is similar to what we showed in appendix~\ref{ap:GalSup} for the Galileon superfluid.

\subsubsection{Two-point function}
We now want to consider the current two-point function in this theory and show both that the structure of mixed anomalies completely fixes its structure and that the spectral decomposition has a massless spin-1 in the spectrum.
The most general form of such a correlator consistent with Lorentz invariance is
\begin{align} 
&\braket{H_{{\mu_1}{\mu_2}|{\mu_3}}(p)*I*_{{\nu_1}{\nu_2}|{\nu_3}}(-p)}= 9\mathcal{Y}_{2,1}\mathcal{Y}_{2,1}\Big[\frac{c_1(p^2)}{8} p^2 \eta _{{\mu_1}{\nu_1} 
 } \eta _{{\mu_2}{\nu_2} } \eta _{{\mu_3}{\nu_3}
 }-\frac{c_2(p^2)}{4}  p^2 \eta
   _{{\mu_1}{\nu_1}   } \eta _{{\mu_2}  {\mu_3} }\eta _{{\nu_2}  {\nu_3} }
   \nonumber\\
   &\phantom{=}\quad\quad -\frac{c_3(p^2)}{8}
  p_{{\mu_1}}p_{{\mu_2}}\eta _{{\mu_3}{\nu_2}
   } \eta
   _{{\nu_1}  {\nu_3} } +\frac{c_4(p^2)}{4} p_{{\mu_1}}p_{{\nu_1}} \eta
   _{{\mu_2}  {\mu_3} } \eta _{{\nu_2}  {\nu_3} }+\frac{c_5(p^2)}{4}
   p_{{\mu_1}}p_{{\nu_1}} \eta _{{\mu_3}{\nu_2} 
   } \eta _{{\mu_2}{\nu_3} }\label{eq:hookhookAnsatz}\\
   &\phantom{=}\quad\quad+\frac{c_6(p^2)}{4}p_{\mu_3} p_{\nu_3} \eta
   _{{\mu_1}{\nu_1} } \eta _{{\mu_2}{\nu_2}   }+\frac{c_7(p^2)}{8} p_{\nu_1} p_{\nu_3} \eta _{{\mu_1}{\nu_2} } \eta
   _{{\mu_2}  {\mu_3} }+\frac{c_8(p^2)}{4}\frac{ 
  \eta _{{\mu_1}{\nu_1}   }p_{\mu_2} p_{\mu_3} p_{\nu_2} p_{\nu_3}}{p^2}\Big]\, ,\nonumber
\end{align}
where $I_{(d-2|d-1)}$ is the current associated to the magnetic biform symmetry and  $\mathcal{Y}_{2,1}\mathcal{Y}_{2,1}$ is the projector on the Young tableau:
\be
 \raisebox{1.25ex}{\Yboxdimx{13.5pt}
\Yboxdimy{13.5pt}\gyoung({{\hspace{.1em}\mu_1}};{{\hspace{.1em}\mu_3}},{{\hspace{.09em}\mu_2}})}~\medotimes~\, \raisebox{1.25ex}{\Yboxdimx{13.5pt}
\Yboxdimy{13.5pt}\gyoung({{\hspace{.075em}\nu_1}};{{\hspace{.05em}\nu_3}},{{\hspace{.075em}\nu_2}})}\,.\label{eq:PY2Y22}
\ee
It is again convenient to consider the generic case and $d=3$ separately.

\vspace{6pt}
\noindent
{\bf In $\boldsymbol{d \neq 3}$ dimensions:}
We can use the conservation equations we just derived to fix the current-current two-point function \eqref{eq:hookhookAnsatz}. When $d\neq 3$, we know that we can choose a traceless background gauge field  so that our current with be both traceless and conserved. The last free coefficient is fixed by the anomalous dual conservation equation \eqref{eq:dimeqsgravity2}.

Requiring both tracelessness and conservation 
\be 
\text{tr}\left(H_{{\mu}\nu|\rho}\right) = H\indices{^{\mu}_{\nu|\mu}} = 0\, , \qquad \partial^\mu H_{\mu\nu|\rho}=0\, ,
\ee
and using the anomalous conservation equation, we obtain 
\be 
c_1 = \frac{2(d-2)}{3g^2(d-3)}\, ,\qquad c_2 =  c_3 = c_4 =-c_6  = \frac{2}{g^2(d-3)}\, ,\qquad c_5 = -\frac{4(d-2)}{3g^2(d-3)}\, ,\qquad c_8 = \frac{2}{g^2}\, .
\ee
We can write the correlator in terms of the projectors that we will derive shortly. We obtain: 
\be 
\braket{H^{{\mu_1}{\mu_2}|{\mu_3}}(p)*I*^{{\nu_1}{\nu_2}|{\nu_3}}(-p)} = \frac{p^2(d-2)}{g^2(d-3)}\Pi_{(2|1)}^{{\mu_1}{\mu_2}{\mu_3}{\nu_1}{\nu_2}{\nu_3}}\, ,
\ee
where the projector is given in equation \eqref{eq:missingnames13}.

\vspace{6pt}
\noindent
{\bf In $\boldsymbol{d = 3}$ dimensions:}
In three dimensions, we can rederive the two-point function with the different set of conservation equations \eqref{eq:dimeqsgravity22}. We obtain the exact same non-local terms, while the local contact terms are different, and in particular have a trace, since we cannot require that the current be tracelessness. Nevertheless, the non-local part of the correlator is exactly the same, and hence we reach the same conclusion after performing a spectral decomposition: there is a massless state in the spectrum, which is the photon.

\subsubsection*{\KL decomposition}

We can now perform a spectral decomposition of the correlator to show that there is a massless spin-1 particle in the spectrum. Using the results from appendix~\ref{ap:KLrep}, we can write the two-point function as 
\be 
\begin{aligned}
\braket{H^{{\mu_1}{\mu_2}|{\mu_3}}(p)*I*^{{\nu_1}{\nu_2}|{\nu_3}}(-p)} = \int_0^\infty \rd s\, \frac{s^2}{p^2+s}& \bigg(\rho_{(1|0)}(s)\tl{\Pi}^{{\mu_1}{\mu_2}{\mu_3}{\nu_1}{\nu_2}{\nu_3}}_{(1|0)}-\rho_{(1|1)}(s)\tl{\Pi}^{{\mu_1}{\mu_2}{\mu_3}{\nu_1}{\nu_2}{\nu_3}}_{(1|1)}\\
&\!\!\!\!\!\!-\rho_{(2|0)}(s)\tl{\Pi}^{{\mu_1}{\mu_2}{\mu_3}{\nu_1}{\nu_2}{\nu_3}}_{(2|0)}+\rho_{(2|1)}(s)\tl{\Pi}^{{\mu_1}{\mu_2}{\mu_3}{\nu_1}{\nu_2}{\nu_3}}_{(2|1)}\bigg) ,\label{eq:missingnames12}
\end{aligned}
\ee
where $\tl{\Pi}^{{\mu_1}{\mu_2}{\mu_3}{\nu_1}{\nu_2}{\nu_3}}_{(1|0)}$ is the (off-shell) projector for a spin-$1$ particle,$\tl{\Pi}^{{\mu_1}{\mu_2}{\mu_3}{\nu_1}{\nu_2}{\nu_3}}_{(1|1)}$ is the projector for a rank-$2$ symmetric tensor particle, $\tl{\Pi}^{{\mu_1}{\mu_2}{\mu_3}{\nu_1}{\nu_2}{\nu_3}}_{(2|0)}$ is the projector a rank-$2$ antisymmetric tensor particle and $\tl{\Pi}^{{\mu_1}{\mu_2}{\mu_3}{\nu_1}{\nu_2}{\nu_3}}_{(2|1)}$ is the projector for a particle with the symmetries of a hook diagram. These projectors are transverse, traceless, and complete in the space of tensors with the symmetries of a traceless hook tableau.
Explicitly, the projectors are given by 
\begin{align}
\Pi^{{\mu_1}{\mu_2}{\mu_3}{\nu_1}{\nu_2}{\nu_3}}_{(1|0)} &= \mathcal{Y}_{(2|1)}^T \mathcal{Y}_{(2|1)}^T\left[\frac{9}{2}\frac{1}{p^2}\frac{d-1}{d-2}\eta^{{\mu_1}{\nu_1}}p^{\mu_2}p^{\mu_3}p^{\nu_2}p^{\nu_3}\right]\, ,\label{eq:missingnames12}\\
\Pi^{{\mu_1}{\mu_2}{\mu_3}{\nu_1}{\nu_2}{\nu_3}}_{(1|1)} &= \mathcal{Y}_{(2|1)}^T \mathcal{Y}_{(2|1)}^T\left[\frac{9}{4p^2}\Big(2\eta^{{\mu_1}{(\nu_1}}\eta^{{\nu_2)}{\mu_2}}p^{\mu_3}p^{\nu_3}-\frac{1}{p^2}\,\eta^{{\mu_1}{\nu_1}}p^{\mu_2}p^{\mu_3}p^{\nu_2}p^{\nu_3}\Big)\right] \, ,\\
\Pi^{{\mu_1}{\mu_2}{\mu_3}{\nu_1}{\nu_2}{\nu_3}}_{(2|0)} &=  \mathcal{Y}_{(2|1)}^T \mathcal{Y}_{(2|1)}^T\left[\frac{3}{4p^2}\Big(2\eta^{{\mu_1}{[\nu_1}}\eta^{{\nu_2]}{\mu_2}}p^{\mu_3}p^{\nu_3}-\frac{9}{p^2}\,\eta^{{\mu_1}{\nu_1}}p^{\mu_2}p^{\mu_3}p^{\nu_2}p^{\nu_3}\Big)\right]\, ,\\
\Pi^{{\mu_1}{\mu_2}{\mu_3}{\nu_1}{\nu_2}{\nu_3}}_{(2|1)} &=  \mathcal{Y}_{(2|1)}^T \mathcal{Y}_{(2|1)}^T\bigg[\frac{3}{4}\Big(\eta^{{\mu_1}{\nu_1}}\eta^{{\mu_2}{\nu_2}}\eta^{{\mu_3}{\nu_3}}-\frac{4}{p^2}\eta^{{\mu_1}{\nu_1}}\eta^{{\mu_2}{\nu_2}}p^{\mu_3}p^{\nu_3}-\frac{2}{p^2}\eta^{{\mu_1}{\nu_2}}\eta^{{\mu_2}{\nu_1}}p^{\mu_3}p^{\nu_3}\Big)\nonumber\\
&\phantom{=}\qquad \qquad \qquad + \frac{3}{4}\frac{6(d-3)}{p^4(d-2)}\eta^{{\mu_1}{\nu_1}}p^{\mu_2}p^{\mu_3}p^{\nu_2}p^{\nu_3}\bigg]\label{eq:missingnames13}\, ,
\end{align}
where $ \mathcal{Y}_{(2|1)}^T \mathcal{Y}_{(2|1)}^T$ is the traceless version of the projector \eqref{eq:PY2Y22}.
As in the other cases, we obtain the tilde tensors that appear in \eqref{eq:missingnames12} from the projectors \eqref{eq:missingnames12}--\eqref{eq:missingnames13} by replacing $p^2\rightarrow -s$. These agree with the projectors on-shell. 
The two-point function of pure Maxwell theory is recovered by taking the spectral densities to be
\be 
\rho_{(1,0)}(s) =\frac{d-2}{g^2(d-1)}\delta(s)\, ,\qquad \qquad \rho_{(1|1)}(s)  =  \rho_{(2|0)}(s)  =\rho_{(2|1)}(s)  =0\, .
\ee
This implies that there is a massless spin-$1$ particle in the spectrum. As in the previous cases, this is independent of the presentation of the anomaly, because this only changes the contact terms appearing in the correlator, to which the spectral decomposition is insensitive.

\newpage
\section{Anomaly structure in linearized gravity}
\label{app:lingravanom}

In this appendix, we elaborate on the structure of mixed anomalies in linearized gravity, and review some technical details required in the main text. In section~\ref{sec:grav2pt}, we focused on a particular presentation of the anomaly, but here we want to explore how the anomaly can be shuffled around into other conditions on the currents.

\subsection{Projectors\label{app:GravProj}}

We begin by constructing the projectors needed to perform the \KL decomposition in the main text. Note that the currents of interest can only fail to be traceless by contact terms---which can be ignored in the spectral decomposition---so we only require traceless projectors to decompose their nonlocal parts. There are three projectors, $\Pi^{(i|j)}$ for $(i,j)\in \lbrace (1,1)\, , (2,1)\, , (2,2)\rbrace$, that are labeled by the Young diagram of the representation that they carry and which have the following properties:
\begin{itemize}
\item \textit{Tracelessness: }
\be 
\eta^{\mu_1\nu_1}\Pi^{(i|j)}_{\mu_1\mu_2\nu_1\nu_2\alpha_1\alpha_2\beta_1\beta_2} = \eta^{\alpha_1\beta_1}\Pi^{(i|j)}_{\mu_1\mu_2\nu_1\nu_2\alpha_1\alpha_2\beta_1\beta_2}=0\, ,
\ee
\item \textit{Orthonormality: }
\be 
\Pi^{(i|j)\,\,\,\,\hspace{5mm}\rho_1\rho_2\sigma_1\sigma_2}_{\mu_1\mu_2\nu_1\nu_2}\Pi^{(k|l)}_{\rho_1\rho_2\sigma_1\sigma_2\alpha_1\alpha_2\beta_1\beta_2}= \delta^{ik}\delta^{jl}\Pi^{(i|j)}_{\mu_1\mu_2\nu_1\nu_2\alpha_1\alpha_2\beta_1\beta_2}\, ,
\ee
\item \textit{Completeness: }They add up to the identity on the space of traceless tensors with the symmetry of the Riemann tensor. This implies 
\be 
\Pi^{(1|1)}_{\mu_1\mu_2\nu_1\nu_2\alpha_1\alpha_2\beta_1\beta_2}+ \Pi^{(2|1)}_{\mu_1\mu_2\nu_1\nu_2\alpha_1\alpha_2\beta_1\beta_2}+\Pi^{(2|2)}_{\mu_1\mu_2\nu_1\nu_2\alpha_1\alpha_2\beta_1\beta_2} = \frac{3}{4}\mathcal{P}\Big[\eta_{\mu_1\alpha_1}\eta_{\mu_2\alpha_2}\eta_{\nu_1\beta_1}\eta_{\nu_2\beta_2}\Big]\, ,
\ee
where the right hand side is the identity on the space of traceless tensors with the symmetries of the Riemann tensor.\footnote{\label{ft:PY2Y2}The projector appearing here is ${\cal P} \equiv {\cal Y}_{(2|2)^T} {\cal Y}_{(2|2)^T} $, which is the Young projector onto the tableau~\eqref{eq:weylweylprojectortraceless}.
}

\end{itemize}

\noindent These conditions are uniquely fix these three projectors, which take the following form: The projector on the symmetric 2 tensor states is
\be 
\Pi^{(1|1)}_{\mu_1\mu_2\nu_1\nu_2\alpha_1\alpha_2\beta_1\beta_2} \equiv  {\cal P}\,\left[\frac{9(d-2)}{8(d-3)} \frac{p_{\mu_1}p_{\nu_1}p_{\alpha_1}p_{\beta_1}}{p^4}\left(\eta_{\mu_2\alpha_2}\eta_{\nu_2\beta_2} + \eta_{\nu_2\alpha_2}\eta_{\mu_2\beta_2}-\frac{2}{d-2}\eta_{\mu_2\nu_2}\eta_{\alpha_2\beta_2}\right)\right]\, .
\label{eq:Pi11defap}
\ee
The projector on hook tensor fields is given by
\be 
\Pi^{(2|1)}_{\mu_1\mu_2\nu_1\nu_2\alpha_1\alpha_2\beta_1\beta_2} \equiv  {\cal P}\left[3 \frac{p_{\mu_1}p_{\alpha_1}\eta_{\nu_1\beta_1}}{p^4}\left(-6p_{\mu_2}p_{\beta_2}\eta_{\nu_2\alpha_2} +p^2 \eta_{\mu_2\alpha_2}\eta_{\nu_2\beta_2}\right)\right]\, .
\label{eq:Pi21defap}
\ee
Finally, the projector on window fields (with the symmetries of Riemann) is given by
\be 
\begin{aligned}
\Pi^{(2|2)}_{\mu_1\mu_2\nu_1\nu_2\alpha_1\alpha_2\beta_1\beta_2} \equiv  {\cal P}&\left[\frac{3}{4}\eta_{\mu_1\alpha_1}\eta_{\mu_2\alpha_2}\eta_{\nu_1\beta_1}\eta_{\nu_2\beta_2} - \frac{3 p_{\mu_1}p_{\alpha_1}\eta_{\mu_2\alpha_2}\eta_{\nu_1\beta_1}\eta_{\nu_2\beta_2}}{p^2}\right.\\
&\left.~~+ \frac{9(d-4)}{d-3}\frac{p_{\mu_1}p_{\mu_2}p_{\alpha_1}p_{\alpha_2}\eta_{\nu_1\beta_1}\eta_{\nu_2\beta_2}}{p^4}\right]\, .
\label{eq:Pi22defap}
\end{aligned}
\ee
The objects appearing in the \KL decomposition are slightly different objects, which coincide with the projectors on shell (when $p^2 = -s$). They are given by replacing $p^2\rightarrow -s$ in the projectors,
\begin{align}
\tl{\Pi}^{(1|1)}_{\mu_1\mu_2\nu_1\nu_2\alpha_1\alpha_2\beta_1\beta_2}(s) &\equiv  {\cal P}\! \left[\frac{9(d-2)}{8(d-3)} \frac{p_{\mu_1}p_{\nu_1}p_{\alpha_1}p_{\beta_1}}{s^2}\left(\! \eta_{\mu_2\alpha_2}\eta_{\nu_2\beta_2} + \eta_{\nu_2\alpha_2}\eta_{\mu_2\beta_2}-\frac{2}{d-2}\eta_{\mu_2\nu_2}\eta_{\alpha_2\beta_2\! }\right)\! \right],
\label{eq:Pi11sdefap}\\
\tl{\Pi}^{(2|1)}_{\mu_1\mu_2\nu_1\nu_2\alpha_1\alpha_2\beta_1\beta_2} (s)&\equiv  {\cal P}\! \left[3 \frac{p_{\mu_1}p_{\alpha_1}\eta_{\nu_1\beta_1}}{s^2}\left(-6p_{\mu_2}p_{\beta_2}\eta_{\nu_2\alpha_2} -s \,\eta_{\mu_2\alpha_2}\eta_{\nu_2\beta_2}\right)\right]\, ,
\label{eq:Pi21sdefap}\\
\tl{\Pi}^{(2|2)}_{\mu_1\mu_2\nu_1\nu_2\alpha_1\alpha_2\beta_1\beta_2} (s)&\equiv  {\cal P}\! \left[\frac{3}{4}\eta_{\mu_1\alpha_1}\eta_{\mu_2\alpha_2}\eta_{\nu_1\beta_1}\eta_{\nu_2\beta_2} + \frac{3 p_{\mu_1}p_{\alpha_1}\eta_{\mu_2\alpha_2}\eta_{\nu_1\beta_1}\eta_{\nu_2\beta_2}}{s}\right.\nonumber\\
&\left.\quad\quad \quad + \frac{9(d-4)}{d-3}\frac{p_{\mu_1}p_{\mu_2}p_{\alpha_1}p_{\alpha_2}\eta_{\nu_1\beta_1}\eta_{\nu_2\beta_2}}{s^2}\right]\, .
\label{eq:Pi22sdefap}
\end{align}
\subsection{Current-current two-point function}

We now describe the systematics of the construction of the current two-point function in linearized gravity and the appearance of anomalies.

\subsubsection{Ansatz\label{sec:Ansatz}}

First, we describe the most general ansatz used to compute the current-current correlator. This object is a tensor with 8 indices that is constructed out of the metric $\eta_{\mu\nu}$ and the momentum $p^\mu$. The general structure of the ansatz is the following: 
\be
\braket{J_{{\mu_1}{\mu_2}{\mu_3}{\mu_4}}* K*_{{\nu_1}{\nu_2}{\nu_3}{\nu_4}}} =\sum_{i=1}^{27}e_i(p^2) T^{(i)}_{{\mu_1}{\mu_2}{\mu_3}{\mu_4}{\nu_1}{\nu_2}{\nu_3}{\nu_4}}\, , \label{eq:GenAns}
\ee
where there are 27 terms that are pairwise antisymmetric (meaning they are antisymmetric in $[\mu_1\mu_2]\, ,[\mu_3\mu_4]\, ,[\nu_1\nu_2]\, ,[\nu_3\nu_4]$).
The coefficients $e_i(p^2)$ are arbitrary functions of $p^2$, and we will not write the tensor structures $T^{(i)}_{{\mu_1}{\mu_2}{\mu_3}{\mu_4}{\nu_1}{\nu_2}{\nu_3}{\nu_4}}$ explicitly, as they are straightforward but tedious to obtain. Nevertheless the counting is the following: First, there are six terms which are built only out of metrics of the schematic form
\be 
T^{(i)}_{{\mu_1}{\mu_2}{\mu_3}{\mu_4}{\nu_1}{\nu_2}{\nu_3}{\nu_4}}\sim p^2 \eta_{{\mu_1}{\mu_3}}\eta_{{\mu_2}{\mu_4}}\eta_{{\nu_1}{\nu_3}} \eta_{{\nu_2}{\nu_4}} + \cdots\, ,
\ee
where the $\cdots$ contain all the contributions needed by symmetry, and we have introduced the factor of $p^2$ since we know this is how the Weyl tensor two-point function scales in linearized gravity. Second, there are eighteen terms with two momenta and three metrics which are schematically given as 
\be 
T^{(j)}_{{\mu_1}{\mu_2}{\mu_3}{\mu_4}{\nu_1}{\nu_2}{\nu_3}{\nu_4}} \sim \eta_{{\mu_1}{\mu_3}}\eta_{{\mu_2}{\mu_4}}\eta_{{\nu_1}{\nu_3}} p_{{\nu_2}}p_{{\nu_4}} + \cdots\, .
\ee
Third, there are three terms with four momenta and two metrics, which are schematically 
\be 
T^{(k)}_{{\mu_1}{\mu_2}{\mu_3}{\mu_4}{\nu_1}{\nu_2}{\nu_3}{\nu_4}} \sim \frac{1}{p^2}\eta_{{\mu_2}{\mu_4}}\eta_{{\nu_2}{\nu_4}}p_{{\mu_1}}p_{{\nu_1}} p_{{\mu_3}}p_{{\nu_3}} + \cdots\, .
\ee
Note that in the main text, we start with an ansatz that has eleven terms. This difference in counting is due to imposing the first algebraic Bianchi identity, which provides eighteen constraints amongst the free coefficients $e_i(p^2)$.

\subsubsection{Einstein gravity Weyl-Weyl two-point function}

Here we describe the calculation of the two-point function of the Weyl tensor in linearized gravity. This expression agrees with the current two-point function obtained in the main text up to local terms, as it must.

The Weyl tensor in linearized gravity is
\be 
W_{\mu\nu\rho\sigma} = R_{\mu\nu\rho\sigma} -\frac{2}{d-2}\left(\eta_{\mu[\rho}R_{\sigma]\nu}-\eta_{\nu[\rho}R_{\sigma]\mu}\right)+ \frac{2}{(d-1)(d-2)}\eta_{\mu[\rho}\eta_{\sigma]\nu}R\, .\label{eq:WeylGR2}
\ee
Using the expression for $R_{\mu\nu\rho\sigma}$ in terms of the metric $h_{\mu\nu}$, which is given in \eqref{eq:LinRiemannT}, it is straightforward to obtain the expression for $W_{\mu\nu\rho\sigma}$ in terms of $h_{\mu\nu}$. We can then use the linearized graviton propagator in de Donder gauge (see e.g. \cite{Hinterbichler:2011tt} for a derivation),
\be 
\braket{h_{\mu\nu}(p)h_{\alpha\beta}(-p)} = \frac{1}{p^2}\left(\eta_{\mu\alpha}\eta_{\nu\beta} + \eta_{\nu\alpha}\eta_{\mu\beta}-\frac{2}{d-2}\eta_{\mu\nu}\eta_{\alpha\beta}\right)\, .
\ee 
Taking the appropriate derivatives, projecting onto the right symmetry type and removing traces, we find
\be 
\braket{W_{\mu\nu\rho\sigma}(p) W_{\alpha\beta\gamma\delta}(-p)} = 9 \mathcal{P}\left[\frac{p_{\mu}p_\rho p_\alpha p_\gamma }{p^2}\braket{h_{\nu\sigma}(p)h_{\beta\delta}(-p)}\right]= 8\frac{(d-3)}{(d-2)}p^2 \Pi^{(1|1)}_{\mu\nu\rho\sigma\alpha\beta\gamma\delta} \, ,
\ee
where the projector $\mathcal{P}$ is defined in Footnote~\ref{ft:PY2Y2}.

\subsubsection{Current-current two-point function in $d>4$}

In the main text, we wrote the current-current two-point function as \eqref{eq:current2ptgenD}. Interestingly, the contact terms are exactly those that allow us to write the two-point function in terms of a different projector:
\begin{align}
\braket{J_{\mu_1\mu_2|\nu_1\nu_2}*\hspace{-1pt}K\hspace{-1pt}*_{\alpha_1\alpha_2|\beta_1\beta_2}}
 & = \frac{d-3 }{4 (d -2)}\bigg[ p^2\Pi^{(1|1)}_{\mu_1\mu_2\nu_1\nu_2\alpha_1\alpha_2\beta_1\beta_2} +{\rm contact~terms}\bigg]\, ,\\
  &= \frac{d-3}{4(d-4)}p^2 \Pi^{(2|2)}_{\mu_1\mu_2\nu_1\nu_2\alpha_1\alpha_2\beta_1\beta_2}\, .
\label{eq:current2ptgenDap}
\end{align}
The fact that the two-point function is exactly proportional to a projector---but that this is not the projector associated to the massless states in the spectral decomposition---is completely analogous to what happens  for a superfluid and for electromagnetism.

\subsubsection{Parameter counting}

We now describe all the different conditions that we can impose on the current-current two-point function. Here we slightly generalize the discussion in the main text, and allow for the currents $J_{\mu\nu|\alpha\beta}$ and $\ast K\ast_{\mu\nu|\alpha\beta}$ to be only antisymmetric in their index pairs, without assuming that the $\mu,\nu$ and $\alpha,\beta$ are symmetric under interchange. We enumerate the possible conditions that we can impose exactly (even at coincident points) and
also give the number of free coefficients $e_i(p^2)$ that are left. In each case, we will write in parentheses the $d=4$ case. In all cases, the number of free coefficients is fewer or equal in $d=4$ to the number in generic dimension. This is a consequence of dimension-dependent identities that imply that some different solutions in generic dimension are degenerate in four dimensions.

\noindent \textbf{One condition:}
Unsurprisingly, we can impose any of the four conditions we like. The parameter counting is as follows: 
\begin{itemize}
\item \textit{Tracelessness:} This yields a 9 (6 in $d=4$) parameter family of solutions.
\item \textit{Conservation: } This yields a 6 (3 in $d=4$) parameter family of solutions.
\item \textit{Dual conservation: }This yields a 3 (3 in $d=4$) parameter family of solutions.
\end{itemize}
 
\noindent \textbf{Two conditions:}
Imposing two conditions further fixes the two-point function. The different cases are:
\begin{itemize}
\item \textit{Conservation \& dual conservation:} This cannot be imposed.
\item \textit{Conservation \& tracelessness:} This yields a 3  (0 in $d=4$) parameter family of solutions. 
\item \textit{Dual conservation \& tracelessness:} This yields a 1 parameter family of solutions. 
\end{itemize}

\noindent
Including the anomalous conditions completely fixes the correlator.
\subsubsection{Four dimensions}

In four dimensions, the most we can simultaneously impose on the current is conservation. Taking account of the anomalies, yields the correlator 
\begin{align}
&\braket{J_{\mu_1\mu_2|\nu_1\nu_2}*\hspace{-1.75pt}K\hspace{-.8pt}*_{\alpha_1\alpha_2|\beta_1\beta_2}}  = {\cal P} \bigg[ -\frac{9}{32}p^2 \eta_{\mu_1\mu_2}\eta_{\nu_1\nu_2}\eta_{\alpha_1\alpha_2}\eta_{\beta_1\beta_2}+\frac{3}{16}p^2 \eta_{\mu_1\alpha_1}\eta_{\mu_2\alpha_2}\eta_{\nu_1\beta_1}\eta_{\nu_2\beta_2}\nonumber\\
& \qquad\qquad+\frac{9}{16} \eta_{\mu_1\mu_2}\eta_{\nu_1\nu_2}\eta_{\alpha_1\alpha_2}p_{\beta_1}p_{\beta_2}-\frac{3}{4}\eta_{\mu_1\alpha_1}\eta_{\mu_2\alpha_2}\eta_{\nu_1\beta_1}p_{\nu_2}p_{\beta_2}-\frac{9}{32}\eta_{\mu_1\mu_2}\eta_{\alpha_1\beta_1}\eta_{\alpha_2\beta_2}p_{\nu_1}p_{\nu_2}\nonumber\\
&\qquad\qquad+\frac{9}{4}\frac{\eta_{\mu_1 \alpha_1}\eta_{\mu_2\beta_2}p_{\nu_1}p_{\beta_1}p_{\nu_2}p_{\alpha_2}}{p^2}-\frac{9}{8}\frac{\eta_{\mu_1 \mu_2}\eta_{\alpha_1\beta_2}p_{\nu_1}p_{\beta_1}p_{\nu_2}p_{\alpha_2}}{p^2}\bigg]\, ,\label{eq:curcur2ptd4}
\end{align}
where $ \mathcal{P}$ is the projector defined in footnote~\ref{ft:PY2Y2}.

We see that in the four-dimensional case, the conditions that we can impose on our two-point function are slightly different from the generic case. One can also check that the allowed possibilities are consistent with electric-magnetic duality in linearized gravity.

\addcontentsline{toc}{section}{References}
\bibliographystyle{utphys}
{\small
\bibliography{gravityanomaly}
}

\end{document}